# Lattice Gas Automata for Reactive Systems

Jean Pierre Boon
Center for Nonlinear Phenomena and Complex Systems
Université Libre de Bruxelles - Campus Plaine - C.P.231
1050 - Bruxelles, Belgium

David Dab
Department of Chemistry
Massachusetts Institute of Technology
Cambridge, Mass. 02139, USA

Raymond Kapral
Chemical Physics Theory Group
Department of Chemistry
University of Toronto
Toronto, Ontario M5S 1A1, Canada

Anna Lawniczak
Department of Mathematics and Statistics
Guelph-Waterloo Program for Graduate Work in Physics
University of Guelph
Guelph, Ontario N1G 2W1, Canada
(October 1995)

Reactive lattice gas automata provide a microscopic approach to the dynamics of spatially-distributed reacting systems. An important virtue of this approach is that it offers a method for the investigation of reactive systems at a mesoscopic level that goes beyond phenomenological reaction-diffusion equations. After introducing the subject within the wider framework of lattice gas automata (LGA) as a microscopic approach to the phenomenology of macroscopic systems, we describe the reactive LGA in terms of a simple physical picture to show how an automaton can be constructed to capture the essentials of a reactive molecular dynamics scheme. The statistical mechanical theory of the automaton is then developed for diffusive transport and for reactive processes, and a general algorithm is presented for reactive LGA. The method is illustrated by considering applications to bistable and excitable media, oscillatory behavior in reactive systems, chemical chaos and pattern formation triggered by Turing bifurcations. The reactive lattice gas scheme is contrasted with related cellular automaton methods and the paper concludes with a discussion of future perspectives.

## Contents

# I. INTRODUCTION

Often systems with a large number of degrees of freedom exhibit macroscopic behavior where the details of the microscopic dynamics are relatively unimportant. This feature makes a general description possible: systems with different microscopic characteristics can be described on the macroscopic scale with a generic set of equations where the specific nature of the components of the system is reflected in the numerical values of a restricted number of coefficients which enter through the constitutive relations. When the macroscopic equations are recast into a non-dimensional form the global behavior of the system depends on a limited number of universal control parameters where the microscopic nature of the elementary constituents does not appear explicitly. The macroscopic level of analysis therefore provides a phenomenological description where (i) the complexity of describing the dynamics of the microscopic degrees of freedom is bypassed, and (ii) many different physical phenomena are grouped into a limited number of classes. Classical fluid mechanics offers a striking example through Reynolds' dynamical similarity law (1883).

Nevertheless, it is obvious that in this process of reasoning the connection between the phenomenology and the underlying microscopic mechanisms has been lost. Consequently, the macroscopic level of description, which employs average quantities, cannot be used to analyze how large scale phenomena are triggered by local and/or transient deviations from these averages: it neglects fluctuations. This is one of the objects of the statistical mechanical approach which establishes the microscopic basis of the properties of many-body systems. We have observed that it is not necessary to have a complete knowledge of the details of the microscopic interactions to understand how macroscopic phenomena emerge. We shall demonstrate that it is possible to develop a mesoscopic approach to macroscopic phenomena by modeling the microscopic dynamics by means of a simplified description, provided the basic requirements of fundamental physics and chemistry are correctly incorporated, i.e. the conservation laws, symmetry properties, reaction mechanisms, etc.

This philosophy has been exploited successfully in hydrodynamics and in statistical hydrodynamics where schematic microscopic models - lattice gas automata (LGA) - were initially developed with the goal to produce simplified models for statistical mechanics [1] and subsequently to provide new computational tools for the study of complicated problems in fluid dynamics [2]. Various LGA have been proposed to model macroscopic physical phenomena such as 2-d and 3-d flows at moderate Reynolds number, immiscible multi-component fluids and flows in porous media [3]. These LGA share a common structure where point-like particles move along the links of a regular lattice and interact on the nodes through collisions conserving particle number (mass), energy and momentum. The key point is that these conservation laws along with some symmetry requirements for the lattice structure suffice as basic ingredients for the emergence of macroscopic fluid behavior in agreement with Navier-Stokes equations.[1]

Quite naturally attempts were also made to construct a "lattice chemistry" in very much the same way that hydrodynamics had been successfully implemented on lattices. The study of the combination of hydrodynamics and chemistry via LGA methods could eventually lead to a simplified approach to highly complex problems such as reactive flows and combustion [4]. The operational feasibility of simulating reactive flows was demonstrated by implementing a simple specific reactive scheme ($A + B \to 2\,C$) in a two-dimensional lattice gas subject to a shear constraint [5].

The goal of the research described here is different in scope. We present a general LGA approach proceeding from micro-dynamical equations governing the dynamics of reactive "particles" to the macroscopic behavior of reaction-diffusion (RD) systems. An important distinction between macroscopic fluid dynamics as described by the Navier-Stokes equations and RD systems is that for the latter there is no unique similarity law. As a consequence, universality cannot be achieved in the same way for reactive phenomena. The LGA scheme for reactive systems therefore has as its goal the construction of a microdynamics for a class of reaction-diffusion systems which exhibit space- and time-dependent solutions corresponding to a broad variety of observed and predicted phenomena in reactive systems. Our applications include bistability, chemical wave and pattern formation processes (for example spiral waves and Turing patterns) and systems that exhibit chemical chaos.

---

[1] Galilean invariance is usually not satisfied at the microscopic level of LGA and this affects the macroscopic behavior. However, Galilean invariant macroscopic behavior can usually be obtained through scaling for quasi-incompressible, single-species flow [2].

Our approach utilizes the concepts of statistical physics which provide a mean to establish the connection between microscopic and macroscopic properties of many-body systems. In this respect, an important feature of the LGA approach is the ability to yield a mesoscopic level of description of the space- and time-dynamics which can be used to investigate the role of fluctuations in spatially-distributed reacting systems. The analysis of fluctuation correlations appears to be crucial in RD systems when macroscopic phenomena are triggered by the amplification of microscopic fluctuations. Such fluctuations are intrinsic to LGA because particles have a discrete nature in this approach and because the rules governing the dynamics are probabilistic. For instance, it is possible that global solutions obtained from the automaton may be at variance with the macroscopic behavior predicted by the phenomenological equations. The origin of such discrepancies stems from the fact that phenomenological descriptions do not incorporate naturally molecular chaos effects. One is then led to analyze the limit of validity of classical macroscopic reactive kinetics and consequently to address fundamental questions underlying the statistical mechanics of reaction-diffusion systems. The simplicity of the microscopic dynamics of the automaton also yields operational advantages with respect to classical approaches using floating point algorithms. From an operational view point, LGA provide "stable" algorithmic prescriptions for the simulation of the class of reactive phenomena considered[2]. Furthermore the LGA method offers interesting perspectives for the investigation of complex macroscopic phenomena which are difficult to treat with classical analytical or computational methods and for situations (a.o. complex media, complicated boundaries, etc.) where quantitative laboratory experiments are difficult.

Of course, other methods have been used to explore the dynamics of spatially-distributed reactive systems from a more microscopic perspective than that provided by macroscopic field equations. Full molecular dynamics gives the most detailed classical description of the reactive dynamics. Simulation of large systems, where macroscopic spatial or temporal structure is possible under far-from-equilibrium conditions, using realistic molecular potentials for the scattering events, is still beyond the scope of existing computational power. However, for model hard sphere systems where the reactive events are basically "coloring" processes there has been some work on the effects of fluctuations on limit cycle oscillations and other aspects of bifurcations in far-from-equilibrium conditions. [6–8] While these simulations are limited to relatively small numbers of particles compared to the automaton simulations, they have the advantage that the energetics of the reactive events are treated in a more sophisticated manner.

The modeling strategy and perspective taken here are also similar to those for random walk model of reacting systems. These random walk models have been used in a number of studies to explore specific reactive systems and the validity of mean field descriptions. [9,10] This class of models is contained within the general framework presented here.

In addition, reactive lattice-gas automaton models have a number of features in common with birth-death master equation models for reactive systems. [11–14] Birth-death master equation descriptions have been and continue to be used to gain an understanding of the role of fluctuations on far-from-equilibrium rate processes and should be viewed as an approach which is complementary to the lattice-gas automaton methods described here. One may also compare reactive lattice-gas automata to kinetic Ising models where one has discrete space-time dynamics with discrete spin variables. [15] Such models, in very simplified contexts, can be used to discribe reaction-diffusion dynamics at a mesoscopic level.

The paper is organized as follows. We first present in Sec. II the physical picture behind the automaton rule construction and an overview of the types of results one may obtain through its use.

The next two sections give a statistical mechanical description of the automaton for a simple one-variable system whose dynamics occurs on a square lattice. This allows us to introduce and illustrate a number of features involved in the model construction and use. Section III on diffusion is devoted to the description of non-reactive dynamics where the passage to a discrete Boltzmann equation, and subsequently to the diffusion equation, is made. Section IV considers the modifications that occur when reactive processes also take place. The combination of diffusion and reaction yields the full automaton dynamics. In this section we describe the passage via the lattice Boltzmann equation to the linearized reaction-diffusion equations and the compatibility with the phenomenological description.

---

[2]By *stable* we mean that the algorithms are exempt of numerical overflow and roundoff errors.

The general reactive lattice-gas automaton rule is stated in Sec. V. After discussing how the considerations of the previous sections can be extended to the multi-species case, a detailed strategy for the construction of the reaction probability matrix, which forms a central part of the rule construction, is given. The combination of this implementation of the reactive step with variants of the diffusion dynamics allows a variety of systems to be treated. This section constitutes a "receipe" for algorithmic implementations of the automaton.

Specific applications are treated in sections VI to IX where we present some of the most typical examples of reaction-diffusion phenomena in bistable and excitable media, Turing bifurcations and chemical chaos. The next section, Sec. X, briefly considers the generalization of the model to cases without exclusion, which allows a broader class of systems to be described. Section XI briefly comments on the relation of the reactive lattice-gas automaton considered here to other cellular automaton methods for reactive systems. The final section of the paper gives some perspectives for future work in the field.

*Hitchhikers guide to the contents:* This article was written with several types of reader in mind; so it is perhaps useful to provide a road map to its contents. For those who wish merely to gain some familarity with the ideas behind the method and its applications to specific problems we suggest reading Secs. II and VI to IX. For those who want to implement the method on a computer for their own purposes we recommend Secs. II and V–X. Readers interested in the statistical mechanics of the method should consult Secs. II-IV. Naturally, dedicated scientists are invited to read the entire paper.

## II. PHYSICAL PICTURE

Consider a reactive system where various species $X, A, B, C, \ldots$ diffuse in a solvent $S$. Suppose that the species $X$ undergoes one or a series of reactions of the form

$$\alpha X + \ldots \rightleftharpoons \beta X + \ldots, \qquad (1)$$

where the "..." represent terms involving the $A, B, C, \ldots$ species. We assume that these $A, B, C, \ldots$ species have their concentrations kept at constant values through external constraints; thus, the system is maintained in a non-equilibrium state. Then the macroscopic dynamics of species $X$ can be described by a reaction-diffusion equation

$$\frac{\partial \rho_X(\mathbf{r},t)}{\partial t} = F(\rho_X(\mathbf{r},t)) + D\nabla^2 \rho_X(\mathbf{r},t), \qquad (2)$$

where $\rho_X(\mathbf{r},t)$ is the density of species $X$ at point $\mathbf{r}$ at time $t$. In this description, the species $A, B, C, \ldots$ do not appear explicitly: their concentrations are incorporated in the reactive rate $F(\rho_X)$. The solvent, whose microscopic role is to scatter molecules through elastic collisions, is also hidden in this description where it manifests itself only through the diffusion term $D\nabla^2 \rho_X$. Now we show that a mesoscopic approach which ignores the dynamics of $A, B, C, \ldots$ and $S$ molecules and retains solely the dynamics of $X$ molecules can be constructed to yield the same macroscopic phenomenology as described in (2).

We consider a $d$-dimensional lattice with coordination number $m_c$ where molecules are modeled as point particles which undergo displacements along the links connecting nodes. Particles move on the lattice with discrete velocities, that is they hop at discrete time steps from a node to one of the neighboring nodes as dictated by the particle velocities. Each node of the lattice possesses $m$ channels where particles can reside, and an exclusion principle forbids more than one particle to reside in any channel. Therefore, the total number of particle at a node is restricted between $0$ and $m$. A velocity is associated to each channel so that a particle has the velocity of the channel on which it resides. In most of the models we consider in this paper, the number of channels $m$ is equal to the coordination numbers of the lattice $m_c$ and the allowed velocities correspond to the jumps from one node to a nearest neighbor node in one discrete time step.

The time evolution of this single species mesoscopic system occurs at discrete time steps and follows from the iterated application of an evolution operator $\mathcal{E}$ (also called the rule of the automaton)

$$[\text{State at time } (k+1)] = \mathcal{E}\,[\text{State at time } k] \qquad (3)$$

which can be conveniently decomposed into three basic operations: *propagation $P$*, *velocity randomization* or *mixing $R$*, and *chemical transformation $C$* [16]

$$\mathcal{E} = C \circ R \circ P. \qquad (4)$$

These operations will now be described in their simplest forms; generalizations are given in Sec. V.

(i) During propagation, each particle hops from its channel to the corresponding channel of a neighbor node as dictated by the particle velocity; it is a free streaming process in which the number of particles and their momenta are conserved.

(ii) In the velocity randomization step, the velocity configuration is randomly shuffled at each node where $X$ particles are redistributed amongst the channels. This operation conserves the number of particles at each node but the momentum is not conserved. The momentum changes can be viewed as elastic collisions between $X$ particles and a solvent not described on the lattice (a stochastic momentum reservoir). In this sense, the role of the velocity randomization is to modify the velocity distribution in much the same way as repeated collisions with solvent molecules would do.

(iii) During the chemical transformation, $X$ particles are created or annihilated at each node in reactions of the form

$$\alpha X \to \beta X. \qquad (5)$$

This operation is performed independently and simultaneously at each node of the lattice where a configuration with $\alpha$ $X$ particles is transformed into a configuration with $\beta$ $X$ particles, with probability $P(\alpha, \beta)$. Except for $\alpha = \beta$ (which leaves the state of the node unchanged), the reaction operator conserves neither particle number nor local momentum.

An illustration of a sequence $C \circ R \circ P$ of the automaton microscopic dynamics is shown in Fig. 1.

FIG. 1. Illustration of the microscopic dynamics of a reactive lattice gas automaton on a square lattice. Arrows indicate the presence of particles with their corresponding velocity vectors. The figure shows the successive transformations corresponding to the evolution operator $\mathcal{E} = C \circ R \circ P$: (a) state at some discrete time $k$; (b) state after propagation $P$; (c) configuration change after velocity randomization $R$; (d) state after chemical transformation $C$: this is the state at discrete time $k + 1$. Boundary conditions are periodic.

Assume, for a moment, that no reaction occurs (i.e. the reactive collision leaves the system unchanged: $C =$ identity). Then, through the repeated application of the operator $R \circ P$, $X$ particles execute random walks; it will be shown in Sec. III that diffusive behavior follows in the macroscopic limit where the system dynamics is well modeled by the diffusion equation

$$\frac{\partial \rho_X(\mathbf{r}, t)}{\partial t} = D \nabla^2 \rho_X(\mathbf{r}, t). \tag{6}$$

Now, when reactions take place, the number of particles is not conserved and a source term enters the diffusion equation to yield (2). If reactions are not frequent ($P(\alpha, \beta) \ll 1$ for $\alpha \neq \beta$), we shall see in Sec. IV that the source term takes the form of a reactive rate

$$F(\rho_X) = f(c_X) = \sum_{\alpha, \beta} (\beta - \alpha) \binom{m}{\alpha} c_X^\alpha (1 - c_X)^{m-\alpha} P(\alpha, \beta), \tag{7}$$

where $c_X = \rho_X/m$ is the $X$-particle density per channel (the channel occupation probability). The physical interpretation of (7) is as follows. When reactive processes are infrequent, particles on the lattice can be assumed to be distributed according to an equilibrium distribution. Under this local equilibrium assumption, channels are independentely populated, and the probability to have $\alpha$ particles (and $m - \alpha$ empty channels) on a node is given by a binomial distribution $\binom{m}{\alpha} c_X^\alpha (1 - c_X)^{m-\alpha}$. Multiplying this expression by $P(\alpha, \beta)$ gives the probability for producing $\beta$ $X$ particles out of $\alpha$ $X$ particles. By averaging the variation of the particle number per node $(\beta - \alpha)$ one obtains the polynomial rate given by (7).

Generally, different probabilities $P(\alpha,\beta)$ produce different reactive rates $F(\rho_X)$, which allows one to model various macroscopic phenomena. Whether a particular reactive rate $F(\rho_X)$ can be realized by an appropriate choice of values for the probabilities $P(\alpha,\beta)$, and how this choice is set, will be discussed in Sec. V. There are restrictions imposed on the types of systems that can be treated with LGA because of the exclusion principle, but as discussed in Secs. X and XI, LGA and CA models without exclusion can be construted to remove some of these restrictions.

We close this section with two illustrations of LGA model results that presage the more extensive developments in Secs. VI- IX. The first example concerns a RD system based on the reaction scheme

$$A \underset{k_{-1}}{\overset{k_1}{\rightleftharpoons}} X ,$$
$$2X + B \underset{k_{-2}}{\overset{k_2}{\rightleftharpoons}} 3X , \qquad (8)$$

known as the Schlögl model [17], one of the best-known examples of a reactive system that gives rise to bistable states. The model is usualy considered under conditions where the concentrations of $A$ and $B$ are maintained at constant values by an appropriate external feed of chemicals. In this situation, the reactive rate for $X$ is a cubic polynomial

$$F(\rho_X) = a_0 + a_1\rho_X + a_2\rho_X^2 + a_3\rho_X^3 , \qquad (9)$$

whose coefficients $a_i$ depend parametricaly on the concentrations of $A$ and $B$ and on the kinetic constants $k_{\pm 1}, k_{\pm 2}$. For appropriate values of the parameters, the rate equation $\dot\rho_X = F(\rho_X)$ shows two stable and one unstable steady solutions. Figure 2 concerns this bistable regime and shows how a system initially prepared in the homogeneous unstable state evolves through domain formation and front propagation.

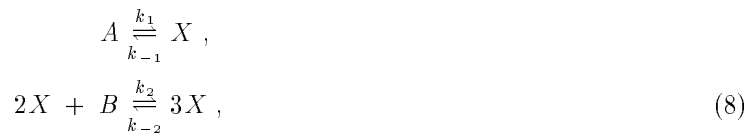

FIG. 2. Evolution from the unstable state for the Schlögl lattice-gas automaton.

Richer varieties of space- and time-dependent phenomena are seen in multi-species systems. Such systems are described by sets of RD equations for which automaton models can be constructed. As an example we show in Fig. 3 spiral wave pair formation as obtained in a mesoscopic simulation of the two-species Selkov model. Lattice gas automaton models for such multi-species systems provide tools for the study of the effects of fluctuations on a variety of chemical pattern formation processes. [18,19]

FIG. 3. Spiral wave evolution in the Selkov lattice-gas automaton.

In the next two sections we pursue the development of the reactive lattice-gas automaton for a single reactive species in order to illustrate the passage to discrete Boltzmann and linearized (reaction)-diffusion equations. The generalization to multi-species systems is outlined in Sec. V.

*Note*: For the sake of keeping with conventional notations, we use the symbol $P$ for the propagator operator and also for the probability matrices, e.g. $P(\alpha, \beta)$. However no confusion should result since all probability matrices are distinguished by their arguments.

## III. DIFFUSION

In a non-reactive cellular automaton ($C$ = identity), each particle executes a discrete random walk through the repeated application of the propagation and velocity randomization operators. Since the random walks performed by different particles are only weakly dependent[3] we can expect the macroscopic behavior of the particle density to be diffusive. Indeed, this can be formalized and demonstrated as will be shown in this section. The approach starts from the microdynamics which is expressed in terms of Boolean random variables. An exact equation is derived for the average particle populations and we show how this leads to a diffusion equation in the macroscopic limit. Fluctuations and small scale dynamics are also considered. For the sake of simplicity, the discussion is restricted to a single chemical species whose dynamics takes place on a square lattice with use of a particular velocity randomization operator.

### A. Space, time and system states

It is convenient to measure the (discrete) time $k$ by the number of applications of the evolution operator $\mathcal{E}$. With this time unit, the time interval between two successive states of the system is equal to one and $k$ takes integer values; without loss of generality we shall assume that $k = 0$ labels the initial state of the system.

The physical space $\mathcal{L}$ considered here is a rectangular subset of a 2-d square lattice. The rectangle has $L_1$ nodes along one of its edges and $L_2$ along the other edge (edges are assumed to be parallel to the main directions of the lattice). For finite values of $L_1$ and/or $L_2$ we assume periodic boundary conditions along the corresponding directions. To label the nodes of $\mathcal{L}$, we introduce a vector $\mathbf{r}$ which is conveniently projected on a basis set made of two orthogonal vectors connecting a node to two of its first neighbors. In such a basis, the components of $\mathbf{r} \equiv (r_1, r_2)$ have integer values. When the system is finite these values are considered modulo $L_1$ and $L_2$ respectively. Each node of the system can be occupied by particles differing only by their velocities which are restricted to the four unit vectors connecting a node to its nearest neighbors: $\{\mathbf{c}_i : i = 1, \ldots 4\}$.

An exclusion principle forbids more than one particle with a given velocity to be at the same node at the same time. The state of a node is therefore completely defined by four Boolean occupation variables whose values specify the states of the four channels, $w(\mathbf{r}, i)$, $i = 1, \ldots, 4$, at the node:

$$\eta_i(\mathbf{r}, k) = \begin{cases} 1 & \text{if at time } k \text{ there is a particle} \\ & \text{with velocity } \mathbf{c}_i \text{ at node } \mathbf{r} \text{ and} \\ & \text{time } k; \\ 0 & \text{if at time } k \text{ there is no particle} \\ & \text{with velocity } \mathbf{c}_i \text{ at node } \mathbf{r} \text{ and} \\ & \text{time } k. \end{cases} \tag{10}$$

When there is a particle with velocity $\mathbf{c}_i$ at node $\mathbf{r}$, the channel $w(\mathbf{r}, i)$ is said to be occupied ($\eta_i(\mathbf{r}) = 1$) and unoccupied otherwise ($\eta_i(\mathbf{r}) = 0$).

The four occupation variables $\{\eta_i(\mathbf{r}, k) : i = 1, \ldots, 4\}$ which determine the configuration of a node will often be denoted as a Boolean vector (or word)

$$\boldsymbol{\eta}(\mathbf{r}, k) = \langle \eta_i(\mathbf{r}, k) \rangle_{i=1,\ldots,4}. \tag{11}$$

Here and in the sequel the angle brackets $\langle \cdots \rangle_i$ are used to denote the elements of a vector. This vector takes its values in the state space $\mathbf{S}$ of all $2^4 (= 16)$ four bit words. The state of the full system at a given time $k$ is determined by the knowledge of all the occupation variables at that time:

---

[3] When two or more particles reside on a node, their velocities cannot be randomized independently because of the exclusion principle: if the velocity changes were independent, different particles would sometimes end up with the same velocity.

$$\boldsymbol{\eta}(\cdot, k) = \{\boldsymbol{\eta}(\mathbf{r}, k) \; ; \; \mathbf{r} \in \mathcal{L}\} \; . \tag{12}$$

The phase space in which this "Boolean field" takes values is denoted by $\Gamma$.

### B. Propagation

In the propagation step, each particle moves in the direction of its velocity vector to the first neighboring node where it occupies the channel with the same velocity label. The resulting state transformation is equivalent to that produced by a free flight of each particle during a unit time step. However we emphasize that we do *not* associate any duration to the propagation operation which is simply defined as a state transformation; this allows us to use more than one propagation operator to define the evolution operator $\mathcal{E}$.

In mathematical terms the propagation operation is defined by its action on the state variable $\mathbf{s}(\cdot)$:

$$P \; : \; \mathbf{s}(\cdot) \to \mathbf{s}^P(\cdot) \; : \; s_i(\mathbf{r}) \to s_i^P(\mathbf{r}) = s_i(\mathbf{r} - \mathbf{c}_i); \tag{13}$$

it is a permutation of the Boolean occupation variables on the lattice $\mathcal{L}$ and its inverse $P^{-1}$ is given by

$$P^{-1} \; : \; \mathbf{s}(\cdot) \to \mathbf{s}^{P^{-1}}(\cdot) \; : \; s_i(\mathbf{r}) \to s_i^{P^{-1}}(\mathbf{r}) = s_i(\mathbf{r} + \mathbf{c}_i) \; . \tag{14}$$

Similar definitions can be written for the microdynamical variables $\eta_i$.

### C. Velocity randomization

A general velocity randomization operator performs a random permutation of the channel occupation variables at each node. Here we consider particular velocity randomization operators where the permutations are selected independentely at each node by a stochastic rule which ignores the node state and its location on the lattice. In addition, we require the selection rule to be invariant under the "node symmetry group" (i.e. the subgroup of channel permutations corresponding to rotations and reflections of the velocities in physical space). The 2-d square lattice model considered here has four channels at each node and the number of different ways to permute them is $4! (= 24)$. Let $\Pi$ be this set of permutations. The velocity randomization operator is fully characterized when a probability $p_\pi$ is assigned to each possible channel permutation in $\Pi$ with the requirement that $\sum_\pi p_\pi = 1$, and $p_{\pi_1} = p_{\pi_2}$ when $\pi_1$ and $\pi_2$ are equivalent to within a node symmetry. To express the velocity randomization operator $R$ in mathematical terms, it is useful to introduce a set of random variables. Let $\tilde{\xi} = \langle \xi_\pi \rangle_{\pi \in \Pi}$ be a random Boolean vector such that:

1. for each realization of $\tilde{\xi}$ there is one and only one component $\pi$ (the randomly selected permutation) such that $\tilde{\xi}_\pi = 1$ ;

2. for each possible permutation $\pi$ the probability of the event $\tilde{\xi}_\pi = 1$ is given by $p_\pi$ .

Each time we need to perform a velocity randomization at a node we can draw a Boolean vector $\xi$ distributed as $\tilde{\xi}$ and select the permutation $\pi$ such that $\xi_\pi = 1$; all selections of $\xi$ are assumed to be mutually independent. Suppose that there is only one instance of $R$ entering into the evolution operator $\mathcal{E}$ as in (4). Then we need one copy of $\xi$ at each node $\mathbf{r}$ and at each time step $k$. Let $\xi(\mathbf{r}, k)$ be that copy. With these notations, the transformation of the state resulting from velocity randomization at time $k$ can be written as

$$s_i^R(\mathbf{r}) = \sum_{\pi \in \Pi} \xi_\pi(\mathbf{r}, k) \sum_j p_{ji}(\pi) s_j(\mathbf{r}) \; , \tag{15}$$

where $s_i$ and $s_i^R$ denote the pre- and post- velocity randomization states, respectively, and where $\{p_{ij}(\pi) : i, j = 1, \ldots 4\}$ is a Boolean matrix coding the channel permutation $\pi$ in the following way:

$$p_{ij}(\pi) = \begin{cases} 1 & \text{if the permutation } \pi \text{ maps channel } i \text{ on channel } j \text{ ;} \\ 0 & \text{otherwise.} \end{cases} \tag{16}$$

A simple example of such a velocity randomization operator is constructed as follows: the four possible rotations $(0, \pi/2, \pi, 3\pi/2)$ are equally probable[4] and all the particles residing on a given node undergo the same rotation, i.e.[5]

$$s_i^R(\mathbf{r}, k) = \begin{cases} s_i(\mathbf{r}, k) & \text{with probability } 1/4 \\ s_{i+1}(\mathbf{r}, k) & \textit{ibid.} \\ s_{i+2}(\mathbf{r}, k) & \textit{ibid.} \\ s_{i+3}(\mathbf{r}, k) & \textit{ibid.} \end{cases} \tag{17}$$

independently at each node $\mathbf{r}$ and at each time step. We term this the *4-equi-rotation velocity randomization*. An expression similar to (15) can be written for microdynamical variables.

Velocity randomization can also be defined in terms of transition probability matrices. Given a velocity randomization operator $R$, we can evaluate for each pair of configurations $\mathbf{s}$ and $\mathbf{s}'$ the probability $R(\mathbf{s}, \mathbf{s}')$ that $\mathbf{s}$ is mapped onto $\mathbf{s}'$ by $R$; these probabilities are collected in a matrix $\mathbf{R}$ with elements $\{R(\mathbf{s}, \mathbf{s}') : \mathbf{s}, \mathbf{s}' \in S\}$, the (local) transition probability matrix of $R$. Note that there are transition probability matrices which do not correspond to any velocity randomization: to represent an acceptable velocity randomization, the matrix $R(\mathbf{s}, \mathbf{s}')$ must be compatible with (i) the particle number conservation, (ii) the "state independent" nature of the permutation selection, and (iii) the node symmetry invariance. When the transition probability matrix $R(\mathbf{s}, \mathbf{s}')$ is obtained we define a matrix whose elements are Boolean random variables:

$$\{\xi_{\mathbf{ss}'} : \mathbf{s}, \mathbf{s}' \in S\} \tag{18}$$

such that:

1. for each configuration $\mathbf{s}$ there is one and only one configuration $\mathbf{s}'$ (the selected post-randomization configuration) such that $\xi_{\mathbf{ss}'} = 1$,

2. the probability of the event $\xi_{\mathbf{ss}'} = 1$ is given by $R(\mathbf{s}, \mathbf{s}')$.

Each time we need to perform a velocity randomization, we can draw a random matrix $[\xi_{\mathbf{ss}'}]$ and replace the node configuration $\mathbf{s}$ by the configuration $\mathbf{s}'$ such that $\xi_{\mathbf{ss}'} = 1$ (all selections are assumed to be mutually independent).

If only one velocity randomization is performed at each time step, then a single copy of $[\xi_{\mathbf{ss}'}]$ is needed at each node $\mathbf{r}$ and time $k$; we denote that copy by $[\xi_{\mathbf{ss}'}(\mathbf{r}, k)]$. With this notation the velocity randomization at time $k$ transforms a system state $\mathbf{s}$ into $\mathbf{s}^R$ given by

$$s_i^R(\mathbf{r}) = \sum_{\mathbf{s}''} \xi_{\mathbf{ss}''}(\mathbf{r}, k) \, s_i'' \;, \tag{19}$$

which we can also write as

$$s_i^R(\mathbf{r}) = \sum_{\mathbf{s}', \mathbf{s}''} \xi_{\mathbf{s}'\mathbf{s}''}(\mathbf{r}, k) \, s_i'' \prod_{j=1}^{4} (s_j(\mathbf{r}, k))^{(s_j')} (1 - s_j(\mathbf{r}, k))^{(1-s_j')} \;, \tag{20}$$

where we make use of the convention that $s^{s'} = 1$ when $s = s'$ ($s$ and $s'$ are Boolean variables)[6].

---

[4]The irrational number $\pi$ here should not be confused with the same symbol used earlier for a permutation.
[5]We recall that the channel indices are defined modulo 4.
[6]In this expression the product

## D. Dynamics and microdynamical equations

The simplest evolution rule we can construct with the propagation $P$ and the velocity randomization $R$ operators is one in which we apply $R$ and $P$ sequentially to make the system evolve from $k$ to $k+1$. This can be done in two different ways:

$$\mathcal{E} = R \circ P, \tag{22}$$

or

$$\mathcal{E} = P \circ R. \tag{23}$$

The corresponding dynamics

$$\boldsymbol{\eta}(\cdot, k+1) = R \circ P \, \boldsymbol{\eta}(\cdot, k), \tag{24}$$

and

$$\boldsymbol{\eta}(\cdot, k+1) = P \circ R \, \boldsymbol{\eta}(\cdot, k), \tag{25}$$

are referred to as $RP-$ and $PR$-dynamics, respectively. They are similar but not strictly equivalent because $R$ and $P$ do not commute. The main difference is that the system is always observed after the propagation step in the $PR$-dynamics, and after the velocity randomization in the $RP-$dynamics [7].

Explicit microdynamical equations are obtained by substitution of the analogs of (13) and (15) for microdynamical variables for $P$ and $R$ in (24) and (25) yielding

$$\eta_i(\mathbf{r}, k+1) = \sum_{\pi \in \Pi} \xi_\pi(\mathbf{r}, k) \sum_j p_{ji}(\pi) \, \eta_j(\mathbf{r} - \mathbf{c}_j, k), \tag{26}$$

for the $RP-$dynamics, and

$$\eta_i(\mathbf{r} + \mathbf{c}_i, k+1) = \sum_{\pi \in \Pi} \xi_\pi(\mathbf{r}, k) \sum_j p_{ji}(\pi) \, \eta_j(\mathbf{r}, k), \tag{27}$$

for the $PR-$dynamics. Alternative expressions follow from using (20) for the randomization, i.e.

$$\eta_i(\mathbf{r}, k+1) = \sum_{\mathbf{s}, \mathbf{s}'} \xi_{\mathbf{ss}'}(\mathbf{r}, k) \, s_i' \prod_{j=1}^4 \eta_j^{s_j}(\mathbf{r} - \mathbf{c}_j, k)(1 - \eta_j(\mathbf{r} - \mathbf{c}_j, k))^{(1-s_j)}, \tag{28}$$

and

$$\eta_i(\mathbf{r} + \mathbf{c}_i, k+1) = \sum_{\mathbf{s}, \mathbf{s}'} \xi_{\mathbf{ss}'}(\mathbf{r}, k) \, s_i' \prod_{j=1}^4 \eta_j^{s_j}(\mathbf{r}, k)(1 - \eta_j(\mathbf{r}, k))^{(1-s_j)}. \tag{29}$$

for $RP-$ and $PR-$dynamics respectively.

Since both dynamics are similar, we will restrict the detailed analysis to the $PR-$dynamics; the development for $RP-$dynamics follows exactly the same lines.

---

$$\prod_{j=1}^4 (s_j)^{(s_j')}(1 - s_j)^{(1-s_j')} \tag{21}$$

indicates whether the equality $\mathbf{s} = \mathbf{s}'$ is true (value 1) or not (value 0) and therefore selects only the term $\mathbf{s} = \mathbf{s}'$ in the sum over $\mathbf{s}'$. Other expressions can be used as equivalent indicators of $(\mathbf{s} = \mathbf{s}')$ but it will be seen that this particular form is the appropriate choice to derive a Boltzmann equation for reactive models.

[7] It is clear that both dynamics are very similar when the sequence $(R \circ P)^n$ is cast into the form $R \circ (P \circ R)^{(n-1)} \circ P$.

### E. Probabilistic approach

Since the random vectors $\xi = \langle \xi_\pi \rangle_{\pi \in \Pi}$ appearing in the microdynamical equations (26)–(27) are (i) equally distributed, (ii) independent of the past evolution of the system and (iii) mutually independent, it follows that the entire evolution process $\{\boldsymbol{\eta}(\cdot, k) \ : \ k = 0, 1, 2 \ldots\}$ is a stationary Markov chain defined in the phase space $\Gamma$. This process is fully characterized by its transition probability matrix with elements

$$\mathcal{E}(\mathbf{s}(\cdot), \mathbf{s}'(\cdot)) = \mathcal{P}\left(\boldsymbol{\eta}(\cdot, k+1) = \mathbf{s}'(\cdot) | \boldsymbol{\eta}(\cdot, k) = \mathbf{s}(\cdot)\right) \ , \tag{30}$$

$$= \mathcal{P}\left(\mathcal{E}\mathbf{s}(\cdot) = \mathbf{s}'(\cdot)\right) \ , \tag{31}$$

which gives for each pair of configurations $\mathbf{s}(\cdot)$ and $\mathbf{s}'(\cdot)$ the probability to find the system in a state $\mathbf{s}'(\cdot)$ at time $k+1$ when the state at the previous time $k$ was $\mathbf{s}(\cdot)$. The Chapman-Kolmogorov equation of the Markov chain

$$\mathcal{P}(\mathbf{s}'(\cdot), k+1) = \sum_{\mathbf{s}(\cdot) \in \Gamma} \mathcal{E}(\mathbf{s}(\cdot), \mathbf{s}'(\cdot)) \mathcal{P}(\mathbf{s}(\cdot), k) \ , \tag{32}$$

with

$$\mathcal{P}(\mathbf{s}(\cdot), k) \equiv \mathcal{P}(\boldsymbol{\eta}(\cdot, k) = \mathbf{s}(\cdot)) \ , \tag{33}$$

governs the evolution of probability measures defined in the phase space and can be viewed as a Liouville equation for the lattice gas automaton. Note that in statistical mechanics the Liouville equation describes the deterministic evolution of a statistical ensemble of initial configurations; in the LGA models considered here, randomness is also present in the dynamics.

The transition probability matrix $\mathcal{E}(\mathbf{s}(\cdot), \mathbf{s}'(\cdot))$ of the full dynamical process can be expressed in terms of the transition probability matrix of the operators $P$ and $R$. Since the propagation $P$ is a deterministic permutation of the channel occupation variables of the entire lattice, its transition probability matrix

$$P(\mathbf{s}(\cdot), \mathbf{s}'(\cdot)) = \mathcal{P}\left(P\mathbf{s}(\cdot) = \mathbf{s}'(\cdot)\right) \ , \tag{34}$$

takes a simple form in which each row and each column has one and only one non-zero ($=1$) element:

$$P(\mathbf{s}(\cdot), \mathbf{s}'(\cdot)) = \begin{cases} 1 & \text{if} \quad s'_i(\mathbf{r}) = s_i(\mathbf{r} - \mathbf{c}_i) \quad \forall \ \mathbf{r} \in \mathcal{L}, \\ & \hspace{3cm} \forall \ i = 1, \ldots 4; \\ 0 & \text{otherwise.} \end{cases} \tag{35}$$

Note that in the argument of $\mathcal{P}$ in (34) $P$ is the propagation operator and should be distinguished from the transition probability matrix on the left hand side.

The randomization operator $R$ is a product of mutually independent local mixing operators acting on single nodes. Each element of its transition probability matrix

$$R(\mathbf{s}(\cdot), \mathbf{s}'(\cdot)) = \mathcal{P}\left(R\mathbf{s}(\cdot) = \mathbf{s}'(\cdot)\right) \tag{36}$$

can therefore be expressed as a product of local transition probabilities $R(\mathbf{s}, \mathbf{s}')$:

$$R(\mathbf{s}(\cdot), \mathbf{s}'(\cdot)) = \prod_{\mathbf{r} \in \mathcal{L}} R(\mathbf{s}(\mathbf{r}), \mathbf{s}'(\mathbf{r})) \ . \tag{37}$$

The transition probability matrix of a particular evolution operator $\mathcal{E}$ defined as a product of $R$ and $P$ operators can be obtained from the matrices $R(\mathbf{s}(\cdot), \mathbf{s}'(\cdot))$ and $P(\mathbf{s}(\cdot), \mathbf{s}'(\cdot))$ by taking their matrix product. For instance, we have

$$\mathcal{E}(\mathbf{s}(\cdot), \mathbf{s}'(\cdot)) = \sum_{\mathbf{s}''(\cdot)} R(\mathbf{s}(\cdot), \mathbf{s}''(\cdot)) P(\mathbf{s}''(\cdot), \mathbf{s}'(\cdot)) \tag{38}$$

$$= \prod_{\mathbf{r} \in \mathcal{L}} R(\mathbf{s}(\mathbf{r}), \mathbf{s}'^{P^{-1}}(\mathbf{r})) \ , \tag{39}$$

for $PR$–dynamics. If we substitute this expression for $\mathcal{E}(\mathbf{s}(\cdot), \mathbf{s}'(\cdot))$ in the Chapman-Kolmogorov equation of the Markov chain we obtain explicit equations for the evolution of a probability measure defined in $\Gamma$:

$$\mathcal{P}(\mathbf{s}'(\cdot), k+1) = \sum_{\mathbf{s}(\cdot) \in \Gamma} \prod_{\mathbf{r} \in \mathcal{L}} R(\mathbf{s}(\mathbf{r}), \mathbf{s}'^{P^{-1}}(\mathbf{r})) \mathcal{P}(s(\cdot), k) \ . \tag{40}$$

### F. Mean values

Physically interesting quantities are defined as the expectations of functions of the Boolean variables $\eta_i(\mathbf{r}, k)$. Among them, the average number of particles per channel (*channel density*)

$$N_i(\mathbf{r}, k) = \mathrm{E}[\eta_i(\mathbf{r}, k)] \ , \tag{41}$$

plays an important role. Note that because of the Boolean nature of the occupation variable $\eta_i(\mathbf{r}, k)$ we can identify $N_i(\mathbf{r}, k)$ with the probability to find a particle with velocity $\mathbf{c}_i$ at the node $\mathbf{r}$ at time $k$ ; the density per channel is therefore equivalent to the reduced one body distribution function in statistical mechanics.

Other important average quantities are: *the local mass density* [8] obtained by summing $N_i(\mathbf{r}, k)$ over the different channels:

$$\rho(\mathbf{r}, k) = \mathrm{E}[\sum_{i=1}^{4} \eta_i(\mathbf{r}, k)] = \sum_{i=1}^{4} N_i(\mathbf{r}, k) \ , \tag{42}$$

and *the local mass current density*

$$\mathbf{j}(\mathbf{r}, k) = \mathrm{E}[\sum_{i=1}^{4} \eta_i(\mathbf{r}, k)\mathbf{c}_i] = \sum_{i=1}^{4} N_i(\mathbf{r}, k)\mathbf{c}_i \ . \tag{43}$$

It is also possible to define a *local mean velocity* by dividing the mass current by the mass density

$$\mathbf{u}(\mathbf{r}, k) = \begin{cases} \mathbf{j}(\mathbf{r}, k)/\rho(\mathbf{r}, k) & \text{if } \rho(\mathbf{r}, k) \neq 0 \\ 0 & \text{if } \rho(\mathbf{r}, k) = 0 \ ; \end{cases} \tag{44}$$

note that this quantity is not the expectation of a simple function of the occupation variables.

More involved average quantities can be defined by considering the expectation of a function involving occupation variables at different positions and times. Among them, the space- and time-dependent *density fluctuation correlation function*

$$G(\mathbf{r}, k; \mathbf{r}', k') = \mathrm{E}\left[ (\hat{\rho}(\mathbf{r}, k) - \mathrm{E}[\hat{\rho}(\mathbf{r}, k)]) \ (\hat{\rho}(\mathbf{r}', k') - \mathrm{E}[\hat{\rho}(\mathbf{r}, k)]) \right] \tag{45}$$

$$= \mathrm{E}\left[ (\hat{\rho}(\mathbf{r}, k) - \rho(\mathbf{r}, k)) \ (\hat{\rho}(\mathbf{r}', k') - \rho(\mathbf{r}, k)) \right] \ , \tag{46}$$

where $\hat{\rho}(\mathbf{r}, k) = \sum_{i=1}^{4} \eta_i(\mathbf{r}, k)$, is most important and will be considered in detail in the applications (see e.g. Sec. VI).

It is sometimes useful to consider the densities $N_i(\mathbf{r}, k)$ as the components of a vector in the Euclidean space $E_0^4$

$$\mathbf{N}(\mathbf{r}, k) = \langle N_i(\mathbf{r}, k) \rangle_{i=1,\ldots 4} \ . \tag{47}$$

---

[8] In the systems considered here the particle mass plays no role and the mass may be taken to be unity so that the mass density is equal to the number density.

In this space, the mass density is interpreted as the scalar product of $\mathbf{N}(\mathbf{r},k)$ with the particular vector whose column form is

$$\mathbf{N}_m = \begin{pmatrix} 1 \\ 1 \\ 1 \\ 1 \end{pmatrix}, \tag{48}$$

which we call *mass vector*. Similarly the two components $j_1(\mathbf{r},k)$ and $j_2(\mathbf{r},k)$ of the current density are the scalar products of $\mathbf{N}(\mathbf{r},k)$ with the particular vectors

$$\mathbf{N}_{j_1} = \begin{pmatrix} 1 \\ 0 \\ -1 \\ 0 \end{pmatrix}, \qquad \mathbf{N}_{j_2} = \begin{pmatrix} 0 \\ 1 \\ 0 \\ -1 \end{pmatrix}, \tag{49}$$

which we call the *momentum vectors in the directions 1 and 2* respectively. The set $\mathbf{N}_m$, $\mathbf{N}_{j_1}$ and $\mathbf{N}_{j_2}$ can be completed with a fourth vector

$$\mathbf{N}_q = \begin{pmatrix} 1 \\ -1 \\ 1 \\ -1 \end{pmatrix}, \tag{50}$$

to obtain a complete orthogonal basis of $E_0^4$. The projection of $\mathbf{N}(\mathbf{r},k)$ along $\mathbf{N}_q$ gives the mean value of the difference in particle number between the pairs of channels $(w(\mathbf{r},1), w(\mathbf{r},3))$ and $(w(\mathbf{r},2), w(\mathbf{r},4))$. We call this quantity the *q-density* $\rho_q(\mathbf{r},k)$:

$$\rho_q(\mathbf{r},k) = \mathrm{E}\Big[\sum_{i=1}^{4}(-1)^{i+1}\eta_i(\mathbf{r},k)\Big] = N_1(\mathbf{r},k) - N_2(\mathbf{r},k) + N_3(\mathbf{r},k) - N_4(\mathbf{r},k). \tag{51}$$

### G. Lattice Boltzmann equations

As is usual in statistical mechanics, one faces in LGA theory the complexity of systems with many degrees of freedom. Use of the Chapman-Kolmogorov equation (32) recursively allows one, at least in principle, to determine the evolution of any initial probability measure defined on $\Gamma$. In practice this is a formidable task for any system with more than a few nodes. However, full knowledge of the information contained in the probability measure is generally neither necessary nor interesting. The knowledge of the one body reduced distribution function is often sufficient; this observation calls for the elaboration of a theory establishing the evolution of the reduced distribution, bypassing the evaluation of the full probability measure.

A formal equation for the evolution of the mean occupation numbers $N_i(\mathbf{r},k)$ is obtained by taking the expectation of the microdynamical equation $\boldsymbol{\eta}(\cdot,k+1) = \mathcal{E}\boldsymbol{\eta}(\cdot,k)$:

$$\mathbf{N}(\cdot,k+1) = \mathrm{E}\Big[\mathcal{E}\,\boldsymbol{\eta}(\cdot,k)\Big] \qquad k = 0,1,\ldots \tag{52}$$

For the $PR$−dynamics this equation reduces to

$$\begin{aligned} N_i(\mathbf{r}+\mathbf{c}_i,k+1) &= \mathrm{E}\bigg[\sum_{\pi\in\Pi}\xi_\pi(\mathbf{r},k)\sum_j p_{ji}(\pi)\,\eta_j(\mathbf{r},k)\bigg] \\ &= \sum_{\pi\in\Pi} p_\pi \sum_j p_{ji}(\pi)\,N_j(\mathbf{r},k), \end{aligned} \tag{53}$$

where we have used the microdynamical equation (27), the independence of the random vectors $\xi$ and the Boolean field $\boldsymbol{\eta}(\cdot, k)$, and $\mathrm{E}[\xi_\pi] = p_\pi$. Expression(53) provides a set of closed equations governing the evolution of the mean occupation numbers. These equations exhibit similarities with the continuous Boltzmann equations of statistical mechanics: they have the same structure and are obtained in a similar way; they are therefore referred to as the *lattice Boltzmann equations* of the model. Note however that lattice Boltzmann equations (53) exhibit two particular features: they are linear and are obtained without any approximation. These features arise from the absence of real interactions between particles: propagation and mixing are state independent channel permutations. Indeed, the only "interactions" between particles are in the velocity randomization because the exclusion principle does not allow independent changes of velocities of different particles on the same node. The correlations so produced do not preclude the derivation of an exact closed equation for the evolution of the mean occupation numbers but such correlations manifest themselves as soon as the evolution of two – or many – body functions is considered. For such many-body functions, it is still possible to obtain an exact closed equation but this cannot be achieved by a simple factorization of the expectation values[9]. It should also be noticed that the Boltzmann equations (53) have been obtained from the microdynamical equations written in a form which makes it clear that the dynamics proceeds by state independent channel permutations (27). When the automaton rules do not reduce to state independent channel permutations, the microdynamical equations must be written as in (29) and approximations must be used to obtain a set of closed equations for the densities $N_i(\mathbf{r}, k)$, a situation which will be discussed in the context of reactive systems.

### H. Macroscopic behavior

To simplify the discussion, we consider the 4-equi-rotation velocity randomization operator (17) which transforms independently and simultaneously each node by performing a random rotation of the velocity configuration. When this velocity transformation is used in the $PR$-dynamics, the Boltzmann equations (53) take the explicit form:

$$
\begin{aligned}
N_1(r_1 + 1, r_2, t + 1) &= \frac{1}{4}\Big(N_1(r_1, r_2, k) + N_2(r_1, r_2, k) + N_3(r_1, r_2, k) + N_4(r_1, r_2, k)\Big), \\
N_2(r_1, r_2 + 1, t + 1) &= \frac{1}{4}\Big(N_2(r_1, r_2, k) + N_3(r_1, r_2, k) + N_4(r_1, r_2, k) + N_1(r_1, r_2, k)\Big), \\
N_3(r_1 - 1, r_2, t + 1) &= \frac{1}{4}\Big(N_3(r_1, r_2, k) + N_4(r_1, r_2, k) + N_1(r_1, r_2, k) + N_2(r_1, r_2, k)\Big), \\
N_4(r_1, r_2 - 1, t + 1) &= \frac{1}{4}\Big(N_4(r_1, r_2, k) + N_1(r_1, r_2, k) + N_2(r_1, r_2, k) + N_3(r_1, r_2, k)\Big).
\end{aligned}
\tag{54}
$$

This set of linear, finite-difference equations is conveniently solved in Fourier space. Introducing a Fourier mode[10]

$$
N_i(\mathbf{r}, k) = N_i \, \exp(i\mathbf{k} \cdot \mathbf{r}) \, A^k, \tag{55}
$$

into (54) we obtain a linear algebraic system for the $N_i$'s:

$$
\sum_{j=1}^{4} M_{ij} N_j = 0, \qquad i = 1, \ldots, 4, \tag{56}
$$

---

[9]Indeed, the BBGKY hierarchy is completely degenerate because equations (26) and (27) are linear in the variables $\eta_i(\mathbf{r}, k)$.

[10]Fourier indices will always be written in bold face and their components will be designated by numerical or roman subscripts; so there should be no confusion with the discrete time symbol $k$.

$$\mathbf{M} = \begin{pmatrix} 1/4 - A\exp(\mathrm{i}k_1) & 1/4 & 1/4 & 1/4 \\ 1/4 & 1/4 - A\exp(\mathrm{i}k_2) & 1/4 & 1/4 \\ 1/4 & 1/4 & 1/4 - A\exp(-\mathrm{i}k_1) & 1/4 \\ 1/4 & 1/4 & 1/4 & 1/4 - A\exp(-\mathrm{i}k_2) \end{pmatrix} . \tag{57}$$

Non-trivial solutions follow from the condition $\det \mathbf{M} = 0$ and we obtain a fourth order polynomial equation for the damping factor $A$:

$$\det \mathbf{M} = A^3 \left(A - \frac{\cos(k_1) + \cos(k_2)}{2}\right) = 0 . \tag{58}$$

The solutions $A^{(j)}(\mathbf{k})(j = 1, ..., 4)$ and the corresponding vectors $\mathbf{N}^{(j)}(\mathbf{k})$ are given by[11]

$$A^{(1)}(\mathbf{k}) = 1/2 \Big(\cos(k_1) + \cos(k_2)\Big) \; ; \mathbf{N}^{(1)}(\mathbf{k}) = \begin{pmatrix} \exp(-\mathrm{i}k_1) \\ \exp(-\mathrm{i}k_2) \\ \exp(\mathrm{i}k_1) \\ \exp(\mathrm{i}k_2) \end{pmatrix} , \tag{59}$$

$$A^{(2)}(\mathbf{k}) = 0 \; ; \mathbf{N}^{(2)}(\mathbf{k}) = \begin{pmatrix} 1 \\ 0 \\ -1 \\ 0 \end{pmatrix} , \tag{60}$$

$$A^{(3)}(\mathbf{k}) = 0 \; ; \mathbf{N}^{(3)}(\mathbf{k}) = \begin{pmatrix} 0 \\ 1 \\ 0 \\ -1 \end{pmatrix} , \tag{61}$$

$$A^{(4)}(\mathbf{k}) = 0 \; ; \mathbf{N}^{(4)}(\mathbf{k}) = \begin{pmatrix} 1 \\ -1 \\ 1 \\ -1 \end{pmatrix} . \tag{62}$$

Since these vectors are linearly independent for all values of $\mathbf{k}$, it follows that a general solution of the Boltzmann equation (54) is given by

$$N_i(\mathbf{r}, k) = \sum_{j=1}^{4} \sum_{k_1=0}^{L_1} \sum_{k_2=0}^{L_2} a^{(j)}(\mathbf{k}) N_i^{(j)}(\mathbf{k}) \exp\left(2\pi\mathrm{i}\frac{k_1}{L_1}r_1\right) \exp\left(2\pi\mathrm{i}\frac{k_2}{L_2}r_2\right) \left(A^{(j)}(\mathbf{k})\right)^k , \tag{63}$$

for a $L_1 \times L_2$ (periodic) lattice, and by

$$N_i(\mathbf{r}, k) = \sum_{j=1}^{4} \int_{-\pi}^{\pi} dk_1 \int_{-\pi}^{\pi} dk_2 \, a^{(j)}(\mathbf{k}) N_i^{(j)}(\mathbf{k}) \exp(\mathrm{i}\mathbf{k} \cdot \mathbf{r}) (A^{(j)}(\mathbf{k}))^k , \tag{64}$$

for an infinite lattice, where the $a^{(j)}(\mathbf{k})$ are constants fixed by the initial condition.

The macroscopic behavior of the system is determined by the dynamics of the long wave length Fourier modes

$$\exp(\mathrm{i}\mathbf{k} \cdot \mathbf{r})(A^{(j)}(\mathbf{k}))^k \mathbf{N}^{(j)}(\mathbf{k}) , \qquad |\mathbf{k}| \to 0 ; \tag{65}$$

i.e., by the asymptotic properties of the damping factors $A^{(j)}(\mathbf{k})$ when $|\mathbf{k}| \to 0$. From the expressions (59)–(62) we have

$$A^{(1)}(\mathbf{k}) = 1 - \frac{1}{4}|\mathbf{k}|^2 + O(\mathbf{k}^4) \,, \tag{66}$$

and

$$A^{(2)}(\mathbf{k}) = A^{(3)}(\mathbf{k}) = A^{(4)}(\mathbf{k}) = 0 \,. \tag{67}$$

---

[11] Here the kernel corresponding to $A(\mathbf{k}) = 0$ is spanned by the vectors (49) and (50), an appropriate choice as the components in this basis have a physical interpretation .

Two different types of temporal behavior appear:

1. The damping factor $A^{(1)}$ converges to 1 when $|\mathbf{k}|$ goes to zero ; i.e., the corresponding solutions relax arbitrarily slowly when the wavelength becomes very large. In the limit $|\mathbf{k}| = 0$, $A^{(1)} = 1$ and the solution is time independent.

2. The other damping factors $A^{(j)}$ ($j = 2, 3, 4$) are equal to zero and the corresponding solutions relax to zero in one time step independently of the value of $|\mathbf{k}|$.

Consequently, when $|\mathbf{k}|$ is sufficiently small, there is one macroscopic mode persisting over long times ($A \simeq 1$), and three microscopic modes ($A = 0$) which are unobservable on macroscopic time scales.

When $|\mathbf{k}| \to 0$, the damping factor $A^{(1)}(\mathbf{k})$ corresponding to the persisting mode is close to one and can be expressed as the exponential of an equivalent continuous damping rate $\omega(\mathbf{k})$

$$A^{(1)}(\mathbf{k}) = \exp(\omega(\mathbf{k})) \,, \tag{68}$$

or

$$\omega(\mathbf{k}) = \ln(A^{(1)}(\mathbf{k})) \,, \tag{69}$$

which, with (66), yields

$$\omega(\mathbf{k}) = -\frac{1}{4}|\mathbf{k}|^2 + O(|\mathbf{k}|^4) \,. \tag{70}$$

Equation (70) is the dispersion relation of the diffusion equation

$$\frac{\partial}{\partial t}\rho(\mathbf{x}, t) = D\frac{\partial^2}{\partial \mathbf{x}^2}\rho(\mathbf{x}, t) \,, \tag{71}$$

with

$$D = \frac{1}{4} \,. \tag{72}$$

In this sense, the long wavelength persisting modes in the LGA are diffusive.

Notice that an isotropic dispersion relation is obtained to order $O(|\mathbf{k}|^2)$ where the orientational discreteness of the lattice is not visible. However, anisotropic terms appear at order $O(|\mathbf{k}|^4)$ where the fine structure of the lattice emerges. This result is not specific to the particular model considered and holds for a large variety of systems when discrete isotropy is present at the microscopic scale.

A physical interpretation of the diffusive modes is obtained from the component of $\mathbf{N}^{(1)}(\mathbf{k})$ in the basis $\{\mathbf{N}_m, \mathbf{N}_{j_1}, \mathbf{N}_{j_2}, \mathbf{N}_q\}$ (48)–(50). Expanding these components to first significant order in $\mathbf{k}$ yields

$$4 + O(|\mathbf{k}|^2) \,, \tag{73}$$

$$-2ik_1 + O(|\mathbf{k}|^3) \,, \tag{74}$$

$$-2ik_2 + O(|\mathbf{k}|^3) \,, \tag{75}$$

and

$$-2\mathrm{i}k_1 + 2\mathrm{i}k_2 + O(|\mathbf{k}|^3) , \tag{76}$$

respectively. We observe that the projection on the mass vector is dominant when $|\mathbf{k}| \to 0$. Hence, the long wavelength persisting mode of the $PR$−dynamics describes the diffusion of the mass density.

It is worth mentioning that the $RP$−dynamics produces slightly different modes because, in that case, the state of the system is always observed after the velocity randomization. Indeed, when the velocity randomization is the 4-equi-rotation operator (17), the mass current density and the $q$−density are projected to zero before they are observed. On the contrary, in $PR$−dynamics the state of the system is always observed after propagation, a step during which the channel densities propagate along the lattice links and produce local $q$−densities and mass current densities where density gradients are present.

### I. Conserved quantities and spurious modes

In the previous subsection, the time evolution of the long wave length modes was determined. Here we obtain a complementary characterization of the dynamics by considering the undamped modes and their spatial properties.

The undamped modes of the $PR$−dynamics with the 4-equi-rotation velocity randomization (17) are obtained from (59)–(62) by solving $|A^{(j)}(\mathbf{k})| = 1$ for $j$ and $\mathbf{k}$. Two solutions are found[12]:

1. For $j = 1$ and $\mathbf{k} = 0$ we have $A^{(1)}(\mathbf{k}) = 1$ and the corresponding eigenvector $\mathbf{N}^{(1)}(\mathbf{k})$ is the mass vector $N_m$ (see (59)). We recover mass conservation.

2. For $j = 1$ and $k_1 = k_2 = \pi$ we have $A^{(1)}(\mathbf{k}) = -1$ and the corresponding eigenvector $\mathbf{N}^{(1)}(\mathbf{k})$ is along the mass vector $N_m$ (see (59)). We have an inhomogeneous mode oscillating with period 2. The corresponding invariant is the difference of mass on odd and even nodes (the parity of a node $\mathbf{r}$ is defined as the parity of the sum of its coordinates $r_1 + r_2$). It is a cyclic invariant which changes sign at each time step. In the lattice gas automata literature, this is known as the *checkerboard parity* invariant. Invariants which, like the checkerboard parity invariant, do not have a counterpart in real systems are called spurious invariants, and the corresponding slowly decaying modes are called *spurious modes* [20], [21].

    It should be noted that the wave vector $\mathbf{k} = (\pi, \pi)$ is not compatible with the boundary conditions when at least one of the system lengths $L_1$ or $L_2$ is odd. In that case the checkerboard parity invariant is absent in the dynamics but very slow decays are observed for modes corresponding to wave vectors close to $\mathbf{k} = (\pi, \pi)$ ; their decay time is of the order of the square of the system size.

A geometric interpretation of the checkerboard parity invariance can be given. If in a system with $L_1$ and $L_2$ even, the lattice nodes are painted as a periodic checkerboard, each particle has a trajectory visiting alternatively a node of each color. As a result two particles on differently colored nodes will never interact and the cellular automaton universe consists of two totally independent subsystems corresponding to alternate colors at alternate time steps.

If $L_1$ or $L_2$ is odd, the lattice cannot be painted globally as a periodic checkerboard, but this can still be done locally. As a result two particles occupying neighboring nodes of different colors cannot interact before their mutual distance reaches at least one half of the smallest odd size of the system. This explains why the checkerboard parity mode becomes a very slow mode, with a decay time depending on the system size, when $L_1$ and/or $L_2$ is odd.

The cherkerboard parity invariance involves a microscopic wave length and therefore its influence on the macroscopic behavior can be thought to be negligible. This is indeed true when fluctuations are ignored and when the Boltzmann equations are linear and obtained without any assumption. In this case, the modes around $\mathbf{k} = (\pi, \pi)$ are weakly excited by macroscopic initial conditions and the Boltzmann equations guarantee that they remain so at any later time. In addition there is no coupling between modes, which protects the diffusive modes from the influence of the other modes.

---

[12]The same solutions are found when $RP$−dynamics is considered.

When reactive processes are included in the dynamics, it is not possible to derive a Boltzmann equation in a rigorous way because reactive interactions couple the $N_i$ dynamics to correlation functions of arbitrarily high order; in addition the Boltzmann equations are then usually non-linear. For these reasons, a mode initially excited with a very small amplitude is not guaranteed to remain so when the system evolves. Indeed, a weakly excited mode can be amplified to the macroscopic level when spontaneous symmetry breaking occurs. In this way, spurious modes can emerge at the macroscopic level even when they are absent from the initial condition; once excited, they can remain in the system for a very long or infinite time. Consider for instance a system with the checkerboard parity invariance, and suppose that this system is initially prepared in a microscopic state compatible with a smoothly varying density field $N_i(\mathbf{r})$. If a spontaneous symmetry breaking occurs, it usually arises differently in the two checkerboard subsystems because they are completely disconnected. As a result, the density field $N_i(\mathbf{r}, k)$ can evolve very differently in the two subsystems which makes the amplitude of the checkerboard parity mode grow from a microscopic fluctuation to the macroscopic level. If the two subsystems are not considered separately, unphysical results can be obtained. If one wants to keep the two checkerboard subsystems unseparated, one can remove the checkerboard parity invariance by introducing at each node additional "channels" for rest particles (i.e. particles which do not propagate but are included in the velocity - or channel occupation - randomization). Indeed, the rest particles couple the checkerboard subsystems with one another, and as a result, the undamped checkerboard mode gives way to a decaying mode with a decay time depending on the number of rest particles per node and the efficiency of the mixing between rest and moving particles; with a small number of rest particles per node this time can be set to the order of an elementary time step.

Since spurious modes can contaminate the physically interesting macroscopic behavior, they must be identified and, as far as possible, eliminated. If some spurious invariants remain in the dynamics, it is important to check that they do not affect the macroscopic behavior. It should be emphasized that spurious invariants are frequent in lattice gas automata because of the discrete nature of the models. As an example, the *2-equi-rotation velocity randomization* involving only *two* different rotation angles

$$s_i^R(\mathbf{r}, k) = \begin{cases} s_{i+1}(\mathbf{r}, k) & \text{with probability } 1/2 \, , \\ s_{i+3}(\mathbf{r}, k) & \text{with probability } 1/2 \, , \end{cases} \tag{77}$$

that was used in the early developments of reactive lattice gas automata, yields up to seven spurious invariants when it is combined with propagation in a $RP-$ or $PR-$dynamics; a detailed description is presented in [22].

Other schemes for removing the checkerboard invariant that do not involve rest particles may also be constructed. For instance, the propagation and velocity randomization operations can be modified to couple the two sublattices or the nature of the lattice can be changed. Some of the simulations reported in the applications (Sec.VI to IX) were carried out on hexagonal lattices to avoid such spurious invariants.

### J. Equilibrium states

The long time behavior of the channel densities $N_i(\mathbf{r}, k)$ obtained in the previous section is an important result but an incomplete characterization of the asymptotic dynamics since it does not provide any information about correlations. Here we show that a complete characterization of the asymptotic statistical state of the system can be obtained when the dynamics is based on the 4-equi-rotation velocity randomization operation and when the system has finite size. Only the main lines of reasoning will be sketched and the discussion will be restricted to a finite rectangular system with even dimensions $L_1$ and $L_2$, and with periodic boundary conditions; a detailed discussion is presented in [22]. The important hypothesis is that the system is finite. One can proceed along the same lines for other geometries, but it should be emphasized that different geometries can produce a different asymptotic behavior because the presence of a global checkerboard parity invariant is very sensitive to the boundary conditions. Here $L_1$ and $L_2$ are both even and therefore the system has the checkerboard parity invariance. The analysis given in [22] is as follows:

1. One first considers the evolution of the automaton at even times
$$\left\{\eta(\mathbf{r},k) \ : k = 0,2,4,6,\ldots\right\} \ ; \tag{78}$$
   this process is a stationary, finite, Markov chain.

2. Noting that the total number of particles remains constant separately in the two checkerboard subsystems, one splits the initial chain into reduced chains corresponding to fixed numbers of particles on odd and even nodes (at even times the two checkerboard subsystems can be identified with odd and even nodes respectively).

3. One shows that the reduced chains so-obtained are aperiodic; this result is obtained by showing that the random sequence $P \circ R \circ P \circ R$ reduces to the identity with non-zero probability.

4. One also shows that the reduced chains are irreducible: starting from a given lattice configuration with $M_e$ and $M_o$ particles on even and odd nodes, respectively, the dynamics can lead to any other lattice configuration with the same numbers of particles in the two subsystems.

5. The above properties allow one to make use of a classical theorem (see for instance [23]) stating that a Markov chain which is finite, aperiodic and irreducible
   (a) has a unique invariant measure ;
   (b) is ergodic ;
   (c) is mixing (in the sense that any initial probability distribution converges towards the invariant measure).

6. One shows that the probability measure which assigns the same weight to any lattice configuration with $M_e$ and $M_o$ particles on even and odd nodes, respectively, is an invariant measure of the corresponding reduced chain. From the abovementioned theorem, this probability measure is the unique invariant measure towards which all initial measures converge.

7. Having obtained the invariant measure at even times one determines what this measure becomes at odd times.

This yields the following result. Suppose we have a dynamics based on the 4-equi-rotation velocity randomization combined with the propagation in a $RP-$ or $PR-$dynamics. Suppose also that the system resides on a finite rectangle of $L_1 L_2$ nodes with $L_1$ and $L_2$ even and with periodic boundary conditions. Suppose finally that the initial configuration $\eta(\cdot, 0)$ has $M_e$ and $M_o$ particles on even and odd nodes respectively. Then, the probability $\mathcal{P}(\eta(\cdot,k) = \mathbf{s}(\cdot))$ to find the system in state $\mathbf{s}(\cdot)$ at time $k$ obeys the following relations

$$\lim_{k \to \infty} \mathcal{P}(\boldsymbol{\eta}(\cdot,2k) = \mathbf{s}(\cdot)) = \begin{cases} \frac{M_e! M_o! ((L_1 L_2/2) - M_e)! ((L_1 L_2/2) - M_o)!}{(L_1 L_2/2)! (L_1 L_2/2)!} & \text{if state } \mathbf{s}(\cdot) \text{ has } M_e \text{ and } M_o \text{ particles on even and odd nodes, respectively ;} \\ 0 & \text{otherwise.} \end{cases} \tag{79}$$

$$\lim_{k \to \infty} \mathcal{P}(\eta(\cdot,2k+1) = \mathbf{s}(\cdot)) = \begin{cases} \frac{M_e! M_o! ((L_1 L_2/2) - M_e)! ((L_1 L_2/2) - M_o)!}{(L_1 L_2/2)! (L_1 L_2/2)!} & \text{if state } \mathbf{s}(\cdot) \text{ has } M_e \text{ and } M_o \text{ particles on odd and even nodes, respectively ;} \\ 0 & \text{otherwise.} \end{cases} \tag{80}$$

This provides a full characterization of the asymptotic behavior of any initial probability measure.

Three consequences of these results are particularly important:

1. When a $RP-$ or $PR-$dynamics is based on the equi-rotation velocity randomization, the dynamics has exactly two independent global invariants: the total mass in each checkerboard subsystem.

2. In each checkerboard subsystem the equilibrium state is fully characterized by a single parameter: the density per channel.

3. In a large system at equilibrium, the occupation Boolean variables $\eta_i(\mathbf{r})$ of a small number $l$ of channels ($l \ll L_1 L_2$) can be considered as almost mutually independent.

### K. Observation of diffusion

In the previous subsections we showed how a probabilistic description of the automaton leads to diffusive behavior: the long wavelength modes which survive over long times are characterized by a dispersion relation whose leading term is diffusive: $\omega(\mathbf{k}) \propto |\mathbf{k}|^2$. Here we briefly discuss the conditions that must be fulfilled to observe diffusive behavior in an automaton simulation.

A first requirement is that the simulation must be performed on a large length scale so that the dispersion relation of the excited modes is close to $\omega(\mathbf{k}) \propto |\mathbf{k}|^2$. Indeed the dispersion relation is never exactly diffusive (see (70)) and the length scale depends on the required level of accuracy. Since the precision is frequently limited by other factors, wavelengths of the order of 20 to 80 lattice links are in practice often sufficient. Of course, long length scales imply long time scales for observing diffusive behavior; this is because, at a constant value of $D$, time scales like the square of the length.

Since it is the density that diffuses (and not the particles which only perform random walks), diffusive behavior is seen only when the densities $N_i(\mathbf{r},k)$ can be inferred from the random variables $\eta_i(\mathbf{r},k)$. The most direct way to do so is to consider an ensemble of $n_r$ independent realizations of the dynamics

$$\{\eta_i^{(h)}(\mathbf{r},k) \ : \ h = 1, 2, \ldots, n_r\} \ , \tag{81}$$

and estimate the probability $N_i(\mathbf{r},k)$ by the mean

$$\frac{1}{n_r} \sum_{h=1}^{n_r} \eta_i^{(h)}(\mathbf{r},k) \ . \tag{82}$$

Unfortunately, this method does not generalize nicely to reactive systems where important phenomena — e.g. spontaneous symmetry breaking — occur in single realizations of the dynamics. In these situations, performing an average over an ensemble of systems is still possible but not very relevant. For those systems the density $\rho$ in the phenomenological reaction-diffusion equation and the lattice Boolean variables $\eta_i(\mathbf{r},k)$ must be connected in single realizations of the dynamics. This is accomplished by spatial and/or temporal averaging of the Boolean variables $\eta_i(\mathbf{r},k)$. However larger systems may be necessary to obtain a correct spatial resolution.[13]

### L. Evaluation of the diffusion coefficient

When performing simulations, it is quite useful to have a general expression for the diffusion coefficient in terms of the probabilities governing the velocity randomization. Then the diffusion coefficient can be given particular values by tuning the probabilities. For general velocity randomization and/or complex lattices, it may be difficult to obtain an explicit expression for $D$ from the Boltzmann equations since this requires the solution of a generalized eigenvalue problem.

---

[13]In addition, a spatio-temporal averaging acts like a filter which modifies the observed diffusion coefficient. This effect can be important when the wavelength is not very large compared to the typical length of averaging.

Here we derive a general expression for the diffusion coefficient by evaluating the time dependence of the mean square displacement of a tagged particle and by using Einstein's formula:

$$D = \lim_{k \to \infty} \frac{1}{2dk}\sigma^2(k) , \tag{83}$$

where $d$ is the dimension of space and $\sigma^2(k)$ is the average squared displacement performed by a tagged particle during a time interval $k$. In a lattice gas automaton one must specify how a particle is tagged in order to use (83). Indeed, different tagging rules are possible and therefore different values of $D$ can be obtained with Einstein's formula. This stems from the fact that the diffusion coefficient associated with a single tagged particle does not necessarily coincide with the diffusion coefficient that is relevant for mass transport in a many-particle system, which is what we are interested in. Therefore we should find a tagging rule such that both diffusion coefficients have the same value. However for the automaton with state independent velocity randomization, the diffusion coefficient is independent of the density and consequently we may as well consider a system with *one* single particle thus bypassing the tagging rule.

In order to illustrate the method, consider the velocity randomization defined by (17) but where the four possible rotations can have different probabilities

$$s_i^R(\mathbf{r}, k) = \begin{cases} s_i(\mathbf{r}, k) & \text{with probability } p_0 \\ s_{i+1}(\mathbf{r}, k) & \text{with probability } p_1 \\ s_{i+2}(\mathbf{r}, k) & \text{with probability } p_2 \\ s_{i+3}(\mathbf{r}, k) & \text{with probability } p_1 \end{cases} \tag{84}$$

with $p_0 + 2\,p_1 + p_2 = 1$. Note that the probabilities for the rotations $\pi/2$ and $3\pi/2$ are equal according to our general assumption that the velocity randomization is invariant under the node symmetry group. Let

$$\{\mathbf{r}_1, \mathbf{r}_2, \mathbf{r}_3, \ldots, \mathbf{r}_k, \ldots\} \tag{85}$$

$$\mathbf{r}_k \in \{\mathbf{c}_i, \quad i = 1\ldots 4\} . \tag{86}$$

be the succesive displacements of the tagged particle. This sequence is a stationary Markov chain whose transition probability measure

$$P_{i,j} = \mathcal{P}(\mathbf{r}_{k+1} = \mathbf{c}_j | \mathbf{r}_k = \mathbf{c}_i); \qquad i,j = 1\ldots 4 , \tag{87}$$

is given by

$$\begin{pmatrix} p_0 & p_1 & p_2 & p_1 \\ p_1 & p_0 & p_1 & p_2 \\ p_2 & p_1 & p_0 & p_1 \\ p_1 & p_2 & p_1 & p_0 \end{pmatrix} . \tag{88}$$

Except for special cases, this Markov chain is irreducible and aperiodic so that it has a single invariant measure towards which any initial measure converges; this is the uniform distribution. When the chain is reducible or aperiodic, a velocity related spurious invariant exists and the dynamics should be modified to remove it. Let

$$\mathbf{d}(k) = \sum_{n=1}^{k} \mathbf{r}_n , \tag{89}$$

be the total displacement after $k$ time steps and assume that the initial condition is the invariant measure

$$\mathcal{P}(\mathbf{r}_1 = \mathbf{c}_i) = 1/4, \qquad \forall i ; \tag{90}$$

by definition, the same distribution applies at any later time. Under these conditions, the expectation of $\mathbf{d}(k)$ is equal to 0 at any time and

$$\text{var}[\mathbf{d}(k)] = \text{E}[\mathbf{d}^2(k)] = k + 2 \sum_{\substack{n,n' \\ n'<n}}^{k} \text{E}[\mathbf{r}_{n'} \cdot \mathbf{r}_n] \,, \tag{91}$$

which reduces further (*cf.* initial conditions and stationarity) to

$$\text{var}[\mathbf{d}(k)] = k + \frac{1}{2} \sum_{n=1}^{k-1}(k-n) \sum_{i=1}^{4} \sum_{j=1}^{4} \mathbf{c}_i \cdot \mathbf{c}_j \ [P_{i,j}]^n \,. \tag{92}$$

This expression can be rewritten as

$$\text{var}[\mathbf{d}(k)] = k + \frac{1}{2} \sum_{n=1}^{k-1}(k-n) \, Tr(\mathbf{GP}^n) \,, \tag{93}$$

where the matrix $\mathbf{G}$ is $G_{i,j} = \mathbf{c}_i \cdot \mathbf{c}_j$. The trace $Tr(\mathbf{GP}^k)$ on the r.h.s. of (93) is easily obtained in a basis where $\mathbf{P}$ and $\mathbf{G}$ are simultaneously diagonal (see Table I). The sum in (93) can then be evaluated and Einstein's formula (83) yields

$$D = \frac{1}{4} \frac{p_0 - p_2 + 1}{p_2 - p_0 + 1} \,. \tag{94}$$

For additional information on diffusion in lattice gas automata see Ref. [24].

| Vectors | Eigenvalues of $\mathbf{P}$ | Eigenvalues of $\mathbf{G}$ |
|---|---|---|
| $(1,1,1,1)$ | 1 | 0 |
| $(1,-1,1,-1)$ | $p_0 + p_2 - 2p_1$ | 0 |
| $(1,1,-1,-1)$ | $p_0 - p_2$ | 2 |
| $(1,-1,-1,1)$ | $p_0 - p_2$ | 2 |

TABLE I. Eigenvalues and eigenvectors of $\mathbf{G}$ and $\mathbf{P}$.

## IV. REACTION-DIFFUSION

In a reactive LGA, the dynamics is built from three basic operators: the velocity randomization operator $R$, the propagation operator $P$, and the chemical transformation operator $C$, as described in Sec. II. At first sight, the chemical transformation operator $C$ can be viewed as a simple extension of the velocity randomization operator $R$: both operators modify each node configuration independently by a stochastic transformation specified by a transition probability matrix. The difference is that $R$ conserves the total number of particles while $C$ does not. However, a subtle feature makes these two operators very different: when the velocity randomization is applied to a node configuration, the new average populations $N_i$ can be predicted solely on the basis of the previous average populations (see (53)). This is not the case for the chemical transformation where the full joint probability distribution of the different channels is needed to predict how the average populations $N_i$ are transformed. The difference arises because the occurrence of a particular reactive event $\alpha \to \beta$ requires exactly $\alpha$ particles on the node before the reaction, a condition whose probability depends not only on the average population in each channel but also on correlations. As a result, a closed equation for the average populations $N_i(\mathbf{r}, k)$ cannot be obtained for a generic reactive LGA unless one considers particular regimes where approximations can be introduced to disconnect the dynamics of average populations from correlations. In this section, we consider the regime where reactions are infrequent and the Boltzmann approximation can be used to derive kinetic equations for the average population variables. The validity of these equations and how they lead to a macroscopic reaction-diffusion equation is discussed.

### A. From microdynamics to Boltzmann equations

For simplicity, we restrict the presentation to the dynamics generated by an evolution operator of the form

$$\mathcal{E} = P \circ C \circ R . \tag{95}$$

Before we begin the discussion of this dynamics, it is important to specify how a change in particle number $\alpha \to \beta$, selected according to the transition matrix $P(\alpha, \beta)$, is translated into a configuration change $\mathbf{s} \to \mathbf{s}'$. Each time particles are created or annihilated by the chemical transformation operator $C$, a modification is also produced in the velocity distribution. Unless models are designed to account for specific changes of the velocity distribution by reactive collisions, the simplest ansatz is to completely randomize the velocity distribution and neglect the reactive contributions to the diffusion coefficient [14]. However, this ansatz makes unnecessary changes in the velocity distribution and, therefore, does not minimize the reactive contributions to the diffusion coefficient. These contributions can be reduced by performing the transitions $\alpha \to \beta$ in the following way:

- When a non-reactive event $\alpha \to \alpha$ is selected, the initial configuration is not modified: $\mathbf{s} \to \mathbf{s}$.

- When a creation event is selected $\alpha \to \beta$ ($\alpha < \beta$), the particles already present before the reaction are unaffected while new particles are created on empty channels (all combinations being equally probable).

- When an annihilation $\alpha \to \beta$ ($\alpha > \beta$) is selected, empty channels remain unaffected and particles are removed from initially occupied channels (all combinations being equally probable).

---

[14] For the particular models discussed in this section, the chemical transformation does not modify the diffusion coefficient because $C$ is always combined with a velocity randomization $R$ which reshuffles completely the velocity distribution. Models where the velocity randomization is not very efficient ($R(\mathbf{s}, \mathbf{s}')$ with high probabilities on the diagonal) should be treated with caution.

These rules determine completely the matrix elements $C(\mathbf{s}, \mathbf{s}')$ in terms of the particle number transition probabilities $P(\alpha, \beta)$. The result is shown in Table II where it can be checked that

$$P(\alpha, \beta) = \sum_{\substack{\mathbf{s}' \in S \\ \sum_{i=1}^{4} s'_i = \beta}} C(\mathbf{s}, \mathbf{s}'), \qquad \forall \, \mathbf{s} \text{ such that } \sum_{i=1}^{4} s_i = \alpha \,. \tag{96}$$

Equation (96) expresses the fact that any configuration $\mathbf{s}$ with $\alpha$ particles has a total probability $P(\alpha, \beta)$ to be transformed into a configuration $\mathbf{s}'$ with $\beta$ particles.

|      | 0000 | 1000 | 0100 | 0010 | 0001 | 1100 | 1010 | 1001 | 0101 | 0011 | 0110 | 1110 | 1101 | 1011 | 0111 | 1111 |
|------|------|------|------|------|------|------|------|------|------|------|------|------|------|------|------|------|
| 0000 | $P_{00}$ | $P_{01}/4$ | $P_{01}/4$ | $P_{01}/4$ | $P_{01}/4$ | $P_{02}/6$ | $P_{02}/6$ | $P_{02}/6$ | $P_{02}/6$ | $P_{02}/6$ | $P_{02}/6$ | $P_{03}/4$ | $P_{03}/4$ | $P_{03}/4$ | $P_{03}/4$ | $P_{04}$ |
| 1000 | $P_{10}$ | $P_{11}$ | 0 | 0 | 0 | $P_{12}/3$ | $P_{12}/3$ | $P_{12}/3$ | 0 | 0 | 0 | $P_{13}/3$ | $P_{13}/3$ | $P_{13}/3$ | 0 | $P_{14}$ |
| 0100 | $P_{10}$ | 0 | $P_{11}$ | 0 | 0 | $P_{12}/3$ | 0 | 0 | $P_{12}/3$ | 0 | $P_{12}/3$ | $P_{13}/3$ | $P_{13}/3$ | 0 | $P_{13}/3$ | $P_{14}$ |
| 0010 | $P_{10}$ | 0 | 0 | $P_{11}$ | 0 | 0 | $P_{12}/3$ | 0 | 0 | $P_{12}/3$ | $P_{12}/3$ | $P_{13}/3$ | 0 | $P_{13}/3$ | $P_{13}/3$ | $P_{14}$ |
| 0001 | $P_{10}$ | 0 | 0 | 0 | $P_{11}$ | 0 | 0 | $P_{12}/3$ | $P_{12}/3$ | $P_{12}/3$ | 0 | 0 | $P_{13}/3$ | $P_{13}/3$ | $P_{13}/3$ | $P_{14}$ |
| 1100 | $P_{20}$ | $P_{21}/2$ | $P_{21}/2$ | 0 | 0 | $P_{22}$ | 0 | 0 | 0 | 0 | 0 | $P_{23}/2$ | $P_{23}/2$ | 0 | 0 | $P_{24}$ |
| 1010 | $P_{20}$ | $P_{21}/2$ | 0 | $P_{21}/2$ | 0 | 0 | $P_{22}$ | 0 | 0 | 0 | 0 | $P_{23}/2$ | 0 | $P_{23}/2$ | 0 | $P_{24}$ |
| 1001 | $P_{20}$ | $P_{21}/2$ | 0 | 0 | $P_{21}/2$ | 0 | 0 | $P_{22}$ | 0 | 0 | 0 | 0 | $P_{23}/2$ | $P_{23}/2$ | 0 | $P_{24}$ |
| 0101 | $P_{20}$ | 0 | $P_{21}/2$ | 0 | $P_{21}/2$ | 0 | 0 | 0 | $P_{22}$ | 0 | 0 | 0 | $P_{23}/2$ | 0 | $P_{23}/2$ | $P_{24}$ |
| 0011 | $P_{20}$ | 0 | 0 | $P_{21}/2$ | $P_{21}/2$ | 0 | 0 | 0 | 0 | $P_{22}$ | 0 | 0 | 0 | $P_{23}/2$ | $P_{23}/2$ | $P_{24}$ |
| 0110 | $P_{20}$ | 0 | $P_{21}/2$ | $P_{21}/2$ | 0 | 0 | 0 | 0 | 0 | 0 | $P_{22}$ | $P_{23}/2$ | 0 | 0 | $P_{23}/2$ | $P_{24}$ |
| 1110 | $P_{30}$ | $P_{31}/3$ | $P_{31}/3$ | $P_{31}/3$ | 0 | $P_{32}/3$ | $P_{32}/3$ | 0 | 0 | 0 | $P_{32}/3$ | $P_{33}$ | 0 | 0 | 0 | $P_{34}$ |
| 1101 | $P_{30}$ | $P_{31}/3$ | $P_{31}/3$ | 0 | $P_{31}/3$ | $P_{32}/3$ | 0 | $P_{32}/3$ | $P_{32}/3$ | 0 | 0 | 0 | $P_{33}$ | 0 | 0 | $P_{34}$ |
| 1011 | $P_{30}$ | $P_{31}/3$ | 0 | $P_{31}/3$ | $P_{31}/3$ | 0 | $P_{32}/3$ | $P_{32}/3$ | 0 | $P_{32}/3$ | 0 | 0 | 0 | $P_{33}$ | 0 | $P_{34}$ |
| 0111 | $P_{30}$ | 0 | $P_{31}/3$ | $P_{31}/3$ | $P_{31}/3$ | 0 | 0 | 0 | $P_{32}/3$ | $P_{32}/3$ | $P_{32}/3$ | 0 | 0 | 0 | $P_{33}$ | $P_{34}$ |
| 1111 | $P_{40}$ | $P_{41}/4$ | $P_{41}/4$ | $P_{41}/4$ | $P_{41}/4$ | $P_{42}/6$ | $P_{42}/6$ | $P_{42}/6$ | $P_{42}/6$ | $P_{42}/6$ | $P_{42}/6$ | $P_{43}/4$ | $P_{43}/4$ | $P_{43}/4$ | $P_{43}/4$ | $P_{44}$ |

TABLE II. Expression for the matrix elements $C(\mathbf{s}, \mathbf{s}')$ in terms of the elements of $P(\alpha, \beta)$ which are written as $P_{\alpha\beta}$ in the Table entries. The first column and row indicate the relevant configurations before and after the chemical transformation, respectively $\mathbf{s}$ and $\mathbf{s}'$, represented as 4-bit words (i.e. $\mathbf{s} \equiv \langle s_1 s_2 s_3 s_4 \rangle$). Configurations with same numbers of particles are grouped together, and the groups are separated by vertical and horizontal lines.

In the evolution operator (95), the velocity randomization and the chemical transformation are applied in sequence and it is convenient to define a collision operator which combines the two operations:

$$\mathcal{R} = R \circ C \ , \tag{97}$$

$$= C \circ R \ . \tag{98}$$

Both $R$ and $C$ modify the lattice configuration by independent node transformations, and this property extends to $\mathcal{R}$ which can be described by a local transition probability matrix with elements

$$\{\mathcal{R}(\mathbf{s},\mathbf{s}') \ : \ \mathbf{s},\mathbf{s}' \in S\} \ , \tag{99}$$

giving for each pair of configurations $\mathbf{s}$ and $\mathbf{s}'$ the probability $\mathcal{R}(\mathbf{s},\mathbf{s}')$ that $\mathbf{s}$ is mapped onto $\mathbf{s}'$ by $\mathcal{R}$. From the definitions (97) and (98) this transition matrix is simply the matrix product of the transition matrices of $C$ and $R$:

$$\mathcal{R}(\mathbf{s},\mathbf{s}') = \sum_{\mathbf{s}'' \in S} R(\mathbf{s},\mathbf{s}'')C(\mathbf{s}'',\mathbf{s}') \ ,$$

$$= \sum_{\mathbf{s}'' \in S} C(\mathbf{s},\mathbf{s}'')R(\mathbf{s}'',\mathbf{s}') \ . \tag{100}$$

To derive the microdynamical equations corresponding to the dynamics generated by (95), we write the evolution operator as

$$\mathcal{E} = P \circ \mathcal{R} \ , \tag{101}$$

and we notice that this is identical to $\mathcal{E} = P \circ R$ except that the transition probability matrix of $R$ must now be replaced by $\mathcal{R}(\mathbf{s},\mathbf{s}')$. We can therefore use the same form of the microdynamical equations as for the $RP-$dynamics in Sec. III (see (29)):

$$\eta_i(\mathbf{r}+\mathbf{c}_i, k+1) = \sum_{\mathbf{s},\mathbf{s}'} \xi_{\mathbf{s}\mathbf{s}'}(\mathbf{r},k) \, s'_i \prod_{j=1}^{4} \eta_j^{s_j}(\mathbf{r},k)(1-\eta_j(\mathbf{r},k))^{(1-s_j)} \ , \tag{102}$$

where we change the probability distribution of the Boolean matrices $[\xi_{\mathbf{s}\mathbf{s}'}(\mathbf{r},k)]$:

1. for each configuration $\mathbf{s}$ there is one and only one configuration $\mathbf{s}'$ such that $\xi_{\mathbf{s}\mathbf{s}'}(\mathbf{r},k) = 1$,

2. the probability of the event $\xi_{\mathbf{s}\mathbf{s}'}(\mathbf{r},k) = 1$ is given by $\mathcal{R}(\mathbf{s},\mathbf{s}')$ ,

3. random matrices $[\xi_{\mathbf{s}\mathbf{s}'}(\mathbf{r},k)]$ at different nodes and/or different times are mutually independent.

The interpretation of the random matrices remains the same: a node configuration $\boldsymbol{\eta}(\mathbf{r},k)$ is transformed by $\mathcal{R}$ into the only configuration $\mathbf{s}'$ such that $\xi_{\boldsymbol{\eta}\mathbf{s}'}(\mathbf{r},k) = 1$.

Consider the identity

$$\eta_i = \sum_{\mathbf{s}} s_i \prod_{j=1}^{4} \eta_j^{s_j}(1-\eta_j)^{(1-s_j)} \ , \tag{103}$$

which follows from the fact that

$$\prod_{j=1}^{4} \eta_j^{s_j}(\mathbf{r},k)(1-\eta_j(\mathbf{r},k))^{(1-s_j)} \ , \tag{104}$$

plays the role of a "matching indicator": its value is equal to one if the configuration at node $\mathbf{r}$ at time $k$ matches the configuration $\mathbf{s}$ (i.e. if $\boldsymbol{\eta}(\mathbf{r},k) = \mathbf{s}$) and zero otherwise. Equation (103) together with $\sum_{\mathbf{s}'} \xi_{\mathbf{s}\mathbf{s}'} = 1$ yields

$$\eta_i = \sum_{\mathbf{s},\mathbf{s}'} s_i \xi_{\mathbf{ss}'} \prod_{j=1}^{4} \eta_j^{s_j}(1-\eta_j)^{(1-s_j)} \ . \tag{105}$$

Combining (102) with (105), we obtain

$$\eta_i(\mathbf{r}+\mathbf{c}_i, k+1) - \eta_i(\mathbf{r}, k) = \sum_{\mathbf{s},\mathbf{s}'}(s_i' - s_i)\, \xi_{\mathbf{ss}'}(\mathbf{r}, k) \prod_{j=1}^{4} \eta_j^{s_j}(\mathbf{r}, k)(1-\eta_j(\mathbf{r}, k))^{(1-s_j)} \ , \tag{106}$$

which is equivalent to (102) and will prove to be a convenient form for later applications.

A formal equation for the evolution of the mean occupation numbers $N_i(\mathbf{r}, k)$ is obtained by taking the expectation of the microdynamical equation (102)

$$N_i(\mathbf{r}+\mathbf{c}_i, k+1) = \mathrm{E}\left[\sum_{\mathbf{s},\mathbf{s}'} \xi_{\mathbf{ss}'}(\mathbf{r}, k)\, s_i' \prod_{j=1}^{4} \eta_j^{s_j}(\mathbf{r}, k)(1-\eta_j(\mathbf{r}, k))^{(1-s_j)}\right] \ . \tag{107}$$

Using the independence of the random matrices $\xi$ and the Boolean field $\boldsymbol{\eta}(\cdot, k)$, and using $\mathrm{E}[\xi_{\mathbf{ss}'}(\mathbf{r}, k)] = \mathcal{R}(\mathbf{s}, \mathbf{s}')$, we simplify this equation to obtain

$$N_i(\mathbf{r}+\mathbf{c}_i, k+1) = \sum_{\mathbf{s},\mathbf{s}'} \mathcal{R}(\mathbf{s}, \mathbf{s}')\, s_i'\, \mathrm{E}\left[\prod_{j=1}^{4} \eta_j^{s_j}(\mathbf{r}, k)(1-\eta_j(\mathbf{r}, k))^{(1-s_j)}\right] \ . \tag{108}$$

This is *not* a closed equation for the mean occupation numbers $N_i(\mathbf{r}, k)$ since the expectation

$$\mathrm{E}\left[\prod_{j=1}^{4} \eta_j^{s_j}(\mathbf{r}, k)(1-\eta_j(\mathbf{r}, k))^{(1-s_j)}\right] \ , \tag{109}$$

involves not only average occupation numbers but higher moments, reflecting correlations between different channels at the same node. This is a consequence of the fact that the matching indicator (104) involves the occupations of *all* channels at node $\mathbf{r}$ at time $k$, so that the knowledge of the complete probability distribution of $\boldsymbol{\eta}(\mathbf{r}, k)$ is required to evaluate the expectation (109) for all the possible values of $\mathbf{s} \in \mathbf{S}$.

At this point it is instructive to consider again the diffusive models presented in Sec. III in order to understand why the dynamics of the average populations is independent of the correlations in these models. There the decoupling arises because for each possible realization of the velocity randomization $R$, the post-randomization state of a channel $w(\mathbf{r}, i)$ at node $\mathbf{r}$ depends on the state of a *single* channel $w(\mathbf{r}, j)$ at the same node. The particular channel $w(\mathbf{r}, j)$ which determines how the channel $w(\mathbf{r}, i)$ is transformed depends, of course, on the realization of $R$ but this is unimportant: the crucial point is that, no matter the result of the random selection, for each realization of $R$ there is only one channel $w(\mathbf{r}, j)$ whose state influences the transformation of the channel $w(\mathbf{r}, i)$. In a reactive model this cannot be expected because the way in which a node is transformed by the chemical operator depends precisely on the number of particles at the node.

In general, the expectation (109) cannot be expressed in terms of the average ocupation numbers $N_i(\mathbf{r}, k)$. However an important exception arises when the variables $\eta_i(\mathbf{r}, k)$ are decorrelated: in this case (109) can be replaced by [15]

$$\prod_{j=1}^{4} N_j^{s_j}(\mathbf{r}, k)(1-N_j(\mathbf{r}, k))^{(1-s_j)} \ . \tag{110}$$

---

[15]Notice that in (109), the power in any of the variables $\eta_i$ is equal to 1 or 0.

Therefore, if we assume that the variables $\eta_i(\mathbf{r}, k)$ are decorrelated before each application of $\mathcal{R}$ – this is the Boltzmann *ansatz* – we can substitute (110) for (109) in the formal equation (108) and obtain a closed equation for the average occupation numbers:

$$N_i(\mathbf{r} + \mathbf{c}_i, k+1) = \sum_{\mathbf{s},\mathbf{s}'} \mathcal{R}(\mathbf{s}, \mathbf{s}') s'_i \prod_{j=1}^{4} N_j^{s_j}(\mathbf{r}, k)(1 - N_j(\mathbf{r}, k))^{(1-s_j)} , \qquad (111)$$

or equivalently from (106)

$$N_i(\mathbf{r} + \mathbf{c}_i, k+1) - N_i(\mathbf{r}, k) = \sum_{\mathbf{s},\mathbf{s}'} \mathcal{R}(\mathbf{s}, \mathbf{s}') (s'_i - s_i) \prod_{j=1}^{4} N_j^{s_j}(\mathbf{r}, k)(1 - N_j(\mathbf{r}, k))^{(1-s_j)} . \qquad (112)$$

This set of nonlinear, finite-difference equations is analogous to the continuous Boltzmann equation of statistical mechanics;[16] for this reason (112) (or equivalently (111)) is referred to as the *lattice-Boltzmann equation* of the lattice gas.

It is usually argued that the Boltzmann approximation should hold in the limit of small gradients and infrequent reactions

$$C(\mathbf{s}, \mathbf{s}') \ll 1 \text{ for } \mathbf{s} \neq \mathbf{s}' , \qquad (113)$$

since then reactive processes occur on nodes which support a near diffusive equilibrium distribution where the velocity channels are uncorrelated. In such a circumstance the rate at which the diffusive equilibrium is perturbed by the reactive process is much slower than the rate at which the system returns to a local diffusive equilibrium. However, the argument should be considered with caution because a *typical* relaxation rate is meaningless for diffusion in an infinite system. Indeed, the diffusion rate

$$|\omega| = D|\mathbf{k}|^2 , \qquad (114)$$

depends on the length scale ($|\mathbf{k}|^{-1}$), and on large enough length scales, the diffusive relaxation is always slower than any reactive process. Therefore large scale phenomena can compromise the validity of the Boltzmann approximation even for vanishing reactive rates.[17] These considerations imply that the reaction-diffusion equations follow only in a suitable weak coupling limit that considers both spatial gradients and the ratio of reactive to non-reactive relaxation times to be small. [26] In addition, it has been shown that the correlations which are omitted in the Boltzmann-level description are responsible for a number of physically relevant phenomena. For instance, the diffusion limit, where the rate constant takes the Smoluchowski form and is no longer kinetically dominated, has its microscopic origin in infinite sequences of ring collision events which are not considered in the Boltzmann equation. [27]

Correlation effects due to incomplete mixing are especially evident in low dimensional systems and numerical simulations based on random walk models [28] have provided the basis for a large number of investigations addressing the question of the validity of the local diffusive equilibrium assumption with the aim of increasing our understanding of the new types of kinetic behavior that emerge when this hypothesis is not applicable [10].

The Boltzmann equation may also be inadequate when the microscopic randomness does not disappear at the macroscopic scale. Typical examples are spontaneous symmetry breaking or nucleation processes; in both cases, microscopic fluctuations trigger persistent large scale phenomena. In spite of these limitations, the Boltzmann equations provide a valid description for many situations of practical interest. Examples where the Boltzmann equations apply and where they do not are discussed in the applications sections. Non-Boltzmann phenomena in lattice gas automata have been investigated in Ref. [29], [30].

---

[16]This is most easily seen by noting that in (112) the r.h.s. is a collisional term and by taking the continuous limit of the l. h. s..

[17]The idea that the local equilibrium hypothesis could be inappropriate for some reactive systems was first suggested around 1950 in the context of exothermic reactions [25].

## B. From Boltzmann equations to reaction-diffusion behavior

An alternative form of the Boltzmann equations, (111) and (112), is obtained when the average population variables $N_i$ are expressed in terms of the mass density $\rho(\mathbf{r}, k)$, the mass current density $\mathbf{j}(\mathbf{r}, k)$, and the q-density $\rho_q(\mathbf{r}, k)$ (see (42), (43) and (51)). This transformation is interesting because it separates variables into slow and fast types. Indeed, the q-density and the mass current are not conserved in the velocity randomization operation and therefore these variables are expected to exhibit fast relaxation (in a few time steps). On the other hand, creation and annihilation of mass occur only *via* reactive processes which are assumed to be infrequent. When gradients are weak the relaxation of the mass density is dominated by reactive processes and takes place on a long time scale. If microscopic time scales are ignored, the fast variables can be eliminated to obtain a description which focuses on the slow dynamics of the mass density. How this elimination leads to a reaction-diffusion equation will not be discussed in full generality; instead, we consider the situation where the Boltzmann equations are linearized around a homogeneous steady state.

### 1. Homogeneous solutions of the Boltzmann equations

We show that the determination of a homogeneous and isotropic solution

$$N_i^0(\mathbf{r}, k) = \rho_s/4 = c_s , \qquad \forall i = 1, \ldots 4 \,; k = 0, 1, 2 \ldots \,; \mathbf{r} \in \mathcal{L} , \tag{115}$$

for the Boltzmann equation (111) is a problem equivalent to solving the algebraic equation

$$f(c_s) = 0 , \tag{116}$$

where $f$ is the macroscopic reactive rate (7) and $c_s$ denotes the stationary value of the density of species $X$ per channel, $c_X = c_s$. Substituting $c_s$ for $N_i(\mathbf{r}, k)$ in the Boltzmann equation (111) yields

$$c_s = \sum_{\mathbf{s},\mathbf{s}'} s_i' \, \mathcal{R}(\mathbf{s}, \mathbf{s}') \prod_{j=1}^{4} c_s^{s_j} (1 - c_s)^{(1-s_j)} . \tag{117}$$

Since $C$ and $R$ are invariant under rotations and reflections of the channels, the same invariance applies to $\mathcal{R}$;[18] therefore the r.h.s. of (117) is independent of the index $i$:

$$c_s = \frac{1}{4} \sum_{\mathbf{s},\mathbf{s}'} \sum_{i=1}^{4} s_i' \, \mathcal{R}(\mathbf{s}, \mathbf{s}') \prod_{j=1}^{4} c_s^{s_j} (1 - c_s)^{(1-s_j)} . \tag{119}$$

Expressing $\mathcal{R}(\mathbf{s}, \mathbf{s}')$ in terms of $C(\mathbf{s}, \mathbf{s}')$ and $R(\mathbf{s}, \mathbf{s}')$ as in (100) we obtain

$$c_s = \frac{1}{4} \sum_{\mathbf{s},\mathbf{s}''} C(\mathbf{s}, \mathbf{s}'') \prod_{j=1}^{4} c_s^{s_j} (1 - c_s)^{(1-s_j)} \sum_{\mathbf{s}' \in S} \left( \sum_{i=1}^{4} s_i' \right) R(\mathbf{s}'', \mathbf{s}') , \tag{120}$$

which reduces to

---

[18] By construction, the collision operator conserves the invariance property of $R$ and $C$:

$$\mathcal{R}(\mathbf{s}, \mathbf{s}') = \mathcal{R}(g\mathbf{s}', g\mathbf{s}) , \qquad \forall \mathbf{s}, \mathbf{s}' \in S , \tag{118}$$

for all node transformation $g$ corresponding to a rotation or a reflection of the channels in real space.

$$c_s = \frac{1}{4} \sum_{\mathbf{s},\mathbf{s}'} \sum_{i=1}^{4} s'_i \, C(\mathbf{s},\mathbf{s}') \prod_{j=1}^{4} c_s^{s_j}(1-c_s)^{(1-s_j)} \,, \qquad (121)$$

where we have used the conservation of mass during the velocity randomization:

$$\sum_{\mathbf{s}'} (\sum_{i=1}^{4} s'_i) R(\mathbf{s}'',\mathbf{s}') = \sum_{i=1}^{4} s''_i \,. \qquad (122)$$

Using the identity

$$N_i = \sum_{\mathbf{s},\mathbf{s}'} C(\mathbf{s},\mathbf{s}') \, s_i \prod_{j=1}^{4} N_j^{s_j}(1-N_j)^{(1-s_j)} \,, \qquad (123)$$

which follows from $\sum_{\mathbf{s}'} C(\mathbf{s},\mathbf{s}') = 1$, and summing the result over the index $i$, we can rewrite (121) as

$$0 = \frac{1}{4} \sum_{\mathbf{s},\mathbf{s}'} \sum_{i=1}^{4} (s'_i - s_i) \, C(\mathbf{s},\mathbf{s}') \prod_{j=1}^{4} c_s^{s_j}(1-c_s)^{(1-s_j)} \,. \qquad (124)$$

Combining the different terms on the r.h.s. which correspond to pairs of configurations $\mathbf{s}$ and $\mathbf{s}'$ with $\alpha$ and $\beta$ particles, respectively, (see (96) or Table II), we find

$$0 = \frac{1}{4} f(c_s) = \sum_{\alpha=0}^{4} \sum_{\beta=0}^{4} \frac{1}{4} (\beta - \alpha) P(\alpha,\beta) \binom{4}{\alpha} c_s^{\alpha}(1-c_s)^{4-\alpha} \,. \qquad (125)$$

This shows that the homogeneous isotropic solutions of the Boltzmann equations coincide with the fixed points of the phenomenological rate equation

$$\frac{dc_X}{dt} = \frac{1}{4} f(c_X) \,, \qquad (126)$$

written here in terms of the density per channel.

<div style="text-align:center"><em>2. Linearized Boltzmann equations</em></div>

If we assume that the system is close to a homogeneous isotropic steady state we can linearize the Boltzmann equation (112) about this state. Defining

$$\delta N_i(\mathbf{r},k) = N_i(\mathbf{r},k) - c_s \,, \qquad (127)$$

where $c_s$ is a solution of $f(c_s) = 0$, we have to first order in $\delta N_i$

$$\delta N_i(\mathbf{r}+\mathbf{c}_i, k+1) - \delta N_i(\mathbf{r},k) = \mathcal{L}_{ij} \delta N_j(\mathbf{r},k) \,, \qquad (128)$$

where

$$\mathcal{L}_{ij} = \frac{\partial}{\partial N_j} \mathcal{F}_i(\mathbf{N}) \bigg|_{\{N_\ell = c_s \,:\, \ell = 1,\ldots,4\}} \,, \qquad (129)$$

with

$$\mathcal{F}_i(\mathbf{N}) = \sum_{\mathbf{s},\mathbf{s}'} \mathcal{R}(\mathbf{s},\mathbf{s}')(s'_i - s_i) \prod_{j=1}^{4} N_j^{s_j}(1-N_j)^{(1-s_j)} \,. \qquad (130)$$

It is sometimes useful to write (128) as

$$\delta N_i(\mathbf{r} + \mathbf{c}_i, k+1) = \mathcal{B}_{ij} \delta N_j(\mathbf{r}, k) ,\qquad(131)$$

with

$$\mathcal{B}_{ij} = \delta_{ij} + \mathcal{L}_{ij} .\qquad(132)$$

The matrix $[\mathcal{B}_{ij}]$ describes how small deviations from $c_s$ evolve when the collision operator $\mathcal{R}$ is applied to a node. The main properties of the linearized collision operator $[\mathcal{L}_{ij}]$ and of the matrix $[\mathcal{B}_{ij}]$ are discussed in the next section.

### 3. Properties of the linearized collision operator

From (129) and (130) we have [19]

$$\mathcal{L}_{ij} = \sum_{\mathbf{s},\mathbf{s}'}(s'_i - s_i)\mathcal{R}(\mathbf{s},\mathbf{s}')(2s_j - 1) \prod_{k,k\neq j} c_s^{s_k}(1-c_s)^{1-s_k} ,\qquad(133)$$

which may also be written in the form

$$\begin{aligned}\mathcal{L}_{ij} = &\sum_{\mathbf{s},\mathbf{s}'} s'_i \mathcal{R}(\mathbf{s},\mathbf{s}') s_j \prod_{k,k\neq j} c_s^{s_k}(1-c_s)^{1-s_k} \\ &- \sum_{\mathbf{s},\mathbf{s}'} s'_i \mathcal{R}(\mathbf{s},\mathbf{s}')(1-s_j) \prod_{k,k\neq j} c_s^{s_k}(1-c_s)^{1-s_k} .\end{aligned}\qquad(134)$$

The first term of the r.h.s. accounts for changes in $N_i$ in a collision when there is a particle in channel $w(\mathbf{r},j)$ before collision. The second term on the r.h.s. is the corresponding quantity when channel $w(\mathbf{r},j)$ is empty before collision.

As a consequence of the invariance of $\mathcal{R}$ under the group of rotations and reflections of a node on the lattice, the matrix $[\mathcal{L}_{ij}]$ has necessarily the form

$$\begin{pmatrix} L_{00} & L_{01} & L_{02} & L_{01} \\ L_{01} & L_{00} & L_{01} & L_{02} \\ L_{02} & L_{01} & L_{00} & L_{01} \\ L_{01} & L_{02} & L_{01} & L_{00} \end{pmatrix} .\qquad(135)$$

and in view of the definition (132) the matrix $[\mathcal{B}_{ij}]$ has the same form

$$\begin{pmatrix} B_{00} & B_{01} & B_{02} & B_{01} \\ B_{01} & B_{00} & B_{01} & B_{02} \\ B_{02} & B_{01} & B_{00} & B_{01} \\ B_{01} & B_{02} & B_{01} & B_{00} \end{pmatrix} .\qquad(136)$$

Now we show how the linearized collision operator $[\mathcal{L}_{ij}]$ is connected to the macroscopic rate $f(c_X)$. Summing the expression (133) over $i$ and using (100) and (122) we obtain

$$\sum_{i=1}^{4} \mathcal{L}_{ij} = \sum_{\mathbf{s},\mathbf{s}'} \sum_{i=1}^{4}(s'_i - s_i)C(\mathbf{s},\mathbf{s}')(2s_j - 1) \prod_{k,k\neq j} c_s^{s_k}(1-c_s)^{1-s_k} .\qquad(137)$$

---

[19]To arrive at (133) account is taken of the Boolean nature of $s_j$.

On the r.h.s. of this expression, we combine the terms corresponding to configurations **s** and **s'** with $\alpha$ and $\beta$ particles respectively (see (96) or Table II). This leads to

$$\sum_{i=1}^{4} \mathcal{L}_{ij} = -\frac{1}{4}\frac{df(c_s)}{dc_s} \equiv -\kappa , \qquad (138)$$

In the sequel, we consider only those values of $c_s$ which correspond to linearly stable fixed points of the rate equation (126), i.e. $\kappa$ is always positive.

### 4. Solutions of the linearized Boltzmann equations

In this subsection, we give the general solutions of the linearized Boltzmann equations for the dynamics based on the velocity randomization performed with the 4-equi-rotation rule defined in (17); in this case the matrix $[\mathcal{B}_{ij}]$ has a form which allows analytical calculations. These manipulations parallel those given in (55) to (70) and are repeated here for the sake of clarity. When the velocity randomization (17) is used in $\mathcal{R}$ (see (97)), all the channels on a node are made statistically equivalent after collision. Therefore the matrix $[\mathcal{B}_{ij}]$ takes the simple form

$$\begin{pmatrix} B & B & B & B \\ B & B & B & B \\ B & B & B & B \\ B & B & B & B \end{pmatrix} , \qquad (139)$$

and the Boltzmann equations read

$$\delta N_i(\mathbf{r} + \mathbf{c}_i, k+1) = B \sum_{j=1}^{4} \delta N_j(\mathbf{r}, k) , \qquad i = 1, \ldots 4 . \qquad (140)$$

Inserting in (140) the Fourier mode

$$\delta N_i(\mathbf{r}, k) = \delta N_i \exp(i\mathbf{k} \cdot \mathbf{r}) A^k \qquad (141)$$

we obtain a linear algebraic set for the $\delta N_i$'s

$$\sum_{j=1}^{4} M_{ij} \delta N_j = 0 , \qquad i = 1, \ldots, 4 , \qquad (142)$$

$$M = \begin{pmatrix} B - A \exp(i\, k_1) & B & B & B \\ B & B - A \exp(i\, k_2) & B & B \\ B & B & B - A \exp(-i\, k_1) & B \\ B & B & B & B - A \exp(-i\, k_2) \end{pmatrix} . \qquad (143)$$

Non-trivial solutions follow from the condition $det\mathbf{M} = 0$. Making explicit use of this condition, we obtain a fourth order polynomial equation for the damping coefficient $A$

$$\det \mathbf{M} = A^3 \left( A - B\frac{\cos(k_1) + \cos(k_2)}{2} \right) = 0 . \qquad (144)$$

The solutions $A^{(j)}(\mathbf{k})(j = 1, ..., 4)$ and the corresponding vectors $\delta \mathbf{N}^{(j)}(\mathbf{k})$ are given by[20]

---

[20] Again the kernel corresponding to $A(\mathbf{k}) = 0$ is spanned by the vectors (49) and (50).

$$A^{(1)}(\mathbf{k}) = 2B\left(\cos(k_1) + \cos(k_2)\right) , \qquad \delta N^{(1)}(\mathbf{k}) = \begin{pmatrix} 1 \\ 1 \\ 1 \\ 1 \end{pmatrix} , \qquad (145)$$

$$A^{(2)}(\mathbf{k}) = 0 , \qquad \delta N^{(2)}(\mathbf{k}) = \begin{pmatrix} 1 \\ 0 \\ -1 \\ 0 \end{pmatrix} , \qquad (146)$$

$$A^{(3)}(\mathbf{k}) = 0 , \qquad \delta N^{(3)}(\mathbf{k}) = \begin{pmatrix} 0 \\ 1 \\ 0 \\ -1 \end{pmatrix} , \qquad (147)$$

$$A^{(4)}(\mathbf{k}) = 0 , \qquad \delta N^{(4)}(\mathbf{k}) = \begin{pmatrix} 1 \\ -1 \\ 1 \\ -1 \end{pmatrix} . \qquad (148)$$

Two different types of temporal behavior appear :

1. The damping factors $A^{(j)}$ ($j = 2, 3, 4$) are equal to zero and the corresponding solutions decay to zero in one time step independently of the value of $|\mathbf{k}|$. This reflects the strong dissipation of the mass current density and the q-density as can be checked by projecting (146)–(148) onto (48)–(50). Notice that the "instantaneous" decay $A^{(j)} = 0$ ($j = 2, 3, 4$) arises because the velocity randomization operators erase all information about the particle velocities in one single time step.

2. The damping coefficient $A^{(1)}$ depends on $\mathbf{k}$ and its value is in general different from zero.

Since

$$4B = \sum_{i=1}^{4} \mathcal{B}_{ij} , \qquad (149)$$

we have from (132) and (138)

$$4B = 1 - \kappa , \qquad (150)$$

so that the damping coefficient reads

$$A^{(1)}(\mathbf{k}) = (1 - \kappa)\left(\frac{\cos(k_1) + \cos(k_2)}{2}\right) . \qquad (151)$$

For small wave numbers and infrequent reactive collisions (i.e. $\kappa \ll 1$), the damping coefficient $A^{(1)}(\mathbf{k})$ is close to one and can be expressed as the exponential of an equivalent continuous damping rate $\omega(\mathbf{k})$

$$A^{(1)}(\mathbf{k}) = \exp(\omega(\mathbf{k})) , \qquad (152)$$

or

$$\omega(\mathbf{k}) = \ln\left(A^{(1)}(\mathbf{k})\right) \qquad (153)$$

$$= \ln(1 - \kappa) - |\mathbf{k}|^2/4 + O(|\mathbf{k}|^4) . \qquad (154)$$

Equation (154) shows that in this regime the dispersion relation is equivalent to that of the linearized reaction-diffusion equation

$$\frac{\partial}{\partial t}\delta c_X(\mathbf{r}, t) = -\kappa\delta c_X(\mathbf{r}, t) + D\nabla^2 \delta c_X(\mathbf{r}, t) , \qquad (155)$$

with $D = 1/4$. In this equation, the field $\delta c_X(\mathbf{r}, t) = c_X(\mathbf{r}, t) - c_s$ can be identified with the fluctuation of the mass density per channel on the lattice because (145) is oriented exactly along the mass vector (48).

## V. REACTIVE LATTICE GAS

The formalism required to properly describe a multi-species automaton dynamics in all generality and to carry out the statistical mechanical analysis which yields the reaction-diffusion equations is rather involved and will not be presented here. (The microdynamical formulation of the multi-species model can be found in [18].) However, it is easy to write a general algorithm suitable for simulation of the reactive dynamics without giving details of the microdynamical theory. This section provides the "recipe" for such an algorithm and can form the starting point for the development of computer codes for implementing the reactive lattice gas automaton.

We consider reactive systems which are described phenomenologically by partial differential equations of the reaction-diffusion type

$$\frac{\partial \boldsymbol{\rho}(\mathbf{r},t)}{\partial t} = \mathbf{F}(\boldsymbol{\rho}(\mathbf{r},t)) + \mathbf{D} \cdot \nabla^2 \boldsymbol{\rho}(\mathbf{r},t) , \qquad (156)$$

where $\mathbf{D}$ is a diagonal matrix with positive diagonal elements. Here, each component $\rho_\tau$ ($\tau = 1, 2, \ldots, n$) of the column vector $\boldsymbol{\rho}$ represents the local concentrations at space point $\mathbf{r}$ at time $t$ of a corresponding species $X_\tau$ involved in the reactive processes described by the reaction rate vector $\mathbf{F}(\boldsymbol{\rho})$.

On a phenomenological level the reactive terms in (156) derive from the mass action rate law

$$\frac{d\boldsymbol{\rho}(t)}{dt} = \mathbf{F}(\boldsymbol{\rho}(t)) , \qquad (157)$$

which follows from the reactive collision processes specified in a reaction mechanism. Therefore, $\mathbf{F}$ contains only polynomial contributions in the species concentrations. The second term on the r.h.s of (156) describes diffusive transport.

### A. Multiple species and exclusion principle

Single-species models generalize easily to allow for different reactive species to move and react on a lattice. As previously, the positions of the particles are restricted to a regular lattice $\mathcal{L}$, the time takes integer values, and the particle velocities take their values in a finite set of vectors compatible with the discretization of space and time [21].

There are many different ways to generalize the exclusion principle to account for the existence of different species. For instance one can allow different particles to be at a same node $\mathbf{r}$ provided they belong to different species and/or have different velocities. With this rule, each species is subject separately to its own single-species exclusion principle and a useful representation of the lattice is in terms of a "stack" of species lattices $\mathcal{L}_\tau$, $\tau = 1, \cdots, n$ where each species $\tau$ resides on its own lattice where the *single species exclusion* rule applies. Thus, the lattice $\mathcal{L}$ may be decomposed into $n$ species lattices $\mathcal{L} = \mathcal{L}_1 \times \ldots \times \mathcal{L}_n$, with identical node labels $\mathbf{r}$. This notional decomposition of $\mathcal{L}$ is useful both conceptually and in the mathematical formulation of the automaton. In this way non-reactive collisions can be carried out separately on each lattice $\mathcal{L}_\tau$ with different frequencies, mimicking the elastic collisions with the solvent, and reactive collisions couple the dynamics on all the species lattices. With this generalization of the exclusion principle, we shall see that the diffusion coefficients of different species can be tuned independently.

It is straigthforward to extend the notion of *channel*. At each node $\mathbf{r}$ we define $nm$ channels $w_\tau(\mathbf{r},i)$, $\tau = 1,\ldots,n$, $i = 1,\ldots,m$ where $n$ and $m$ are respectively the number of different species and the number of different velocities in the model. A channel $w_\tau(\mathbf{r}, i)$ is said to be occupied if there is a particle $X_\tau$ with velocity $\mathbf{c}_i$ on node $\mathbf{r}$ (otherwise the channel is said to be unoccupied).

---

[21] In some circumstances, it may be convenient to associate different sets of velocities to different species or restrict species to move on sublattices of $\mathcal{L}$ (some of which can be identical and some can be $\mathcal{L}$ itself). We will not consider such generalizations here.

A more restrictive exclusion principle is obtained when a channel cannot be occupied simultaneously by particles with the same velocity independently of their species (*global exclusion* principle). [31] It is also possible to apply the latter rule separately on different subsets of species: if two particles belong to the same subset the restrictive rule applies while if they belong to different subsets they are allowed to occupy the same channel. All the models so constructed are conceptually similar but differ mainly by (i) the class of reactive rates $\mathbf{F}(\boldsymbol{\rho}(t))$ to which they lead, and (ii) the flexibility they offer to tune independently the diffusion coefficients of different species. Here we restrict ourselves to the single species exclusion principle (In sec. X we consider some models without exclusion principle).

### B. Automaton rule

The automaton rule is built from the composition of propagation, velocity randomization and chemical transformation operators:

- **propagation operator**, $P_\tau$: Moves particles of species $\tau$ a specified number of lattice units in the directions determined by their velocities to other nodes of the lattice.

- **velocity randomization operator**, $R_\tau$: Randomizes the particle velocity configuration on the lattice $\mathcal{L}_\tau$ at each node independently of the others. The mixing operations, governed by the operators $R_\tau$, conserve the number of particles of species $\tau$; however, their momentum is not conserved locally.

- **chemical transformation operator**, $C$: Is responsible for local chemical reactions among the species. Denoting by $\boldsymbol{\alpha} = \langle \alpha_1 \alpha_2 \cdots \alpha_n \rangle$ and $\boldsymbol{\beta} = \langle \beta_1 \beta_2 \cdots \beta_n \rangle$ the vectors specifying the particle occupancy at a node, where $0 \leq \alpha_\tau, \beta_\tau \leq m$ for each $\tau = 1, \cdots, n$, the reaction occurs with probability $P(\boldsymbol{\alpha}, \boldsymbol{\beta})$, which depends only on the occupancy of the nodes $\mathbf{r}$ on $\mathcal{L}$ and not on the velocity configurations. In order to fully specify the reactive transformation, the matrix $\mathbf{P} = [P(\boldsymbol{\alpha}, \boldsymbol{\beta})]$ must be completed by a rule detailing how particle number changes are implemented into channel occupation number changes. The precise knowledge of this implementation is important to properly formulate the automaton dynamics. However we will see that this level of description is not required to determine the phenomenological rate law $\mathbf{F}$ of an automaton dynamics; the knowledge of $\mathbf{P}$ is sufficient for that purpose. In this section we will therefore ignore how particle number changes are actually translated into node configuration changes. However we will make the assumption that the configuration of a species $X_\tau$ remains unchanged in any reaction which does not change the number of particles of species $X_\tau$ ($\alpha_\tau = \beta_\tau$). In this way, we can neglect reactive contributions to the diffusion constants when reactions are unfrequent (see Sec. IV).

Neglecting reactions for the moment, the (non-reactive) dynamics of the system is given by the composition of free streaming and elastic collisions, $P_\tau \circ R_\tau$. To account for the possibly different elastic collision frequencies we may apply this composition of operators different numbers of times for each species in one total automaton time step. The general expression for the non-reactive automaton evolution is

$$[\text{State at time } (k+1)] = \mathcal{E} \, [\text{State at time } k] \tag{158}$$

with

$$\mathcal{E} = \prod_{\tau=1}^{n} (P_\tau \circ R_\tau)^{\ell_\tau} , \tag{159}$$

and $\ell_\tau$ an integer. Combining this non-reactive dynamics with the chemical transformation $C$, we obtain an evolution rule for the full reactive dynamics:

$$\mathcal{E} = \left( \prod_{\tau=1}^{n} (P_\tau \circ R_\tau)^{\ell_\tau} \right) \circ C . \tag{160}$$

## C. Reaction probability matrix

In order to complete the "recipe" for the rule construction some guidelines need to be given for the specification of the reaction probability matrix. We give a general strategy for the determination of **P**. In most circumstances the macroscopic equations of motion provide a good description of the basic phenomenology of non-equilibrium chemical rate processes. If the system is spatially homogeneous (well stirred) the appropriate equations of motion are the laws of mass action kinetics, while if the system is inhomogeneous diffusion terms can be used to augment the description leading to reaction-diffusion equations. Since the mass action rate law provides a mean field level of description of the local kinetics, it is natural to demand that the mean field approximation to the automaton dynamics correspond to the phenomonological rate law. We make use of this correspondence below. It is not difficult to derive an expression for the time rate of change of the average species concentrations from a mean field description of the automaton dynamics. We now summarize and formalize what was presented as a simple physical picture in Sec. II.

Suppose that the system is spatially homogeneous and correlations are neglected. Then the probability to find a node in a configuration $\boldsymbol{\alpha}$ is given by

$$\prod_{\kappa=1}^{n} \binom{m}{\alpha_\kappa} c_\kappa^{\alpha_\kappa} (1-c_\kappa)^{m-\alpha_\kappa} \;, \tag{161}$$

since the particles are binomially distributed. In this equation $c_\tau = \rho_\tau/m$, where $m$ is the number of channels per node. Given that the probability of a transition from a configuration $\boldsymbol{\alpha}$ to a configuration $\boldsymbol{\beta}$ is $P(\boldsymbol{\alpha},\boldsymbol{\beta})$ the rate of change of the average concentration of species $\tau$ per node, $\rho_\tau$, is

$$\rho_\tau(k+1) - \rho_\tau(k) = \sum_{\boldsymbol{\alpha},\boldsymbol{\beta}} (\beta_\tau - \alpha_\tau) P(\boldsymbol{\alpha},\boldsymbol{\beta}) \prod_{\kappa=1}^{n} \binom{m}{\alpha_\kappa} c_\kappa^{\alpha_\kappa} (1-c_\kappa)^{m-\alpha_\kappa} \tag{162}$$

$$= \sum_{\boldsymbol{\alpha}} P_\tau(\boldsymbol{\alpha}) \prod_{\kappa=1}^{n} \binom{m}{\alpha_\kappa} c_\kappa^{\alpha_\kappa} (1-c_\kappa)^{m-\alpha_\kappa} \;. \tag{163}$$

Here

$$P_\tau(\boldsymbol{\alpha}) = \sum_{\boldsymbol{\beta}} (\beta_\tau - \alpha_\tau) P(\boldsymbol{\alpha},\boldsymbol{\beta}) \;, \tag{164}$$

is the average value of the particle number change of species $\tau$ when the input configuration is $\boldsymbol{\alpha}$. The rate law (163) describes the mean field reactive dynamics of the automaton particles. We now establish a correspondence between the fictitious automaton universe and the phenomenological description of reacting systems.

Consider a general reaction mechanism consisting of $r$ elementary steps

$$\nu_1^j X_1 + \nu_2^j X_2 + \cdots \nu_n^j X_n \underset{k_{-j}}{\overset{k_j}{\rightleftharpoons}} \bar{\nu}_1^j X_1 + \bar{\nu}_2^j X_2 + \cdots \bar{\nu}_n^j X_n, \tag{165}$$

where $j = 1, \cdots, r$ labels each step in the mechanism, which is completely characterized by the sets of stoichiometric coefficients $\boldsymbol{\nu}^j = \langle \nu_1^j, \nu_2^j, \cdots, \nu_n^j \rangle$, $\bar{\boldsymbol{\nu}}^j = \langle \bar{\nu}_1^j, \bar{\nu}_2^j, \cdots, \bar{\nu}_n^j \rangle$ and the rate coefficients $\{k_{\pm j} \; : \; j = 1, \cdots, r\}$. The mass action rate law can be written in the form

$$\frac{d\rho_\tau}{dt} = \sum_{j=1}^{r} (\nu_\tau^j - \bar{\nu}_\tau^j) \left\{ -k_j \prod_{\kappa=1}^{n} \rho_\kappa^{\nu_\kappa^j} + k_{-j} \prod_{\kappa=1}^{n} \rho_\kappa^{\bar{\nu}_\kappa^j} \right\} \;. \tag{166}$$

In order to relate the mean field automaton equation (163) and the mass action rate law (166) we assume that the concentration units in these two equations are the same and express the time in terms of the automaton time unit, $h$. In these units the discrete-time version of the continuous-time mass action law (166) may be written as

$$\rho_\tau(k+h) - \rho_\tau(k) = h \sum_{j=1}^{r} (\nu_\tau^j - \bar{\nu}_\tau^j) \left\{ -k_j \prod_{\kappa=1}^{n} \rho_\kappa^{\nu_\kappa^j} + k_{-j} \prod_{\kappa=1}^{n} \rho_\kappa^{\bar{\nu}_\kappa^j} \right\} + \mathcal{O}(h^2) . \tag{167}$$

We now determine the conditions on the reaction probability matrix that ensure the descriptions of the reactive dynamics given in (163) and (167) agree to order $h$. These equations can be made to have the same forms either by expressing each $\rho_\tau$ in (167) in the basis $\{c_\tau^i(1-c_\tau)^{m-i} \; : \; i = 1, \ldots, m\}$ or, alternatively, expanding these factors in (163) to yield powers of $\rho_\tau$.[22] Choosing to expand in the $\{c_\tau^i(1-c_\tau)^{m-i} \; : \; i = 1, \ldots, m\}$ basis, we obtain

$$\rho_\tau(k+h) - \rho_\tau(k) = h \sum_{j=1}^{r} (\nu_\tau^j - \bar{\nu}_\tau^j) \sum_{l_1=0}^{m} \cdots \sum_{l_n=0}^{m}$$

$$\left\{ -k_j \, m^{\sum_{\kappa=1}^{n} \nu_\kappa^j} \binom{m - \nu_1^j}{l_1 - \nu_1^j} \binom{m}{l_1}^{-1} \cdots \binom{m - \nu_n^j}{l_n - \nu_n^j} \binom{m}{l_n}^{-1} \right.$$

$$\left. + k_{-j} \, m^{\sum_{\kappa=1}^{n} \bar{\nu}_\kappa^j} \binom{m - \bar{\nu}_1^j}{l_1 - \bar{\nu}_1^j} \binom{m}{l_1}^{-1} \cdots \binom{m - \bar{\nu}_n^j}{l_n - \bar{\nu}_n^j} \binom{m}{l_n}^{-1} \right\}$$

$$\times \prod_{\kappa=1}^{n} \binom{m}{l_\kappa} c_\kappa^{l_\kappa} (1 - c_\kappa)^{m - l_\kappa} + \mathcal{O}(h^2) . \tag{168}$$

Identification of the right hand sides of (163) and (168) yields a set of constraint conditions on the automaton reaction probabilities:

$$P_\tau(\boldsymbol{\alpha}) = h \sum_{j=1}^{r} (\nu_\tau^j - \bar{\nu}_\tau^j) \left\{ -k_j \prod_{\kappa=1}^{n} m^{\nu_\kappa^j} \binom{m - \nu_\kappa^j}{\alpha_\kappa - \nu_\kappa^j} \binom{m}{\alpha_\kappa}^{-1} + \right.$$

$$\left. k_{-j} \prod_{\kappa=1}^{n} m^{\bar{\nu}_\kappa^j} \binom{m - \bar{\nu}_\kappa^j}{\alpha_\kappa - \bar{\nu}_\kappa^j} \binom{m}{\alpha_\kappa}^{-1} \right\} , \tag{169}$$

where $\binom{b}{a} = 0$ if $a > b$. These conditions on $P_\tau(\boldsymbol{\alpha})$ ensure that the mean field description of the spatially homogeneous automaton dynamics is the same as the mass action rate law. Since $P_\tau(\boldsymbol{\alpha})$ is an average value of the particle number change, (169) fixes only this particular moment; higher moments which reflect correlations of particle number changes are not determined by the mean field mass action rate law. There is no guarantee that the full automaton dynamics will yield such a mass action rate law. Indeed, one of the main interests in the automaton dynamics is that it contains correlations and fluctuations that go beyond such mean field descriptions. Nevertheless, this scheme is very useful for tuning the reactive lattice gas parameters so that the automaton is likely to yield macroscopic behavior of a given type, an interesting feature as it is obtained from a mesoscopic description.

Consider the case where the rate law is characterized by the set of rate constants $\{k_{\pm j} \; : \; j = 1, \cdots, r\}$. Since $P(\boldsymbol{\alpha}, \boldsymbol{\beta})$ are transition probabilities, $P_\tau(\boldsymbol{\alpha})$ as given by (164) is bounded: $-\alpha_\tau \leq P_\tau(\boldsymbol{\alpha}) \leq m - \alpha_\tau$. In view of the time scaling factor $h$ in (169) the values of the $P_\tau(\boldsymbol{\alpha})$ may be scaled to lie arbitrarily close to zero; however, their signs cannot be changed. Consider, for example, the extreme cases where $\alpha_\tau = 0$ or $m$. Then, from (164), we have $P_\tau(\boldsymbol{\alpha}) \geq 0$ if $\alpha_\tau = 0$ and $P_\tau(\boldsymbol{\alpha}) \leq 0$ if $\alpha_\tau = m$. The kinetic parameters on the right hand side of (169) must be such that these sign conditions are met. If the sign conditions are met the scaling factor $h$ can always be chosen such that $P(\boldsymbol{\alpha},\boldsymbol{\beta})$ is a probability. If the sign conditions are not met then the kinetic scheme characterized by a particular set of rate constants $\{k_{\pm j} : j = 1, \cdots, r\}$ cannot be simulated directly by the automaton.[23]

---

[22]The number of channels per node $m$ must be equal to, or larger than the highest power in $\rho_\tau$ in the mass-action rate law.

Additional insight into the sign conditions on $P_\tau(\boldsymbol{\alpha})$ can be obtained by examining the fluxes at the surfaces of the hypercube $0 \leq \rho_\tau \leq m$ within which the species densities must lie as a consequence of the exclusion principle. The fluxes on the boundaries of this cube must always be directed to its interior (or be tangent to the boundary). This follows from (163) and the sign conditions determined above for $P_\tau(\boldsymbol{\alpha})$. Furthermore, one sees from the right hand side of (169) for $\alpha_\tau = 0$ or $m$ that one obtains the same sign conditions on $P_\tau(\boldsymbol{\alpha})$. Thus, no additional constraints on the reaction probability matrix are implied by the flux condition.

The elements of $\mathbf{P}$ are not uniquely determined by the set of macroscopic rate constants and the sign conditions and additional physical considerations that account for correlations in particle number changes can be used to construct a series of automaton models. This flexibility in the choice of $\mathbf{P}$ can be exploited to construct an automaton dynamics that goes beyond the mean field level.

These formal considerations suggest a strategy for the construction of the reaction probability matrix.

- Check that the set of rate constants in the mass action rate law is admissible by calculating the species fluxes at the boundaries.

- Express the order $h$ term on the right-hand side of (167) in the basis

$$\left\{ \prod_{\kappa=1}^{n} \binom{m}{\alpha_\kappa} c_\kappa^{\alpha_\kappa} (1 - c_\kappa)^{m-\alpha_\kappa} : \alpha_\kappa = 0, \ldots, m \right\} \tag{170}$$

and identify each component with the corresponding $P_\tau(\boldsymbol{\alpha})$. This yields (169).

- For each value of $\boldsymbol{\alpha}$, find a set of positive values (possibly larger than 1) of $P(\boldsymbol{\alpha}, \boldsymbol{\beta})$ satisfying (164). Physical considerations concerning correlations should be used to select the elements of $\mathbf{P}$.

- Considering all $\boldsymbol{\alpha}$, find the largest value of

$$\sum_{\substack{\boldsymbol{\beta} \\ \boldsymbol{\beta} \neq \boldsymbol{\alpha}}} P(\boldsymbol{\alpha}, \boldsymbol{\beta}) . \tag{171}$$

Select a value for the scaling factor $h$ in (169) such that the sum in (171) is less than unity. For each value of $\boldsymbol{\alpha}$ the diagonal elements are then given by

$$P(\boldsymbol{\alpha}, \boldsymbol{\alpha})) = 1 - \sum_{\substack{\boldsymbol{\beta} \\ \boldsymbol{\beta} \neq \boldsymbol{\alpha}}} P(\boldsymbol{\alpha}, \boldsymbol{\beta}) . \tag{172}$$

By construction $\mathbf{P}$ is a transition probability matrix.

This completes the specification of the algorithm for the construction of reactive lattice-gas automaton rules. There is a variety of ways in which the propagation and velocity randomization steps can be carried out and there is flexibility in the construction of the reaction probability matrix, as we have seen above.

---

[23] In such a circumstance reactive lattice gas models without exclusion can be used; these are briefly discussed in Sec. X. For some specific reactive schemes, the problem can also be solved by using the same lattice $\mathcal{L}$ for all species with an exclusion principle that forbids more that one particle to be in the same channel at the same time (instead of a stack of lattices $\mathcal{L}_\tau$ and an exclusion principle for each lattice). An important drawback of this latter method is the difficulty to assign different diffusion coefficients to different species [31].

### D. Applications

In sections VI to IX we discuss applications to specific chemical systems and examples of the construction of rules will be given.

We have argued that reactive lattice-gas automata can be used to investigate the dynamics of spatially-distributed chemically reacting systems at the mesoscopic level. The examples in the folowing sections have been chosen to illustrate a number of important spatial and temporal structures that are commonly observed in reacting media. The emphasis in these examples is on the role of fluctuations on the far-from-equilibrium dynamics and on the schemes that are used to construct the automaton models. The phenomena that we consider include pattern formation and wave propagation in bistable chemical systems, spiral waves in excitable and oscillatory media, Turing pattern formation and the dynamics of chemical systems with deterministic chaos. For each case simulation results are shown and analysed.

In a coarse-grained view of the dynamics the magnitude of the local particle number fluctuations depends on the size of the fluid volume element and the time scale on which the fluctuations are measured. The automaton reaction probability matrix **P** determines the rate at which chemical reactions take place at the nodes of the lattice and, therefore, controls the local particle number fluctuations that arise from such reactions. Thus the magnitude of the local particle number fluctuations and the correspondence between the automaton space and time scales and those of the real system are related topics which are intimately connected with the construction of the reaction probability matrix. In the course of discussing the applications we shall also describe how automaton models can be constructed that correspond to different space-time coarse grainings of the real reactive dynamics. Basically, if the automaton elementary cell volume represents a small fluid element and the automaton time step corresponds to a short time interval, the details of the reaction mechanism will be important and will determine the nature of the local fluctuations. However, if the automaton elementary cell represents a larger fluid volume and/or the automaton time step corresponds to a longer time interval, many reactions occur in a local cell in a given time interval and only the macroscopic kinetics embodied in the mass action rate law will be important; the details of the mechanism are then inessential, and indeed many reaction mechanisms are consistent with a given mass action law. We shall show how the rule construction determines the level of coarse graining at which the system is viewed.

## VI. BISTABLE SYSTEMS AND FLUCTUATIONS

### A. The Schlögl model

Bistable chemical systems show a number of characteristic types of wave propagation processes and it is interesting to examine this relatively simple class of systems from the perspective of reactive lattice-gas models, not only to illustrate the methods used to construct the automaton rules but also because fluctuations can have important effects on the dynamics.

One of the best-known examples of a model reaction scheme that gives rise to bistable states is the Schlögl model. [17] The chemical mechanism of this model is

$$A \underset{k_{-1}}{\overset{k_1}{\rightleftharpoons}} X \;,$$
$$2X + B \underset{k_{-2}}{\overset{k_2}{\rightleftharpoons}} 3X \;. \tag{173}$$

Letting the concentrations of $A$, $X$ and $B$ be $\rho_1$, $\rho_2$ and $\rho_3$, respectively, the corresponding mass action rate law is

$$\frac{d\rho_1}{dt} = -k_1\rho_1 + k_{-1}\rho_2 \;,$$
$$\frac{d\rho_2}{dt} = k_1\rho_1 - k_{-1}\rho_2 + k_2\rho_2^2\rho_3 - k_{-2}\rho_2^3 \;,$$
$$\frac{d\rho_3}{dt} = -k_2\rho_2^2\rho_3 + k_{-2}\rho_2^3 \;. \tag{174}$$

We shall use the general formalism of Sec. V to construct several variants of **P** for the Schlögl model that correspond to different coarse grainings of the physical system under different constraint conditions.

#### 1. Closed system

Consider the case where all reactive chemical species are treated on an equal footing and only the solvent molecules are taken to be uniformly distributed over the lattice. We suppose the system is closed so that there are no flows of reagents into or out of the system. The system will relax to a spatially homogeneous equilibrium state and no complex behavior is observed. Nevertheless, this case serves as a simple illustration of the methods used to construct the automaton rule.

Given the mass action rate law (174) and using (169) it follows that

$$\begin{aligned}
P_1(\boldsymbol{\alpha}) &= -r_1^+(\boldsymbol{\alpha}) + r_1^-(\boldsymbol{\alpha}) \\
P_2(\boldsymbol{\alpha}) &= r_1^+(\boldsymbol{\alpha}) - r_1^-(\boldsymbol{\alpha}) + r_2^+(\boldsymbol{\alpha}) - r_2^-(\boldsymbol{\alpha}) \\
P_3(\boldsymbol{\alpha}) &= -r_2^+(\boldsymbol{\alpha}) + r_2^-(\boldsymbol{\alpha}) \;.
\end{aligned} \tag{175}$$

where

$$r_1^+(\boldsymbol{\alpha}) = hk_1\alpha_1 \;, \quad r_2^+(\boldsymbol{\alpha}) = hk_2\frac{m}{(m-1)}\alpha_2(\alpha_2 - 1)\alpha_3 \;,$$
$$r_1^-(\boldsymbol{\alpha}) = hk_{-1}\alpha_2 \;, \quad r_2^-(\boldsymbol{\alpha}) = hk_{-2}\frac{m^2}{(m-1)(m-2)}\alpha_2(\alpha_2 - 1)(\alpha_2 - 2) \;. \tag{176}$$

Note that $\sum_\tau P_\tau(\boldsymbol{\alpha}) = 0$ since the total number of particles is conserved in the mass action law. At this stage, additional considerations are needed to completely specify the reaction probability matrix **P**.

If the automaton node is taken to correspond to a very small fluid volume element and the automaton time unit corresponds to a small real time interval, the automaton dynamics can be chosen to mimic the individual reactive collision events in the chemical mechanism. Then, the only non-zero elements that appear in **P** are those that correspond to reactions in the mechanism (173). This restriction, along with the mean field constraints, allows one to completely specify **P**. We have for $\boldsymbol{\alpha} \neq \boldsymbol{\beta}$

$$
\begin{aligned}
P(\boldsymbol{\alpha},\boldsymbol{\beta}) &= \delta_{\beta_1\alpha_1+1}\delta_{\beta_2\alpha_2-1}\delta_{\beta_3\alpha_3}(r_1^-(\boldsymbol{\alpha}) - r_1^+(\boldsymbol{\alpha})\delta_{\alpha_2 m})(1-\delta_{\alpha_1 m}) \\
&+ \delta_{\beta_1\alpha_1-1}\delta_{\beta_2\alpha_2+1}\delta_{\beta_3\alpha_3}(r_1^+(\boldsymbol{\alpha}) - r_1^-(\boldsymbol{\alpha})\delta_{\alpha_1 m})(1-\delta_{\alpha_2 m}) \\
&+ \delta_{\beta_1\alpha_1-1}\delta_{\beta_2\alpha_2}\delta_{\beta_3\alpha_3}(r_1^+(\boldsymbol{\alpha}) - r_1^-(\boldsymbol{\alpha}))\delta_{\alpha_1 m}\delta_{\alpha_2 m} \\
&+ \delta_{\beta_1\alpha_1}\delta_{\beta_2\alpha_2+1}\delta_{\beta_3\alpha_3-1}(r_2^+(\boldsymbol{\alpha}) - r_2^-(\boldsymbol{\alpha})\delta_{\alpha_3 m})(1-\delta_{\alpha_2 m}) \\
&+ \delta_{\beta_1\alpha_1}\delta_{\beta_2\alpha_2-1}\delta_{\beta_3\alpha_3+1}(r_2^-(\boldsymbol{\alpha}) - r_2^+(\boldsymbol{\alpha})\delta_{\alpha_2 m})(1-\delta_{\alpha_3 m}) \\
&+ \delta_{\beta_1\alpha_1}\delta_{\beta_2\alpha_2}\delta_{\beta_3\alpha_3-1}(r_2^-(\boldsymbol{\alpha}) - r_2^+(\boldsymbol{\alpha}))\delta_{\alpha_2 m}\delta_{\alpha_3 m} ,
\end{aligned}
\qquad (177)
$$

where $\delta_{\beta\alpha}$ is the Kronecker delta and

$$
P(\boldsymbol{\alpha},\boldsymbol{\alpha}) = 1 - \sum_{\boldsymbol{\beta}\neq\boldsymbol{\alpha}} P(\boldsymbol{\alpha},\boldsymbol{\beta}) . \qquad (178)
$$

For the input state $\boldsymbol{\alpha} = (m,m,m)$ the condition $P_\tau((m,m,m)) \leq 0$ for $\tau = 1,2,3$ leads to $k_1 = k_{-1}$ and $k_2 = k_{-2}$. Furthermore, for the input states, listed in the argument, the following conditions $P_1((m,m,\alpha_3)) = P_2((m,m,m)) = P_3((\alpha_1,m,m)) = 0$ imply for the output states $\boldsymbol{\beta}$ that

$$
\begin{aligned}
P((m,m,\alpha_3),\boldsymbol{\beta}) &= 0, \quad \beta_1 \neq m , \\
P((m,m,m),\boldsymbol{\beta}) &= 0, \quad \beta_2 \neq m , \\
P((\alpha_1,m,m),\boldsymbol{\beta}) &= 0, \quad \beta_3 \neq m .
\end{aligned}
\qquad (179)
$$

Since this scheme allows only those reactions that appear in the mechansim (173) (plus those needed to obtain the mean field rate law because of exclusion), this level of description is perhaps the closest that the automaton dynamics can be made to correspond to the true microscopic reactive dynamics since the fictitious automaton particles undergo reactive collisions like those of the real chemical species.

If one imagines that the automaton elementary cell corresponds to a larger fluid volume element or the automaton time corresponds to a longer real time interval many reactions can occur in the volume element during one automaton time step. In this circumstance the details of the mechanism will not play as important a role and the overall kinetics embodied in the rate law will suffice to determine the local reactive events in the automaton. It is now a simple matter to write an expression for **P** that is consistent with the overall kinetics and the constraint conditions. We have for $\boldsymbol{\alpha} \neq \boldsymbol{\beta}$

$$
P(\boldsymbol{\alpha},\boldsymbol{\beta}) = \sum_{\tau=1}^{3} \left[ p_\tau^+(\boldsymbol{\alpha})\delta_{\beta_\tau\alpha_\tau+1} + p_\tau^-(\boldsymbol{\alpha})\delta_{\beta_\tau\alpha_\tau-1} \right] \prod_{\kappa\neq\tau} \delta_{\beta_\kappa\alpha_\kappa} , \qquad (180)
$$

where

$$
\begin{aligned}
p_\tau^+(\boldsymbol{\alpha}) &= q_\tau^+(\boldsymbol{\alpha})(1-\delta_{\alpha_\tau m}) , \\
p_\tau^-(\boldsymbol{\alpha}) &= q_\tau^-(\boldsymbol{\alpha}) - q_\tau^+(\boldsymbol{\alpha})\delta_{\alpha_\tau m} ,
\end{aligned}
\qquad (181)
$$

and

$$
\begin{aligned}
q_1^\pm(\boldsymbol{\alpha}) &= r_1^\mp(\boldsymbol{\alpha}) , \\
q_2^\pm(\boldsymbol{\alpha}) &= r_1^\mp(\boldsymbol{\alpha}) + r_2^\pm(\boldsymbol{\alpha}) , \\
q_3^\pm(\boldsymbol{\alpha}) &= r_2^\mp(\boldsymbol{\alpha}) .
\end{aligned}
\qquad (182)
$$

### 2. Open system

The reaction dynamics of open chemical systems is far richer than that of closed systems close to equilibrium. It is not difficult to extend the automaton dynamics to this situation. Suppose the reaction is forced out of equilibrium by flows of one or more of the reagents into the system through the boundaries. In the Schlögl model we suppose $A$ and $B$ are fed into the system from reservoirs of these reagents. In this case such flow terms must be appended to (174):

$$\frac{d\rho_1}{dt} = k_A(\rho_1^0 - \rho_1) - k_1\rho_1 + k_{-1}\rho_2 \;,$$
$$\frac{d\rho_2}{dt} = k_1\rho_1 - k_{-1}\rho_2 + k_2\rho_2^2\rho_3 - k_{-2}\rho_2^3 \;,$$
$$\frac{d\rho_3}{dt} = -k_2\rho_2^2\rho_3 + k_{-2}\rho_2^3 + k_B(\rho_3^0 - \rho_3) \;, \qquad (183)$$

where a superscript 0 is used to denote a constant reservoir concentration. This is the usual form of the rate law for a reaction in a continuously-stirred tank reactor (CSTR). These flow terms can be represented in the mechanism by adding reactions of the form[24]

$$A_0 \underset{k_A}{\overset{k_A}{\rightleftharpoons}} A \;,$$
$$B_0 \underset{k_B}{\overset{k_B}{\rightleftharpoons}} B \;, \qquad (184)$$

to the Schlögl mechanism (173). If there is one reservoir containing the $A$ and $B$ species then the feed rates are equal and $k_A = k_B = k_f$. If we let $P_\tau^f(\boldsymbol{\alpha})$ denote $P_\tau(\boldsymbol{\alpha})$ with feed terms then

$$P_1^f(\boldsymbol{\alpha}) = P_1(\boldsymbol{\alpha}) + k_A(\rho_1^0 - \alpha_1) \;,$$
$$P_2^f(\boldsymbol{\alpha}) = P_2(\boldsymbol{\alpha}) \;,$$
$$P_3^f(\boldsymbol{\alpha}) = P_3(\boldsymbol{\alpha}) + k_B(\rho_3^0 - \alpha_3) \;. \qquad (185)$$

The reaction probability matrix may now be constructed for this modified reaction mechanism and one of its possible forms is

$$P^f(\boldsymbol{\alpha},\boldsymbol{\beta}) = P(\boldsymbol{\alpha},\boldsymbol{\beta})$$
$$+ \delta_{\beta_1\alpha_1+1}\delta_{\beta_2\alpha_2}\delta_{\beta_3\alpha_3}hk_A\rho_{01}(1-\delta_{\alpha_1 m})$$
$$+ \delta_{\beta_1\alpha_1-1}\delta_{\beta_2\alpha_2}\delta_{\beta_3\alpha_3}hk_A(\alpha_1 - \rho_{01}\delta_{\alpha_1 m})$$
$$+ \delta_{\beta_1\alpha_1}\delta_{\beta_2\alpha_2}\delta_{\beta_3\alpha_3+1}hk_B\rho_{03}(1-\delta_{\alpha_3 m})$$
$$+ \delta_{\beta_1\alpha_1}\delta_{\beta_2\alpha_2}\delta_{\beta_3\alpha_3-1}hk_B(\alpha_3 - \rho_{03}\delta_{\alpha_3 m}) \;. \qquad (186)$$

Note that in this model of the far-from-equilibrium system all three chemical species, $A$, $X$ and $B$ fluctuate.

### 3. One-variable model

An even simpler description of the reaction is possible by assuming that the constrained $A$ and $B$ species are uniformly distributed over the lattice and their numbers do not fluctuate. This leads to a one-variable description of the reaction dynamics. The results for this case may be obtained from those given above in the following way: on the r.h.s. of (176) $\alpha_1$ is replaced by $\rho_1^0$ and $\alpha_3$ is replaced by $\rho_3^0$ where $\rho_1^0$ and $\rho_3^0$ are now constant concentrations in the system. In (175) only $P_2(\alpha_2)$ is relevant with $\boldsymbol{\alpha}$ replaced by $\alpha_2$. Given this value of $P_2(\alpha_2)$ the reaction probability matrix $P(\alpha_2, \beta_2)$ is easily constructed. For example, a form for $P$ that includes only those reactions that appear in the mechanism (and taking into account exclusion) is for $\alpha_2 \neq \beta_2$

$$P(\alpha_2, \beta_2) = [(r_1^-(\alpha_2) + r_2^-(\alpha_2))\delta_{\alpha_2-1\beta_2} + (r_1^+(\alpha_2) + r_2^+(\alpha_2))\delta_{\alpha_2+1\beta_2}](1 - \delta_{\alpha_2 m})$$
$$+ [-r_1^+(\alpha_2) + r_1^-(\alpha_2) + r_2^+(\alpha_2) - r_2^-(\alpha_2)]\delta_{\alpha_2 m} , \qquad (187)$$

which is the analog of (177). Naturally, it is possible to construct other versions of $P(\alpha_2, \beta_2)$ for this one-variable case. Now, of course, only fluctuations in the chemical intermediate $X$ are possible.

---

[24] We assume the reactions involving $A$ and $B$ that produce species $X$ are catalysed and do not occur in the reservoirs.

*4. Simulation results*

*One-variable model*
We begin our discussion of the LGA simulation results by considering the simple one-variable model for the Schlögl reaction and use (187) for the reaction probability matrix. [32] In earlier studies a different reaction probability matrix was used. [16] We dispense with the subscript 2 for the species $X$ since this is the only species whose dynamics is followed. The reaction-diffusion equation is

$$\frac{\partial \rho(\mathbf{r}, t)}{\partial t} = k_1 a - k_{-1}\rho(\mathbf{r}, t) + k_2 b \rho^2(\mathbf{r}, t) - k_{-2}\rho^3(\mathbf{r}, t) + D\nabla^2 \rho(\mathbf{r}, t)$$
$$= -\frac{\delta \mathcal{F}[\rho(\mathbf{r}, t)]}{\delta \rho(\mathbf{r}, t)} , \qquad (188)$$

where $a = \rho_1^0$ and $b = \rho_3^0$. In the second equality above we introduced the local free energy functional

$$\mathcal{F} = -\int d\mathbf{r} \left\{ -V(\rho) + \frac{1}{2}D|\nabla \rho|^2 \right\} , \qquad (189)$$

where the potential $V(\rho)$ is defined as

$$V(\rho) = -k_1 a \rho + \frac{k_{-1}}{2}\rho^2 - \frac{k_2 b}{3}\rho^3 + \frac{k_{-2}}{4}\rho^4 . \qquad (190)$$

Equation (188) is just the time-dependent Ginzburg-Landau equation for a non-conserved order parameter field which has been studied often in other contexts. [33] If a random noise term is appended to (188) we obtain Model A of critical phenomena [34]:

$$\frac{\partial \rho(\mathbf{r}, t)}{\partial t} = -\frac{\delta \mathcal{F}[\rho(\mathbf{r}, t)]}{\delta \rho(\mathbf{r}, t)} + \xi(\mathbf{r}, t) , \qquad (191)$$

where $\xi(\mathbf{r}, t)$ is a Gaussian white noise process,

$$\langle \xi(\mathbf{r}, t)\xi(\mathbf{r}', t') \rangle = 2\delta(\mathbf{r} - \mathbf{r}')\delta(t - t') . \qquad (192)$$

These considerations place the Schlögl model investigation within a class of well-studied models for both the deterministic and stochastic dynamics. Some of the main features of interest in the investigation of these systems are: (*i*) the form of the interface separating the stable states, which is known analytically for the deterministic model in view of the fact that the dynamics derives from a one-dimensional potential; (*ii*) the scaling of the dynamic structure factor on long distance and time scales and its space and time dependence on short distance and time scales; and (*iii*) nucleation and growth dynamics. Some of these features depend crucially on the existence of fluctuations; for example, nucleation-induced growth and the short distance and time dynamics, while others, such as the behavior in the scaling regime, do not. Thus, the model allows one to determine the ability of the automaton dynamics to accurately reproduce the behavior in the deterministic macroscopic regime and to explore the effects of fluctuations on certain aspects of the dynamics.

The automaton calculations were carried out on square $N \times N$ lattices with $N = 512$ and periodic boundary conditions using (187) for the reaction probability matrix. The two checkerboard sublattices were treated separately in the data analysis. The length scale of the local inhomogeneities in the automaton simulations is determined by the relative magnitudes of diffusion and reaction. In this study we have taken (cf. (160)) $\ell = 6$ ($D = 3/2$ in automaton units of lattice units squared per time step). This choice eliminates most small-scale, short-time reactive recollision events which can lead to significant correlation corrections to the steady state average concentrations. [16]

The homogeneous steady states of (188) are given by the solutions of $k_1 a - k_{-1}\rho + k_2 b \rho^2 - k_{-2}\rho^3 = 0$, and are sketched in Fig. 4 as a function of $k_{-1}$ for fixed values of the other parameters (cf. Fig. 4 caption).

FIG. 4. Schlögl steady states as a function of $k_{-1}$. The solid line is the deterministic solution while the points are the automaton simulation results. The system parameters are: $k_1 a = 0.001$, $8k_2 b/3 = 0.095$ and $16k_{-2} = 0.245$.

The two stable steady states will be called $\rho_1^*$ and $\rho_2^*$ while the unstable steady state will be called $\rho_0^*$. Since the dynamics derives from the potential $V(\rho)$ the stability of the coexisting states can be deduced from the relative depths of the potential minima. While the less-stable state is metastable, its lifetime can be very long and fluctuation-induced transitions will occur rarely in simulations on finite systems. The deterministic steady state solutions are compared with finite-duration automaton simulations in Fig. 4. The initial condition was taken to be a random distribution of particles over the nodes of the lattice with mean concentrations corresponding to the deterministic fixed point values. The mean concentration was determined from an average over the lattice and over a time $T = 2000$ steps, following a transient period of 1000 time steps. The figure shows that the average concentration corresponds closely to that of the deterministic system, with small deviations in the interiors of the bistable domain and larger deviations near its boundaries which are due to local fluctuations in the $X$ concentration. On the time scale of the simulations transitions between steady state branches occur only at points very close to the edges of the bistable domain.

Chemical consumption fronts can be generated by taking initial conditions where each half of the lattice has average concentration equal to one of the deterministic stable states. The wave front connecting these two stable states will move in such a way that the more-stable phase consumes the less-stable phase. The point of zero wave velocity corresponds to equistability of the two phases. [17,35] The automaton wave front connecting two steady states with the same stability is shown in Fig. 5.

FIG. 5. Interfacial profile separating two equally-stable states for $k_{-1} = 0.0195$. The wave velocity is zero. The profile was obtained by averaging over the $y$-coordinate on the lattice and is plotted as a function of the $x$ coordinate. The solid line is the deterministic prediction.

The interfacial profile follows easily from the solution of (188) and for the above equistability case it is given by

$$\rho(x) = \rho_0^* + \frac{Q}{\sqrt{k_{-2}}} \tanh \frac{Qx}{\sqrt{2D}} , \qquad (193)$$

where $\rho_0^* = k_2 b/3k_{-2}$, and $Q = \{(k_2 b)^2/3k_{-2} - k_{-1}\}^{1/2}$. From Fig. 5 one sees that the automaton simulations agree well with the predictions of the deterministic model. In fact, it is also possible to derive an expression for the interfacial profile when space is considered as a discrete variable, a situation that is more directly applicable to the automaton simulations. In this case the interfacial profile is determined by a conservative two-dimensional mapping and points on the interface are given by alternate intersections of the stable and unstable manifolds connecting the hyperbolic fixed points that correspond to the temporally stable steady states. [36] Concerning the discrete nature of the automaton, one should also keep in mind that the behavior of a very sharp front can be anisotropic because the isotropy may not yet been recovered at the lenght scale of the width of the front.

It is also interesting to study the analog of a critical quench where the system evolves from the unstable state. The initial decay from the unstable state is strongly influenced by fluctuations; once well-defined domains of the two stable phases have formed a deterministic description should be appropriate. In this long-time regime the dynamics is driven by the curvature of the boundaries separating the stable phases. In an infinite system, although the average order parameter ($\phi = \rho - \rho_0^*$) is zero, domains of arbitrarily large size exist in the system. In finite systems different realizations of the evolution process lead to pure $\rho_1^*$ or $\rho_2^*$ phases (or mixtures of these phases separated by planar interfaces) but averaged over realizations the order parameter will again be zero. The evolution of the system during phase separation may be characterized by the nonequilibrium correlation function

$$C(\mathbf{r}, t) = \langle N^{-2} T^{-1} \sum_{t'=1}^{T} \sum_{\mathbf{r}' \in \mathcal{L}} \phi(\mathbf{r}' + \mathbf{r}, t' + t) \phi(\mathbf{r}', t') \rangle , \qquad (194)$$

whose space Fourier transform is the *intermediate scattering function* $S(k, t)$. Here the angle bracket signifies an average over different realizations of the evolution process, and $\phi(\mathbf{r}, t)$ is now $\sum_{i=1}^{4} \eta_i(\mathbf{r}, t) - \rho_0^*$. If the domain size $R(t)$ is the only characteristic length in the system and its evolution is governed by interfacial curvature, $R(t) \sim t^{1/2}$, the intermediate scattering function will satisfy the scaling equation [33]

$$S(k, t) \sim t F(k t^{1/2}) , \qquad (195)$$

in two dimensions with $F$ a universal scaling function.

An example of phase separation following a critical quench is shown in Fig. 6. In the simulation $k_{-1} = 0.0195$ (the equistable point) and the system was uniformly seeded with $X$ particles with average concentration corresponding to the deterministic unstable steady state. Sharp boundaries form as time increases and slowly deform due to diffusive motion of the interfaces. The final panel in Fig. 6 shows stable phases separated by a nearly planar interface. The intermediate scattering function has been computed using reactive lattice gas dynamics and the scaling relation (195) was found to be verified for the late stage dynamics. [32] Since this reactive dynamics has a non-conserved order parameter one expects $R(t) \sim t^{1/2}$ scaling. This should be contrasted with conserved order parameter growth where the interface dynamics couples to that of the bulk and gives rise to a $t^{1/3}$ growth law. [33]

FIG. 6. One realization of the evolution from the unstable state following a critical quench.

*Three-variable model*

We next briefly consider the effects of treating the reactive dynamics of all three chemical species in the Schlögl model. [37] Experimental studies of pattern formation processes in chemical systems are usually carried out under well-controlled conditions using continuously-fed-unstirred reactors (CFUR). [38] In these reactors material is fed into the reactor (usually containing a gel or other medium which is designed to suppress fluid flow) by well-stirred reservoirs of chemicals whose concentrations are constant. One may then envisage a quasi two-dimensional reacting medium, typically a thin layer of a gel or porous medium, whose surfaces are in contact with reservoirs containing uniformly distributed chemicals with constant concentrations. The reaction probability matrices in Sec. VI A 2 were constructed to treat such situations.

The steady state bifurcation structure depends on the flow rates $k_A$ and $k_B$ in (184). The steady states of (183) are given by the solutions of the equations,

$$\rho_1^* = \frac{k_{-1}\rho_2^* + k_A \rho_1^0}{k_1 + k_A}, \tag{196}$$

$$\rho_3^* = \frac{k_{-2}\rho_2^{*3} + k_B \rho_3^0}{k_2 \rho_2^{*2} + k_B},$$

where $\rho_2^*$ can be obtained from the roots of

$$-\left(k_{-1}k_2 k_A + k_1 k_{-2} k_B + k_{-2} k_A k_B\right) \rho_2^{*3} + \tag{197}$$
$$\left(k_1 k_2 k_A \rho_1^0 + k_2 k_B (k_A + k_1)\rho_3^0\right) \rho_2^{*2} - \left(k_{-1} k_A k_B\right) \rho_2^* + k_1 k_A k_B \rho_1^0 = 0.$$

In the automaton model $k_1 = k_{-1}$ and $k_2 = k_{-2}$. For $k_A = k_B = k_f$, the steady state bifurcation structure is shown in Fig. 7 as a function of $k_1$ for two values of the feed rate constant $k_f$.

FIG. 7. Three-variable Schlögl steady states ($c_\tau = \rho_\tau / m$) as a function of $k_1$ for $k_f = 0.2$ and $k_f = 0.02$. The solid line is the deterministic solution while the points are the automaton simulation results. The light solid line represents the one-variable model results. The remaining system parameters are: $8k_2 = 0.0153$, $\rho_1^0 = 0.666$ and $\rho_3^0 = 2.32$.

For low feed rates one sees that the bistability regime is significantly altered; it is flattened, shifted to higher $k_1$ values and spans a larger range of $k_1$. The three-variable automaton results, indicated by the points in these figures, are in close accord with the deterministic steady state structure for all feed rates studied. The simulation results were obtained by a space average over a $128 \times 128$ triangular lattice and a time average after a transient period. For very high feed rates, $k_f = 1.0$, the one-variable and three-variable results are indistinguishable. The hysteresis loops for the $A$ and $B$ species are compressed to a nearly line-like form and closely approximate the constant values in the one-variable model.

Further insight into the character of the dynamics can be obtained by examining the interface that separates two stable states in the spatially-distributed system. A planar interface was constructed as above by seeding each half of the lattice with particles (now $A$, $B$ and $X$ particles) whose average concentrations corresponded to the stable steady states. Two examples of interfacial profiles at equistability are shown in Fig. 8, one for an intermediate feed rate ($k_f = 0.2$) and the other for a low feed rate ($k_f = 0.02$).

FIG. 8. Three-variable Schlögl interfacial profiles at the equistable point for two feed rates, (a) $k_f = 0.2$ and (b) $k_f = 0.02$. The solid line is computed from the deterministic one-variable model and the fluctuation lines are the automaton simulation results.

In the first case the interfacial profile for the $X$ species closely approximates that from the deterministic one-variable model (193) and the $A$ and $B$ concentrations do not show noticeable systematic variations across the profile. However, for low feed rates no comparison with the one-variable model is possible since in the parameter range where the three-variable model exhibits equistability there are no real stable roots for the one-variable model. One can see that all three species display well-developed interfacial structure, indicating a complete breakdown of the one-variable model. The domain growth and scaling structure for early times also show differences when compared with the one-variable model that suppresses fluctuations in the $A$ and $B$ species. [37] These results show that the automaton dynamics is able to accurately describe the fluctuating dynamics of bistable chemical systems and, in regimes where fluctuations are unimportant, automaton results are in accord with deterministic predictions.

## B. Fluctuations and second moment constraints

In previous sections we pointed out that lattice gas automata provide a microscopic approach to cooperative phenomena and that by construction the automaton possesses intrinsically spontaneous fluctuations. Therefore the LGA method should be well suited to investigate the role of fluctuations on the macroscopic behavior of complex systems. This indeed is an important point at least for two reasons: (i) reaction-diffusion systems exhibit macroscopic and mesoscopic space- and time-dependent behavior which often is triggered or influenced by fluctuations; (ii) these fluctuations are usually not easily accessible to other methods. Now it is crucial that the simplification introduced in the microscopic LGA description should manifest neither at the macroscopic level nor at smaller (mesoscopic) scales where the fluctuations can play an important role. So in order to obtain relevant information, we should have some guarantee that LGA fluctuations capture the essential aspects of actual fluctuations [25]. This problem can be addressed at least within the limits of the linear theory of reaction-diffusion systems [39]. For simplicity we consider a one-variable system.

A standard approach for non-equilibrium steady states (far from a bifurcation point) is the Landau method where a process-independent noise is added to the phenomenonlogical RD equation linearized around a steady state ($\rho = \rho^* + \phi$)

$$\frac{\partial \phi(\mathbf{r},t)}{\partial t} = -\kappa \phi(\mathbf{r},t) + D\nabla^2 \phi(\mathbf{r},t) + \xi_D(\mathbf{r},t) + \xi_R(\mathbf{r},t), \tag{198}$$

with $\kappa$ the linearized reaction rate coefficient and where the diffusive noise $\xi_D$ and the reactive noise $\xi_R$ are taken to be additive and whose mean values and correlations are given by

$$\begin{aligned}
<\xi_D(\mathbf{r},t)> &= 0, \\
<\xi_R(\mathbf{r},t)> &= 0, \\
<\xi_D(\mathbf{r},t)\phi(\mathbf{r},t)> &= 0, \\
<\xi_R(\mathbf{r},t)\phi(\mathbf{r},t)> &= 0, \\
<\xi_D(\mathbf{r},t)\xi_R(\mathbf{r},t)> &= 0, \\
<\xi_D(\mathbf{r},t)\xi_D(\mathbf{r}',t')> &= -A_D \delta(t-t')\nabla^2\delta(\mathbf{r}-\mathbf{r}'), \\
<\xi_R(\mathbf{r},t)\xi_R(\mathbf{r}',t')> &= A_R \delta(t-t')\delta(\mathbf{r}-\mathbf{r}').
\end{aligned} \tag{199}$$

Now the main problem is to characterize the statistical properties of $\phi$. In general, these properties depend on the initial condition which may be stochastic; however, in the long time regime all details of the initial condition fade away and it becomes possible to characterize $\phi$ without reference to any particular initial state. In this regime – the only regime considered here – the first moment of $\phi(\mathbf{r},t)$ is zero:

$$<\phi(\mathbf{r},t)> = 0, \tag{200}$$

which follows from (199) and the linearity of (198). A less trivial characterization of $\phi$ is given by the density fluctuation correlation function

$$G(\mathbf{r},t) = \langle \phi(\mathbf{r}',t')\phi(\mathbf{r}'+\mathbf{r},t'+t) \rangle . \tag{201}$$

Here we are mainly interested in the static correlation function $G(r) \equiv G(r, t=0)$ which can be evaluated from (198) and (199):

---

[25] For instance the thermal lattice gas model constructed by Grosfils, Boon and Lallemand [40] has been shown to exhibit spontaneous fluctuations whose correlation function is in agreement with the dynamic structure factor measured in real fluids.

$$G(r) = \frac{A_D}{2D}\delta(r) + \frac{1}{4\pi}\left(\frac{A_R}{\kappa} - \frac{A_D}{D}\right) H_d\left(r\sqrt{\frac{\kappa}{D}}\right) \tag{202}$$

where the function $H_d(z)$ depends on the space dimension $d$. For $d = 1, 2, 3$ we have

$$\begin{aligned} H_1(z) &= \pi(\frac{\kappa}{D})^{1/2} e^{-z}; \\ H_2(z) &= \frac{\kappa}{D} K_0(z) \quad \text{(modified Bessel function)}; \\ H_3(z) &= \frac{1}{2}(\frac{\kappa}{D})^{3/2} \frac{e^{-z}}{z} \end{aligned} \tag{203}$$

respectively [39]. In the automaton, the correlation function $C(\mathbf{r}, t)$ is defined as in (201):

$$C(\mathbf{r}, t) = \langle \phi(\mathbf{r}', t') \phi(\mathbf{r}' + \mathbf{r}, t' + t) \rangle , \tag{204}$$

but here $\mathbf{r}$ and $t$ take discrete values and $\phi(\mathbf{r}, t) = \sum_{i=1}^{4} \eta_i(\mathbf{r}, t) - \rho_0^*$. In a simulation, the static correlation function $C(\mathbf{r}) \equiv C(\mathbf{r}, 0)$ can be evaluated by space and time averaging

$$C(\mathbf{r}) = N^{-2} T^{-1} \sum_{t'=t}^{T+t} \sum_{\mathbf{r}' \in \mathcal{L}} \phi(\mathbf{r}' + \mathbf{r}, t') \phi(\mathbf{r}', t') , \tag{205}$$

where $t$ must be large to reach the long time regime, and where $T$ must also be large to have a good estimate of $C$; averaging over different realizations can also be performed.

In order to compare (205) with the predictions of the Landau approach, we need to evaluate the factors $A_D$ and $A_R$ appearing in (199). To obtain the value of $A_D$, we consider a node in an equilibrium configuration at the steady state density $\rho^*$, and we evaluate the variance of the number of particles on that node

$$\text{Var}[\sum_{i=1}^{m} \eta_i] = \rho^*(1 - \rho^*/m) . \tag{206}$$

Then, we evaluate a similar variance in the Landau approach by integrating $G(\mathbf{r})$ over a domain of volume $V$ corresponding to a node of the lattice. In the limit of infrequent reactions $\kappa \to 0$, we obtain

$$\text{Var}[\rho] = V A_D / 2D . \tag{207}$$

Indentifying the two variances (206) and (207), we have

$$A_D = 2\frac{D}{V}\rho^*(1 - \rho^*/m) . \tag{208}$$

The value of the factor $A_R$ depends on the probability matrix $P$: a small value is obtained when the changes in particle number $(\beta - \alpha)$ are concentrated around their average value; conversely $A_R$ has a large value when the changes $(\beta - \alpha)$ are dispersed around their average value. Now the phenomenological rate

$$F(\rho) = \sum_{\alpha, \beta} (\beta - \alpha) \binom{m}{\alpha} (\rho/m)^\alpha (1 - \rho/m)^{m-\alpha} P(\alpha\,\beta) , \tag{209}$$

is the *average* change in particle number produced by the action of the operator $C$; analogously $A_R$ is given by the *variance* of the change in particle number [26]

---

[26] The variance is calculated with respect to the grand canonical ensemble distribution. The microcanonical distribution gives the same result in the strong-diffusion limit [22].

$$A_R(\rho) = \frac{1}{Vh} \sum_{\alpha,\beta} \left((\beta - \alpha) - F(\rho)\right)^2 \binom{m}{\alpha} (\rho/m)^\alpha (1 - \rho/m)^{m-\alpha} P(\alpha\,\beta) \ . \tag{210}$$

This is an important relation which allows one to control the level of the reactive noise in the automaton. Indeed, for a given macroscopic rate $F(\rho)$, it is possible to independently prescribe the function $A_R(\rho)$, at least within some limits [39].

Figure 9 shows the static density correlation function $G(r)$ for different microscopic dynamics corresponding to a fixed linear macroscopic rate law. Notice that the system exhibits negative spatial correlations when the reactive noise amplitude is minimized ($A_R\kappa^{-1} < A_D D^{-1}$) and positive correlations in the opposite case. The agreement between simulation data and theoretical predictions is seen to be excellent. The effect of dimensionality is also shown in Fig. 9: clearly the range of spatial correlations decays with the dimension of the system as intuitively expected.

FIG. 9. Left panel: Spatial fluctuations correlation function $G(r)$ in a two-dimensional reactive lattice gas; parameters : $A_D/D = 0.1$ and $A_R/\kappa = 0.05, 0.1$, and $0.225$ (see text for explanation). Error bars are smaller than the size of the symbols. Right panel: Static density fluctuations correlation function $G(r)/G(1)$ in one- (triangles), two- (circles) and three- (crosses) dimensional lattice gases. Symbols are simulation data and solid curves are obtained from (202) and (203) .

These observations show that the steady state fluctuations in reactive LGA are consistent with the predictions obtained from the Landau theory (at least within the limits of linear theory). This is an important result in that two very different approaches yield the same results.

## VII. EXCITABLE MEDIA AND SPIRAL WAVES

Chemical waves in excitable media are some of the most important types of wave processes seen in physical systems because they play a major role in the operation of the nervous system, the heart as well as a variety of other biological processes. [42] In addition, they have been shown to be involved in catalytic oxidation processes on metal surfaces [43] and other chemical systems. The Belousov-Zhabotinsky reaction [44] can be carried out under excitable conditions and typically displays such chemical waves. [45,46]

A chemical system is excitable if it has a stable fixed point and responds to perturbations in the following way: initial conditions obtained from small perturbations of the fixed point give rise to trajectories that make small excursions in phase space or return directly to the resting state in a short time. Perturbations that exceed a threshold value give rise to trajectories that make a large excursion in phase space before returning to the resting state. During such long excursions the system is refractory and insensitive to perturbation. If the system has spatial extent, the excitable medium will respond to local perturbations by producing waves of excitation with various geometries that travel through the system. The spatial characteristics of the wave reflect the temporal dynamics of excitability described above: the wave front corresponds to the system in its excited state while the tail of the wave is refractory as it recovers to the fixed point state. As noted above many chemical systems show such excitable behavior and we shall focus on one such system below: the Selkov model [47].

Our aim in this subsection is to give a simple illustration of the fact that reactive lattice gas automata can be used to study excitable media and to examine how concentration fluctuations can give rise to the spontaneous formation of waves in excitable media and influence their propagation.

### A. The Selkov Model

The Selkov reaction [47]

$$A \underset{k_{-1}}{\overset{k_1}{\rightleftharpoons}} X,$$

$$X + 2Y \underset{k_{-2}}{\overset{k_2}{\rightleftharpoons}} 3Y,$$

$$Y \underset{k_{-3}}{\overset{k_3}{\rightleftharpoons}} B, \tag{211}$$

was originally constructed to model certain aspects of glycolysis. Here we are not concerned with its basis in real chemistry but simply use it to illustrate wave propagation in excitable media. The mass action rate law of the above reaction is

$$\frac{d\rho_1}{dt} = k_1 a - k_{-1}\rho_1 - k_2\rho_1\rho_2^2 + k_{-2}\rho_2^3,$$
$$\frac{d\rho_2}{dt} = k_{-3}b - k_3\rho_2 + k_2\rho_1\rho_2^2 - k_{-2}\rho_2^3, \tag{212}$$

where $\rho_1$ and $\rho_2$ refer to the concentrations of the intermediates $X$ and $Y$, respectively. The reversible version of the reaction was investigated in the well-stirred limit by Rehmus et al. [48] and was shown to display a wide variety of different attractors, ranging from fixed points of various types to bistable states and limit cycles. We assume that the concentrations $\rho_A = a$ and $\rho_B = b$ of the $A$ and $B$ species are held fixed by external constraints and are taken to be the control parameters along with the rate constants in the model. They are treated in the same mean field approximation as the $A$ and $B$ species in the one-variable Schlögl model. It is possible to choose these parameters such that the system is excitable. The excitability of the dynamics is demonstrated in Fig. 10 which shows the evolution from two initial states that are above the threshold.

FIG. 10. Two phase plane trajectories for the Selkov model in the excitable region. Both trajectories start from initial conditions (open circles) that exceed the threshold for excitability and undergo long excursions in phase space before return to the fixed point, which is indicated by a heavy dot. The concentrations are scaled by $(k_2/k_3)^{1/2}$ and the time is scaled by $k_3$.

Note that the system does indeed make long excursions in phase space before return to the fixed point. Initial conditions that lie close to the fixed point relax directly back to it and do not make such large phase space excursions.

One can apply the general formalism of Sec. V to construct a variety of reaction probability matrices for the reactive dynamics, as illustrated above for the Schlögl model. From the mass action rate law (212) and (169) we obtain the relations

$$P_1(\boldsymbol{\alpha}) = r_1^+(\boldsymbol{\alpha}) - r_1^-(\boldsymbol{\alpha}) - r_2^+(\boldsymbol{\alpha}) + r_2^-(\boldsymbol{\alpha}) ,$$
$$P_2(\boldsymbol{\alpha}) = r_2^+(\boldsymbol{\alpha}) - r_2^-(\boldsymbol{\alpha}) + r_3^+(\boldsymbol{\alpha}) - r_3^-(\boldsymbol{\alpha}) , \tag{213}$$

where

$$r_1^+(\boldsymbol{\alpha}) = hk_1 a , \qquad\qquad r_1^-(\boldsymbol{\alpha}) = hk_{-1}\alpha_1 ,$$
$$r_2^+(\boldsymbol{\alpha}) = \frac{hk_2 m}{(m-1)}\alpha_1\alpha_2(\alpha_2 - 1) , \quad r_2^-(\boldsymbol{\alpha}) = \frac{hk_{-2}m^2}{(m-1)(m-2)}\alpha_2(\alpha_2-1)(\alpha_2-2) , \tag{214}$$
$$r_3^+(\boldsymbol{\alpha}) = hk_{-3}b , \qquad\qquad r_3^-(\boldsymbol{\alpha}) = hk_3\alpha_2 .$$

If we construct $P(\boldsymbol{\alpha},\boldsymbol{\beta})$ so that only the steps in the Selkov mechanism are allowed (subject to exclusion constraints) we find for $\boldsymbol{\alpha} \neq \boldsymbol{\beta}$ [18]

$$\begin{aligned}P(\boldsymbol{\alpha},\boldsymbol{\beta}) = \delta_{\beta_2\alpha_2}\Big\{ &\big[r_1^+(\boldsymbol{\alpha})\delta_{\beta_1\alpha_1+1} + r_1^-(\boldsymbol{\alpha})\delta_{\beta_1\alpha_1-1}\big](1-\delta_{\alpha_1 m}) \\
&+ (r_1^-(\boldsymbol{\alpha}) - r_1^+(\boldsymbol{\alpha}))\delta_{\alpha_1 m}\Big\} \\
+ \Big\{ &\big[r_2^+(\boldsymbol{\alpha})\delta_{\beta_1\alpha_1-1}\delta_{\beta_2\alpha_2+1} + r_2^-(\boldsymbol{\alpha})\delta_{\beta_1\alpha_1+1}\delta_{\beta_2\alpha_2-1}\big](1-\delta_{\alpha_1 m})(1-\delta_{\alpha_2 m}) \\
&+ (r_2^+(\boldsymbol{\alpha}) - r_2^-(\boldsymbol{\alpha}))(\delta_{\alpha_1 m} - \delta_{\alpha_2 m})\Big\} \\
+ \delta_{\beta_1\alpha_1}\Big\{ &\big[r_3^-(\boldsymbol{\alpha})\delta_{\beta_2\alpha_2-1} + r_3^+(\boldsymbol{\alpha})\delta_{\beta_2\alpha_2+1}\big](1-\delta_{\alpha_2 m}) \\
&+ (r_3^-(\boldsymbol{\alpha}) - r_3^+(\boldsymbol{\alpha}))\delta_{\alpha_2 m}\Big\} \\
+ \delta_{\alpha_1 m}&\delta_{\alpha_2 m}\delta_{\beta_2\alpha_2-1}\delta_{\beta_1\alpha_1-1}(r_2^+(\boldsymbol{\alpha}) - r_2^-(\boldsymbol{\alpha})) .\end{aligned} \tag{215}$$

If instead we simply demand that the mass action rate law be satisfied but the individual steps in the mechanism need not be taken into account, one of the possible choices for the non-diagonal elements ($\boldsymbol{\alpha} \neq \boldsymbol{\beta}$) of $P(\boldsymbol{\alpha},\boldsymbol{\beta})$ is

$$P(\boldsymbol{\alpha},\boldsymbol{\beta}) = \sum_{\tau=1}^{2} \big[p_\tau^+(\boldsymbol{\alpha})\delta_{\beta_\tau\alpha_\tau+1} + p_\tau^-(\boldsymbol{\alpha})\delta_{\beta_\tau\alpha_\tau-1}\big] \prod_{\kappa(\kappa\neq\tau)}\delta_{\beta_\kappa\alpha_\kappa} , \tag{216}$$

where

$$\begin{aligned}p_\tau^+(\boldsymbol{\alpha}) &= q_\tau^+(\boldsymbol{\alpha})(1-\delta_{\alpha_\tau m}) , \\
p_\tau^-(\boldsymbol{\alpha}) &= q_\tau^-(\boldsymbol{\alpha}) - q_\tau^+(\boldsymbol{\alpha})\delta_{\alpha_\tau m} ,\end{aligned} \tag{217}$$

and

$$q_1^\pm(\boldsymbol{\alpha}) = r_1^\pm(\boldsymbol{\alpha}) + r_2^\mp(\boldsymbol{\alpha}) \;,$$
$$q_2^\pm(\boldsymbol{\alpha}) = r_2^\pm(\boldsymbol{\alpha}) + r_3^\pm(\boldsymbol{\alpha}) \;. \tag{218}$$

### B. Simulation results

Simulations of the Selkov excitable medium were carried out using (215) on square lattices with $m = 4$ with the 4-equi-rotation rule to scramble the velocities. [18] Kinetic parameters were chosen to lie in the excitable region and all diffusion coefficients were selected to be equal. Their magnitudes were controlled by varying $\ell_1$ and $\ell_2$ which were taken to be equal (see *Automaton rule* in section V). In the simulations described below the system was first allowed to relax to the fixed point. If diffusion is sufficiently strong the system will be spatially homogeneous on average, exhibiting only small local fluctuations away from the fixed point value. A growing ring of excitation can be initiated in the automaton by selecting the average concentration within a disk of radius $R$ to differ from the steady state concentration. Provided the concentration differs sufficiently from that of the steady state and the radius is larger than some critical value $R_c$, this local concentration perturbation will excite nodes in the perimeter of the disk leading to a large concentration change before relaxation back to the fixed point concentration occurs. This process will generate a traveling wave of excitation that moves out from the disk. The chemical wave has a refractive tail and leaves behind unexcited medium. An example of this ring growth is shown in Fig. 11.

FIG. 11. Lattice gas automaton growth of a ring of excitation for the Selkov model. The system parameters are: $k_1 a = 0.00008385$, $k_{-1} = 0.1 k_3$, $k_3 = 0.0005$, $k_2 = k_{-2} = 0.01$ and $k_{-3} b = 0.000002326$ with $D_1 = D_2 = 3/4$. Note that the values of the $k_r$'s are indicative of the sensitivity of the system behavior to the parameter values. The simulations were carried on a $1024 \times 1024$ square lattice.

Notice that apart from small fluctuations, the rings are circular in shape, while simple cellular automaton rules often produce square waves with sharp corners. [49,50]

At this point is worth stressing that the mass action rate law follows by construction from the automaton rules. As a result all wave characteristics like wave dispersion and velocity are automatically determined once the mass action rate law, or better, the reaction mechanism is specified. This makes the construction of lattice gas models that reflect the true reactive dynamics and wave propagation processes a simple and straightforward task for any reactive system. Complex automaton rules that are designed to incorporate such features have been constructed to mimic the details of excitable kinetics. [51,52]

In the above simulation the spontaneous fluctuations have a spatial correlation length that is small compared to the critical radius $R_c$. However, if we reduce $D$, the critical radius – which scales as $R_c \approx$ $D/v_c \sim D^{1/2}$ – will decrease until spontaneous fluctuations can produce perturbations that exceed $R_c$. Here $v_c$ is the velocity of a planar chemical wave, which depends on the diffusion coefficient as $v_c \sim D^{1/2}$. [53] Once fluctuations can produce threshold-exceeding perturbations, the medium will spontaneously generate waves of excitation. This is shown in Fig. 12. The various frames in this figure show how fluctuations can continuously produce waves of excitation that spread over the surface of the 2-d system.

FIG. 12. Spontaneous nucleation of rings of excitation. Parameters are the same as Fig. 11 except that $D_1 = D_2 = 1/4$.

The generic form for a chemical wave in an excitable medium is a spiral since a planar or circular wave front fragmenting its free ends will curl to form the core of a spiral wave, provided the medium is sufficiently excitable. There is an extensive literature on the conditions for spiral wave formation in excitable media. [54] In addition, the dynamics of the spiral core itself need not be simple and can undergo bifurcations to periodic, quasiperiodic and perhaps chaotic patterns. [46] In three-dimensional media the spiral core, a topological defect, is drawn out into a filament and 3-d spirals such as scroll waves are formed. [54] Reactive lattice gas automata allow one to investigate such phenomena as well how they are affected by fluctuations. While detailed, quantitative investigations of the effects of fluctuations on spiral waves in excitable media have not been carried out, preliminary studies indicate some of the phenomena to be expected, such as for example, the irregular wave fronts seen in the simulations described below.

Here we simply demonstrate that the reactive lattice gas automaton can yield spiral waves with the correct qualitative features. A spiral wave can be generated from a ring of excitation by shearing the ring to produce free ends which can form the cores of two counter rotating spiral waves.

Figure 13 shows the formation and evolution of a spiral wave pair.

FIG. 13. Formation and evolution of a pair of counter rotating spiral waves produced by removing one half of a growing ring. Parameters are the same as in Fig. 11.

In the simulation the chemical wave was sheared by simply randomly reseeding one half of the lattice with particles whose average concentration corresponds to the steady state. This produces a semi-circular ring with two free ends. As expected the spiral wave pair continually regenerates itself after the wave fronts collide.

## VIII. TURING PATTERNS

In a Turing bifurcation [55] a homogeneous steady state loses its stability and an inhomogeneous state is formed. While these bifurcations are believed to be relevant in many biological pattern formation processes [53], it is only recently that Turing pattern formation has been observed in gel reactors [56,57]. These experimental observations have led to increased research activity on both Turing pattern formation and more complex scenarios involving interactions between Turing and other bifurcations. [58,59] It is now widely believed that immobilization of some species by the gel is responsible for the diffusion coefficient differences that drive the Turing instability. [60,61] One of the characteristic features of Turing patterns is that a regular or quasi-regular structure with a given (set of) wavelength(s) can develop in unconfined systems with no imposed characteristic macroscopic length (in contrast to most typical situations in hydrodynamic instabilities). The wavelengh of the Turing pattern is an intrinsic property of the system: it depends on the kinetic parameters and on diffusion coefficients.

Turing bifurcations are usually described in terms of the activator-inhibitor kinetics of a two-variable reaction-diffusion equation. [53] Suppose $\boldsymbol{\rho} = <\rho_1, \rho_2>$ is the vector of concentration variables, $\mathbf{F}(\boldsymbol{\rho})$ describes the kinetics and $\mathbf{D}$ is a diagonal diffusion coefficient matrix, $D_{\tau\tau'} = D_\tau \delta_{\tau\tau'}$. The conditions for a Turing bifurcation are as follows. Suppose $\mathbf{A} = (\partial \mathbf{F}/\partial \boldsymbol{\rho})_{\boldsymbol{\rho}=\boldsymbol{\rho}^*}$ is the Jacobian matrix evaluated at the steady state $\boldsymbol{\rho}^*$, and $\mathbf{B} = \mathbf{A} - k^2 \mathbf{D}$ is the matrix that governs the linearized evolution of the Fourier transform of the concentration fields. If $A_{11} > 0$ and $A_{22} < 0$ then species $X = X_1$ is the activator and species $Y = X_2$ is the inhibitor. A Turing bifurcation will occur if $det\mathbf{B} = 0$, $Tr\mathbf{B} > 0$ and $A_{11}D_2 + A_{22}D_1 > 0$. The wavenumber at the bifurcation is

$$k_c = \left(\frac{det\mathbf{A}}{D_1 D_2}\right)^{\frac{1}{4}} . \tag{219}$$

Furthermore, since the bifurcation must occur from a stable homogeneous steady state we must have $D_1/D_2 < 1$, i.e., the diffusion coefficient of the inhibitor is greater than that of the activator. The critical diffusion ratio at the bifurcation is

$$\frac{D_1}{D_2} = A_{22}^{-1} \left(det\mathbf{A} - A_{12}A_{21} + 2(A_{12}A_{21}det\mathbf{A})^{1/2}\right) . \tag{220}$$

The unstable wavevectors are determined from the dispersion relation $\sigma(k) = \Re(\lambda)$, where

$$\lambda(k) = \frac{1}{2}Tr\mathbf{B} \pm \left((Tr\mathbf{B})^2 - 4det\mathbf{B}\right)^{1/2} . \tag{221}$$

### A. Selkov model simulations

It is not difficult to realize the above conditions for the Selkov model. The system parameters may be adjusted to lie in the region where the stable fixed point of the spatially-homogeneous system is a focus and is close to the Hopf bifurcation point. For a given set of rate coefficients and control $A$ and $B$ concentrations the diffusion coefficient ratio can be adjusted to yield a Turing bifurcation in the reaction-diffusion equation. A set of rate constant parameters that achieves this is $k_1 a = 0.002656673$, $k_3 = 10 k_{-1} = 0.00665$, $k_2 = k_{-2} = 0.015$, and $k_{-3}b = 0.000531334$. With this set of parameters the critical diffusion coefficient ratio for the onset of a Turing bifurcation is $D_2/D_1 = 16.2$.

The lattice gas simulations of this bifurcation [18] were carried out on a pair of coupled square lattices using the 4-equi-rotation rule to scramble the velocities and the reaction probability matrix (215), just as in the excitable medium simulations. Only the kinetic parameters in (215) were changed to correspond to the Turing region. The diffusion coefficient ratio was changed by carring out different numbers of propagation and rotation steps for the $X_1$ and $X_2$ species so that $D_2/D_1 = \ell_2/\ell_1$. When the Selkov automaton dynamics is carried out with $D_1 = D_2$ and the above set of rate constants no macroscopic pattern formation was observed. The system exhibited small local fluctuations about an average concentration corresponding to the steady state. When the system was started from a uniform random initial state with average concentration different from that of the steady state, the average concentration was observed to decay in an oscillatory fashion to the steady state. These simulations confirm the stability of the steady state and its focal character. When the same type of simulation was carried out with the diffusion ratio selected to lie above the predicted Turing instability the system was observed to evolve to an inhomogeneous steady state. Figure 14 shows the results of a simulation with $D_1 = 1/2$ and $D_2 = 25/2$.

FIG. 14. Formation and evolution of a Turing pattern. The system size is $1024 \times 1024$ and the parameter values are given in the text. Here $D_1 = 1/2$ and $D_2 = 25/2$. The figure shows a realization of the evolution corresponding to one of the two checkerboard sublattices.

Spatial inhomogeneities develop quickly from the random initial state. The system undergoes bulk oscillations during the evolution to the inhomogeneous steady state. The final state is a hexagonal pattern of spots distorted by molecular fluctuations. The magnitudes of these local fluctuations are controlled by the system size and the magnitudes of the diffusion coefficients, which determine the local diffusive length scales. By maintaining the same diffusion coefficient ratio but reducing the magnitudes of both $D_1$ and $D_2$ one can effectively destroy the Turing pattern by fluctuations. The characteristic wavelength of the pattern can be made sufficiently small so that the number of "molecules" in a volume with characteristic size corresponding to this wavelength is small enough to give rise to fluctuations that destroy the pattern. This is illustrated in Fig. 15. While the system still shows strong inhomogeneities in the stationary regime, now the pattern is far more irregular. Thus the reactive lattice-gas automaton simulations can be used to probe both system size and fluctuation effects on Turing pattern formation.

FIG. 15. Formation and evolution of a Turing pattern. Same as Fig. 14 except that $D_1 = 1/4$ and $D_2 = 25/4$.

Often autocatalytic reactions take place in geometries with small linear dimensions. Biological cells are perhaps the most familiar examples where the cell dimensions can be quite small, of order 0.1 $\mu m$. In such small geometries fluctuations are likely to be quite important and it is interesting to speculate as to whether concentration inhomogeneites can develop within the cell as a result of the operation of a Turing or Turing-like mechanism. Lattice-gas methods provide a tool for the study of such problems. [62]

### B. Maginu model simulations

Another model which exhibits Turing structures is the Maginu model [63]. This mathematical model is defined by a set of PDE's that belong to the class of RD equations:

$$\frac{\partial x_1}{\partial t} = x_1 - x_1^3/3 - x_2 + D_1 \nabla^2 x_1 \,,$$
$$\frac{\partial x_2}{\partial t} = (x_1 - kx_2)/c + D_2 \nabla^2 x_2 \,, \qquad (222)$$

with $c > 0$ and $0 < k < 1$. It should be noticed that the corresponding rate law

$$\frac{dx_1}{dt} = x_1 - x_1^3/3 - x_2 \,,$$
$$\frac{dx_2}{dt} = (x_1 - kx_2)/c \,. \qquad (223)$$

is not a mass action rate law because each rate $dx_i/dt$ ($i = 1, 2$) can be negative when $x_j = 0$. As a consequence, the set of equations (222) cannot be simulated directly by a lattice gas automaton. However, it is possible to perform a linear transformation on the variables $(x_1, x_2)$ so that a lattice gas automaton can be constructed for the transformed equations. This can be done in various ways, e.g.

$$\rho_1 = 1/2 + x_1/\sqrt{12(1 + 1/k)} \,,$$
$$\rho_2 = 1/2 + kx_2/\sqrt{12(1 + 1/k)} \,, \qquad (224)$$

which yields to a transformed rate law

$$\frac{d\rho_1}{dt} = -4a_1\rho_1^3 + 6a_1\rho_1^2 - a_2\rho_1 + a_3(1 - \rho_2) \,,$$
$$\frac{d\rho_2}{dt} = a_4(\rho_1 - \rho_2) \,, \qquad (225)$$

with

$$a_1 = (1 + 1/k)/c \,,\ a_2 = (2 + 3/k)/c \,,\ a_3 = 1/kc \,,\ a_4 = k/c \,. \qquad (226)$$

Using the general formalism of Sec. V to construct a reaction probability matrix $P$ we obtain from (225)

$$P_1(\boldsymbol{\alpha})/h = -4a_1\alpha_1(\alpha_1 - 1)(\alpha_1 - 2)\frac{m^2}{(m-1)(m-2)} +$$
$$6a_1\alpha_1(\alpha_1 - 1)\frac{m}{(m-1)} - a_2\alpha_1 + a_3(1 - \alpha_2) \,,$$
$$P_2(\boldsymbol{\alpha})/h = a_4(\alpha_1 - \alpha_2) \,, \qquad (227)$$

where the scaling factor $h$ has been incorporated. These relations, together with the mean field condition (169), do not specify a unique matrix $P$. Additional requirements about higher moments must be provided in order to fully determine the matrix. The following rules are used:

1. for any initial configuration $\boldsymbol{\alpha}$, the changes in particle numbers are performed independentely for each species:

$$P(\boldsymbol{\alpha}, \boldsymbol{\beta}) = \prod_i P_i(\alpha_i, \beta_i) ; \qquad (228)$$

2. in any reaction, the number of particles of each species changes at most by one unit:

$$P_i(\alpha_i, \beta_i) = 0 , \qquad \text{if } |\alpha_i - \beta_i| > 1 ; \qquad (229)$$

3. for any initial configuration $\boldsymbol{\alpha}$, creation and anihilation transitions are exclusive within each species:

$$P_i\big(\alpha_i, (\alpha_i - 1)\big) \neq 0 \text{ is not compatible with } P_i\big(\alpha_i, (\alpha_i + 1)\big) \neq 0 . \qquad (230)$$

The purpose of these conditions is to reduce the number of different reactions occuring on the lattice, and to avoid large fluctuations around the mean field rates. The above rules specify completely the matrix $P$ except for the value of the scaling factor $h$ which remains free at this stage.

Now we turn to the definition of the diffusive part of the dynamics. The goal is to construct a lattice gas rule with enough flexibility to allow for smooth changes in the value of the different diffusion coefficients $D_\tau$. This is important because a small change in the value of a diffusion coefficient can trigger the Turing instability. To obtain this flexibility, we used an automaton rule

$$\mathcal{E} = \left( \prod_\tau \left( P_\tau \circ R_\tau(q_\tau) \right)^{\ell_\tau} \right) \circ C \qquad (231)$$

where the strength of the shuffling performed by each velocity randomization $R_\tau$ is controlled by a corresponding parameter $q_\tau$. Here, $R_\tau(q_\tau)$ is the velocity randomization operator (84) with $p_0 = 1 - 3q_\tau$ and $p_1 = p_2 = p_3 = q_\tau$. Using the result (94) and taking into account the value of $\ell_\tau$, we find the following expression for the diffusion coefficient expressed in automaton units (lattice units squared per time step)

$$D_\tau = \ell_\tau \left( -\frac{1}{4} + \frac{1}{2q_\tau} \right) . \qquad (232)$$

Large values of $D_\tau$ ( $> 5$ in automaton units) are often needed to simulate Turing structures. From (232) we see that two parameters ($\ell_\tau$ and $q_\tau$) can be used to assign a given value to $D_\tau$. A good strategy is to choose the largest possible value for $\ell_\tau$, and keep the flexibility provided by the continuous parameter $q_\tau$ for the fine tuning of the diffusion coefficient; in this way the diffusive behavior emerges at a length scale of a few lattice units. Indeed, the opposite strategy ($\ell_\tau = 1$, and a small value for $q_\tau$) should be avoided because in this case the discrete lattice structure is more likely to appear in the Turing structure which will then be different from the prediction of the RD equations. [22] The only way to avoid this is to make sure that the typical length scale of the Turing structure is larger than the scale required for the emergence of a good level of isotropy in the diffusion. This requires larger lattices and more updates. It should be stressed that the parameters $\{\ell_\tau : \tau = 1, \ldots n\}$ *must* have the same parity in order to maintain the two checkerboard subsystems uncoupled. Neglecting this requirement usually produces strong artifacts. [22]

In some circumstances, the discreteness of time can also become apparent in a Turing structure obtained with an automaton; in particular, the order in which the basic operators $P_\tau$, $R_\tau$ and $C$ are combined to construct $\mathcal{E}$ can be important. For instance, if the automaton rule is (231), the system is always observed *after* the diffusive phase of the dynamics and *before* the reactive transformation; in some sense the Turing structure is observed through a diffusive filter. On the other hand, the equaly acceptable automaton rule

$$\mathcal{E} = C \circ \left( \prod_\tau P_\tau \circ R_\tau \right)^{\ell_\tau} \qquad (233)$$

produces sharper Turing structures. The differences between the two automaton rules should vanish when the densities vary smoothly over large enough space and time scales.

The simulations presented here were carried out on square $N \times N$ lattices with $N = 256$ and periodic boundary conditions. The two checkerboard subsystems were always considered separately (see Sec. III). Linear stability analysis of the Maginu equations shows that the homogeneous steady state can become unstable by spatial destabilization when $\sqrt{D_1/D_2} \leq (1 - \sqrt{1-k})\sqrt{c}/k$ for $0 < c < k$ [64] in which regime the simulations presented here were performed.

Figure 16 shows a wormlike Turing pattern obtained from the destabilization of a system initially prepared in the homogeneous steady state (according to the rate equations (225)).

FIG. 16. Turing pattern ($\rho_1$ in one of the two checkerboard subsystems) obtained from the destabilization of the homogeneous unstable steady state in the Maginu model (225). Lattice size: $256 \times 256$; $k = 0.9$, $c = 0.45$, $h = 1/40$, $\ell_1 = 1$, $\ell_2 = 9$, $q_1 = 1$, $q_2 = 0.6429$, ($D_1 = 0.25$, $D_2 = 4.75$); state after 14000 time steps.

Figure 17 shows a perodic Turing pattern obtained from a perodic initial condition. A quantitative comparison of this structure with the prediction of the phenomenological RD equations shows very good agreement [22].

FIG. 17. Turing pattern ($\rho_1$ in one of the two checkerboard subsystems) obtained from a periodic initial condition. Lattice size: $256 \times 256$; $k = 0.9$, $c = 0.45$, $h = 1/60$, $\ell_1 = 1$, $\ell_2 = 9$, $q_1 = 0.4$, $q_2 = 0.21176$, ($D_1 = 1$, $D_2 = 19$).

# IX. OSCILATIONS AND CHAOS

## A. Limit cycle oscillations

Under certain conditions, the Maginu rate equations (223) (or equivalently (225)) exhibit oscillatory behavior. A set of parameters corresponding to this regime is ($k = .9$ and $c = 2$). Simulations of the automaton were carried out for these values with $D_1 = D_2$. The automaton is prepared in a homogeneous state and the spatial average of the density of each species is recorded during the system evolution. Phase trajectories in the ($\rho_1, \rho_2$)-plane are obtained in this way (separately for each checkerboard subsystem). Figure 18 shows how a trajectory is initially attracted by the limit cycle predicted by the phenomenological rate equations, then shrinks progressively to a smaller cycle whose amplitude depends on the value of the scaling parameter $h$ while its frequency remains locked to the predicted phenomenological value. When $h$ is large, reactive fluctuations are strong, and diffusion can no longer maintain spatial coherence over the whole system. As a result, different zones of the lattice become progressively phase shifted with respect to each other. As time evolves, the phase decoherence increases up to a point where a balance is reached between the source of decoherence (fluctuations) and the homogenization due to diffusion. So the observed limit cycle is contracted due to the combined effects (resulting from spatial averaging) of many out-of-phase local limit cycles with normal phase space size.

The values of the frequency of the limit cycle obtained from the power spectrum of the LGA simulation data $(\rho_1(t), \rho_2(t))$ are in agreement with the phenomenologically predicted value, Eqs. (222). However there is a slight discrepancy for large values of $h$ which can be interpreted as the result of the strong gradients and could also be the consequence of insufficient local diffusive equilibrium when the ratio of reactive collisions to elastic collisions becomes large.

FIG. 18. Phase trajectories of limit cycle behavior in the Maginu model. Left panel: LGA simualtions (one of the two checkerboard subsystems). The difference between the outer ring and the inner ring shows the shrinking effect when the ratio of reactive collisions to elastic collisions is increased, i.e. $h = 10$ versus $h = 2$. Lattice size: $64 \times 64$. Right panel: limit cycle obtained by numerical integration of the Maginu model (225).

## B. Oscillations and chaos

Chemically reacting systems provide some of the best-characterized examples of deterministic chaos and a considerable amount of both experimental and theoretical research has been devoted to an elucidation of its properties and the mechanisms responsible for its appearance. [65] In its simplest form deterministic chemical chaos is manifested in well-stirred systems with three or more chemical species. All of the theoretical investigations of such temporal chaos have been carried out using the ordinary differential equations of mass action kinetics. In this section we examine "deterministic" chemical chaos from the mesoscopic point of view of the reactive lattice gas automaton.

This more microscopic picture of the chaotic dynamics will allow us to consider a number of fundamental questions concerning these systems. We shall show how it is possible to construct a mesoscopic reactive dynamics whose mean field description yields deterministic chaos. [66] On the basis of this dynamics we can then investigate how large the system must be before one obtains a recognizable attractor. The interplay between system size, diffusion and reaction and how they affect the structure of the attractor will be examined.

### 1. The Willamowski-Rössler model

The Willamowski-Rössler model [67] is an example of a chemical mechanism that gives rise to a chaotic attractor. Since it is based on a scheme with mass action kinetics it is especially suitable for the investigation of the microscopic basis of chemical chaos.

The Willamowski-Rössler reaction mechanism is

$$A_1 + X \underset{k_{-1}}{\overset{k_1}{\rightleftharpoons}} 2X, \quad X + Y \underset{k_{-2}}{\overset{k_2}{\rightleftharpoons}} 2Y,$$
$$A_5 + Y \underset{k_{-3}}{\overset{k_3}{\rightleftharpoons}} A_2, \quad X + Z \underset{k_{-4}}{\overset{k_4}{\rightleftharpoons}} A_3, \qquad (234)$$
$$A_4 + Z \underset{k_{-5}}{\overset{k_5}{\rightleftharpoons}} 2Z \; ,$$

where $\mathbf{k} = \{k_{\pm i} : i = 1, \cdots, 5\}$ is a set of forward and reverse rate constants. In (234) $\mathbf{A} = \{A_j : j = 1, \cdots, 5\}$ are a set of species whose concentrations are fixed by constraints. The corresponding mass action rate law for the concentrations of remaining species $X$, $Y$ and $Z$ is ($\tau = X, Y, Z = 1, 2, 3$) is

$$\frac{d\rho_1}{dt} = \kappa_1 \rho_1 - \kappa_{-1} \rho_1^2 - \kappa_2 \rho_1 \rho_2 + \kappa_{-2} \rho_2^2 - \kappa_4 \rho_1 \rho_3 + \kappa_{-4} \; ,$$
$$\frac{d\rho_2}{dt} = \kappa_2 \rho_1 \rho_2 - \kappa_{-2} \rho_2^2 - \kappa_3 \rho_2 + \kappa_{-3} \; ,$$
$$\frac{d\rho_3}{dt} = -\kappa_4 \rho_1 \rho_3 + \kappa_{-4} + \kappa_5 \rho_3 - \kappa_{-5} \rho_3^2 \; . \qquad (235)$$

In (235) the constants $A_i$ have been incorporated into the new set of rate coefficients $\{\kappa\} = \{\kappa_{\pm i} : i = 1, \cdots, 5\}$.

One can recognize in the first three reactions in (234) a Lotka-Volterra oscillator involving the autocatalytic species $X$ and $Y$. This Lotka-Volterra oscillator is coupled via $X$ to a "switch" between the autocatalytic species $Z$ and $X$. A suitable coupling between the Lotka-Volterra subsystem and the switch element leads the observed oscillations of the full Willamowski-Rössler system. [68]

The reaction probability matrix corresponding to the Willamowski-Rössler model used in this study is [66]

$$P(\boldsymbol{\alpha}, \boldsymbol{\beta}) = \left[ p_1^+(\boldsymbol{\alpha}) \delta_{\beta_1 \alpha_1 + 1} + p_1^-(\boldsymbol{\alpha}) \delta_{\beta_1 \alpha_1 - 1} \right] \delta_{\beta_2 \alpha_2} \delta_{\beta_3 \alpha_3}$$
$$+ \left[ p_2^+(\boldsymbol{\alpha}) \delta_{\beta_2 \alpha_2 + 1} + p_2^-(\boldsymbol{\alpha}) \delta_{\beta_2 \alpha_2 - 1} \right] \delta_{\beta_1 \alpha_1} \delta_{\beta_3 \alpha_3} \qquad (236)$$
$$+ \left[ p_3^+(\boldsymbol{\alpha}) \delta_{\beta_3 \alpha_3 + 1} + p_3^-(\boldsymbol{\alpha}) \delta_{\beta_3 \alpha_3 - 1} \right] \delta_{\beta_1 \alpha_1} \delta_{\beta_2 \alpha_2} \; ,$$

for $\boldsymbol{\alpha} \neq \boldsymbol{\beta}$, and

$$P(\boldsymbol{\alpha}, \boldsymbol{\alpha}) = 1 - \sum_{\boldsymbol{\beta} \neq \boldsymbol{\alpha}} P(\boldsymbol{\alpha}, \boldsymbol{\beta}) \ . \tag{237}$$

The explicit expressions for $p_\tau^\pm(\boldsymbol{\alpha})$ are given below:

$$\begin{aligned}
p_1^+(\boldsymbol{\alpha}) &= q_1^+(\boldsymbol{\alpha})(1 - \delta_{\alpha_1 m}), & p_1^-(\boldsymbol{\alpha}) &= q_1^-(\boldsymbol{\alpha}) - q_1^+(\boldsymbol{\alpha})\delta_{\alpha_1 m} \ , \\
p_2^+(\boldsymbol{\alpha}) &= q_2^+(\boldsymbol{\alpha})(1 - \delta_{\alpha_2 m}), & p_2^-(\boldsymbol{\alpha}) &= q_2^-(\boldsymbol{\alpha}) - q_2^+(\boldsymbol{\alpha})\delta_{\alpha_2 m} \ , \\
p_3^+(\boldsymbol{\alpha}) &= q_3^+(\boldsymbol{\alpha})(1 - \delta_{\alpha_3 m}), & p_3^-(\boldsymbol{\alpha}) &= q_3^-(\boldsymbol{\alpha}) - q_3^+(\boldsymbol{\alpha})\delta_{\alpha_3 m} \ ;
\end{aligned}$$

with

$$\frac{q_1^+}{h} = \kappa_1 \alpha_1 + \frac{m\kappa_{-2}}{m-1}\alpha_2(\alpha_2 - 1) + \kappa_{-4}, \qquad \frac{q_1^-}{h} = \frac{m\kappa_{-1}}{m-1}\alpha_1(\alpha_1 - 1) + (\kappa_2\alpha_2 + \kappa_4\alpha_3)\alpha_1 \ ,$$

$$\frac{q_2^+}{h} = \kappa_2 \alpha_1 \alpha_2 + \kappa_{-3}, \qquad \frac{q_2^-}{h} = \frac{m\kappa_{-2}}{m-1}\alpha_2(\alpha_2 - 1) + \kappa_3\alpha_2 \ ,$$

$$\frac{q_3^+}{h} = \kappa_5 \alpha_3 + \kappa_{-4}, \qquad \frac{q_3^-}{h} = \kappa_4 \alpha_1 \alpha_3 + \frac{m\kappa_{-5}}{m-1}\alpha_3(\alpha_3 - 1) \ .$$

When this reaction probability matrix is substituted into the automaton mean field equations one obtains the Willamowski-Rössler mass action rate law. Since the full automaton dynamics is not mean field, we can now use the reactive automaton to investigate the dynamics of this reacting system from a mesoscopic point of view that incorporates fluctuations.

### 2. Spatially homogeneous system

Before examining the dynamics of the Willamowski-Rössler system using the full automaton dynamics it is interesting to first examine a simpler situation where the system remains fully mixed or well stirred throughout the evolution. In view of the description given earlier, the particles are binomially distributed in the well-stirred limit. In this case a Markov chain equation can be written for the time evolution of the probability $P(\mathbf{n}, k)$ that there are $\mathbf{n} = (n_1, n_2, n_3)$ particles of species $X_1$ on $\mathcal{L}_1$, $X_2$ on $\mathcal{L}_2$, and $X_3$ on $\mathcal{L}_3$. The Markov chain reads,

$$P(\mathbf{n}, k+1) - P(\mathbf{n}, k) = \sum_{\mathbf{n}', \mathbf{n}' \neq \mathbf{n}} [W(\mathbf{n}|\mathbf{n}')P(\mathbf{n}', k) - W(\mathbf{n}'|\mathbf{n})P(\mathbf{n}, k)] \ , \tag{238}$$

The transition probability $W(\mathbf{n}'|\mathbf{n})$ from $\mathbf{n}$ to $\mathbf{n}'$ can be written as

$$W(\mathbf{n}'|\mathbf{n}) = \prod_{\tau=1}^{3} \left[ \sum_{\substack{n_{\tau+}, n_{\tau-} \\ (n_{\tau+} - n_{\tau-} = n'_\tau - n_\tau)}} P_\tau(n_{\tau+}, n_{\tau-}|\rho(\mathbf{t})) \right] \ , \tag{239}$$

where $P_\tau(n_{\tau+}, n_{\tau-}|\rho)$ is the probability that there are $n_{\tau+}$ particle transformations on lattice $\mathcal{L}_\tau$ that increase the particle number of species $X_\tau$ by one and $n_{\tau-}$ particle transformations on lattice $\mathcal{L}_\tau$ that decrease the particle number of species $X_\tau$ by one. Here we have used the fact that in the Willamowski-Rössler mechanism the reactions involve particle number changes of $\pm 1$. The explicit form for $P_\tau(n_{\tau+}, n_{\tau-}|\rho)$ is

$$P_\tau(n_+, n_-|\mathbf{n}) = c(n_+, n_-, \mathcal{N})(P_\tau^+)^{n_+}(P_\tau^-)^{n_-}(1 - P_\tau^+ - P_\tau^-)^{\mathcal{N} - n_+ - n_-} \ , \tag{240}$$

where $\mathcal{N} = N^2$. The factor

$$c(n_+, n_-, \mathcal{N}) = \frac{\mathcal{N}!}{n_+! n_-! (\mathcal{N} - n_+ - n_-)!} , \qquad (241)$$

takes into account the number of different ways of assigning the $n_+$ transitions $\alpha_\tau \to \alpha_\tau + 1$ and $n_-$ transitions $\alpha_\tau \to \alpha_\tau - 1$ to possible positions on the lattice. The $P_\tau^+$ and $P_\tau^-$ are transition probabilities averaged over a binomial distribution on $\mathcal{L}_\tau$ and are given by

$$P_\tau^\pm = \sum_{\boldsymbol{\alpha}} p_\tau^\pm(\boldsymbol{\alpha}) \prod_{\kappa=1}^n \binom{m}{\alpha_\kappa} c_\kappa^{\alpha_\kappa} (1 - c_\kappa)^{m - \alpha_\kappa} . \qquad (242)$$

One may again obtain the mass action rate by computing the average value of $\rho_\tau(k)$ from the master equation (238) in the mean field limit:

$$E[\Delta \rho_\tau] = \frac{1}{\mathcal{N}} \sum_{\substack{n_+, n_- \\ n_+ + n_- \leq \mathcal{N}}} (n_+ - n_-) P_\tau(n_+, n_- | \boldsymbol{\rho}) = P_\tau^+ - P_\tau^- = h R_\tau(\boldsymbol{\rho}) . \qquad (243)$$

We may also compute the standard deviation of the net particle number change per node on $\mathcal{L}_\tau$, $\sigma(\Delta \rho_\tau)$, in the same approximation and find

$$\sigma^2(\Delta \rho_\tau) = \sum_{\substack{n_+, n_- \\ n_+ + n_- \leq \mathcal{N}}} \left[ \frac{n_+ - n_-}{\mathcal{N}} - E(\Delta \rho_\tau) \right]^2 P_\tau(n_+, n_- | \boldsymbol{\rho})$$

$$= \frac{1}{\mathcal{N}} \left[ P_\tau^+ + P_\tau^- - (P_\tau^+ - P_\tau^-)^2 \right] . \qquad (244)$$

This result shows that the fluctuations scale as $\mathcal{N}^{-1/2}$ as expected.

One way to simulate this well-stirred dynamics is to introduce a new automaton rule where the propagation and velocity randomization steps are replaced by a mixing operator $B$ that restores the particle distribution to a binomial distribution each automaton time step:

$$T_B = C \circ B . \qquad (245)$$

This automaton rule will allow us to explore the effects of fluctuations on chaos in well-stirred systems where spatial pattern formation does not complicate the interpretation of the results.

*3. Simulation results*

The Willamowski-Rössler model has a chaotic attractor that arises by a period-doubling cascade. In all of the calculations presented below the rate constant parameters were fixed at the values: $\kappa_1 = 31.2$, $\kappa_{-1} = 0.2$, $\kappa_{-2} = 0.1$, $\kappa_3 = 10.8$, $\kappa_{-3} = 0.12$, $\kappa_4 = 1.02$, $\kappa_{-4} = 0.01$, $\kappa_5 = 16.5$, and $\kappa_{-5} = 0.5$. The bifurcation diagram as a function of $\kappa_2$ for fixed values of the other parameters is shown in Fig. 19. The lattice-gas automaton has been used to explore the mesoscopic dynamics that underlies the subharmonic cascade leading to chaos. [66]

FIG. 19. Bifurcation diagram as a function of $\kappa_2$ showing period-doubling and reverse period-doubling cascades to chaos. The ordinate shows the concentration of species $Y$ in a Poincaré section plane $\{\rho_1 = const., \forall \rho_2 \geq 0, \forall \rho_3 \geq 0\}$. The results were obtained by solving the mass action rate law (235).

*Well-stirred system*
Internal noise can have important effects on the dynamics when the system lies in or near the chaotic regime. From earlier investigations on the influence of external noise on nonlinear dynamical systems [69,70] one knows that noise can alter the positions of bifurcation points, destroy bifurcation sequences because periodic orbits cannot be resolved, induce periodic behavior and destroy the structure of chaotic attractors. All of these effects depend on the region in parameter space where the system lies and the amplitude of the noise and its statistical properties. Internal noise is somewhat more subtle to consider since it is generated by the system itself: the same dynamics that gives rise to the periodic or chaotic behavior on the macroscopic level is also responsible for the fluctuations. Therefore, the noise amplitude is not fully under control of the investigator and the possibility exits that the same mechanisms that are responsible for deterministic chaos may lead to an enhancement of the effects of fluctuations. It has been suggested that such effects can lead to the breakdown of the mean field macroscopic descriptions in the chaotic regime. [71] This is a rather paradoxical situation since the very equations that are typically used to analyse deterministic chaos may not be valid when the system is chaotic. The extent of this breakdown is a matter of debate. [72,73]

The reactive lattice gas automaton can be used to explore these questions in a rather clear fashion. The automaton dynamics, while idealized, is based on the reactive dynamics as embodied in the reaction mechanism. Also, the mean field limit of the automaton dynamics is the mass action rate law. Therefore, the full automaton simulations which include fluctuations can be compared with the mean field approximation to test its validity and the extent of its breakdown.

It is especially interesting to study the dynamics for parameter values in the period doubling regime. Here one can see the effects of noise on limit cycle attractors of increasing complexity as one moves closer to and into the chaotic regime. Studies of this type have been carried out in Refs. [66] and [74]. We noted above that the fluctuations scale as $\mathcal{N}^{-1/2}$ but their actual magnitude depends on the system dynamics for the given control parameters. A series of automaton calculations has been carried out for parameters lying in the period doubling cascade for different system sizes. The two examples shown in Figs. 20 and 21 illustrate the qualitative features of the internal noise effects. The period-two case is shown in Fig. 20. In this figure the period-2 attractors obtained from the automaton simulation for two different system sizes (middle and right panels) are compared with the deterministic period-2 orbit (left panel).

FIG. 20. Projection of the 3-d phase-space trajectories on the $(\rho_1,\rho_2)$-plane for $\kappa_2 = 1.48$ for the deterministic system (left panel) and for well-stirred automaton dynamics with $\mathcal{N} = (512)^2$ (middle panel) and $\mathcal{N} = (128)^2$ (right panel). Concentrations were obtained by averaging over all the nodes on the lattice.

For small system sizes, below about $\mathcal{N} = (300)^2$, the noisy attractor bears little resemblance to the deterministic period-2 orbit (cf. right panel). For large system sizes (middle panel) the noisy attractor is simply a thick version of the deterministic period-2 attractor in the left panel.

The effects of fluctuations on the chaotic attractors are considered in Fig. 21 where the deterministic strange attractor is compared with the well-stirred automaton dynamics for two different systems sizes. For small system sizes ($\mathcal{N} = (1024)^2$, top panel) one observes that some of the fine structure of the chaotic attractor has been destroyed by the noise: all but the last the chaotic bands have merged and the dynamics has lost some of its phase coherence. In the middle panel results for $\mathcal{N} = (4096)^2$ are shown where additional fine structure of the attractor is resolved.

FIG. 21. Chaotic attractors for two different system sizes: $\mathcal{N} = (1024)^2$ (upper panel) and $\mathcal{N} = (4096)^2$ (middle panel). Deterministic system (lower panel). **P** indicates Poincaré section plane.

In spite of these marked effects on the strange attractor, the gross structure of the chaotic phase space flow is preserved. This is shown in Fig. 22 where the Poincaré maps are displayed for the deterministic attractor along with those for two different system sizes. The noisy chaotic attractor has a line-like structure but is not banded for very small system sizes ($\mathcal{N} = (128)^2$, right panel of Fig. 22).

FIG. 22. Poincaré maps showing the well-stirred automaton dynamics in the chaotic region, $\kappa_2 = 1.568$, for $\mathcal{N} = (128)^2$ (right panel), $\mathcal{N} = (1024)^2$ (middle panel) and deterministic system (left panel).

Since the bands are no longer distinct, these map regions are no longer visited in a definite order on successive intersections of the phase space trajectory with the Poincaré plane. As the system size increases a critical size is reached where the internal noise becomes sufficiently small that the chaotic bands are resolved and the noisy chaotic attractor appears to be a slightly perturbed version of the deterministic chaotic attractor (middle panel of Fig. 22). Only the structure on the smallest scales of the attractor hierarchy are destroyed by the noise.

The noise scaling properties in the period doubling regime have been studied extensively for one-dimensional maps where a renormalization group treatment is possible. [69,75] Given equivalent points in the period doubling bifurcation sequence, for example the superstable points for period $2^n$, one can ask how much the noise amplitude needs to be reduced to observe one additional period doubling. From numerical and renormalization group studies one finds that the noise amplitude must be reduced by about a factor of 6.7. Extensive calculations using the well-stirred automaton dynamics have verified that the number of particles must be increased by the square of this factor. So a similar scaling applies to both external and internal noise. [74] These results suggest that for macroscopic systems containing a mole of particles one should be able to resolve up to period 256 or 512 if no other sources of noise were present. This implies that in most chemical experiments external noise destroys the period-doubling structure before internal noise effects become important. However, for small systems this may not be the case. In addition, effects arising from spatial inhomogeneities can also lead to situations where internal noise effects can play a greater role. This will be discussed in more detail below.

*Spatially distributed system*

In an unstirred system diffusion is the only mechanism for removing local inhomogeneities in the chemical concentrations. If the system size is small enough and diffusion is strong enough, the spatial distribution of the concentration field will be nearly uniform and the results will be identical to those for the well-stirred system. If these conditions are not met, diffusion will be unable to homogenize the concentration distribution and local rather than global concentration fluctuations will be most relevant for the dynamics. It is just such local fluctuations that are responsible for nucleation-induced pattern formation processes.

The results of a full automaton simulation of a small ($\mathcal{N} = (100)^2$) system are shown in Fig. 23 (left panel) for $\kappa_2 = 1.572$, corresponding to the deterministic chaotic attractor in the right panel. (All simulations were carried out on triangular lattices.) Diffusion is able to smooth the concentration fluctuations arising from reactions occurring on the time scale $\tau_{chem}$ over a length $L = (D\tau_{chem})^{1/2}$. For this automaton simulation $L \sim 43$, which is comparable to $N = 100$; thus, the system will maintain a roughly homogeneous spatial concentration distribution. This homogeneity was confirmed by direct observation of the distribution of chemical species over the lattices as a function of time. The full automaton dynamics is equivalent to well-stirred dynamics for this case. One observes in Fig. 23 that the noisy attractor has a similar gross structure to the corresponding deterministic chaotic attractor but differs in a number of respects. Internal fluctuations cause the phase space trajectories to explore a greater volume of phase space leading to a larger size for the attractor. Furthermore, since the system size is small, fluctuations are sufficiently strong to cause merging of the chaotic bands as noted above for well-stirred dynamics in small systems. Nevertheless, the density is non-uniform on the attractor so that regions of high probability density correspond to the underlying deterministic bands. Again this band merging is accompanied by dephasing of the dynamics when viewed in the Poincaré plane so that the bands (or high density regions) are no longer visited in the same order.

FIG. 23. Chaotic attractor. Parameters are: $\kappa_1 = 31.2$, $\kappa_{-1} = 0.2$, $\kappa_2 = 1.572$, $\kappa_{-2} = 0.1$, $\kappa_3 = 10.8$, $\kappa_{-3} = 0.12$, $\kappa_4 = 1.02$, $\kappa_{-4} = 0.01$, $\kappa_5 = 16.5$, $\kappa_{-5} = 0.5$. The dashed box which contains the attractor has one corner at the origin of the coordinate system, $(x, y, z) = (-.02599, -.05498, -.08193)$, and corners lying along the $x$, $y$ and $z$ axes at $2.51039$, $2.25439$ and $4.77463$, respectively. The Poincaré section is labeled $P$. Automaton simulation (left panel) and deterministic system (right panel).

If the system is larger or the diffusion coefficient is smaller, diffusion may be insufficient to maintain spatial homogeneity over all of space. The simplest manifestation of such lack of diffusive mixing is the desynchronization of the chaotic oscillations in different spatial regions: local volumes (areas) of space containing many nodes oscillate in phase but the phase differs from local volume to volume. This is the analog of phase turbulence [76] for a periodic state but now the underlying oscillation is aperiodic. This effect is shown in Fig. 24. The left panel shows the chaotic phase space trajectory projected in the $(\rho_1, \rho_2)$-plane. As usual the concentrations $\rho_\tau(k)$ were obtained by averaging over all nodes on the species lattices at each time $k$. However, the spatial distribution of particles is highly non-uniform as can be seen in the right panel. As a result, the trajectory occupies a small volume in phase space since the average over the lattice (real space) implies an average over different phases of the aperiodic attractor. In cases where spatial homogeneity no longer obtains, the effects of internal fluctuations are even more subtle since both diffusion and reaction combine to influence the magnitude of the local fluctuations.

FIG. 24. Projection of the phase space trajectory in the $(\rho_1 \rho_2)$-plane for the automaton dynamics with $\mathcal{N} = (100)^2$ and $D = 1/40$ (left panel). Spatial distribution of the $Z = X_3$ species concentration is coded as gray levels (right panel).

Provided the size of the well-stirred system is large enough, the noisy attractor, even in the chaotic regime, closely resembles that obtained from the deterministic rate law: only structure on the smallest scales is obscured by the noise. If the chaotic attractor has a hierarchical banded structure, as is the case for the Willamowski-Rössler attractor, there are critical values of the system size beyond which the structure at a given level in the hierarchy cannot be resolved. If the system size is small fluctuations can be so large that not only is the chaotic attractor significantly modified (cf. Fig. 23 – note that even in this case the gross geometrical structure of the attractor is preserved) but also the bifurcation sequence and locations of the bifurcation points can change. For example, noise can lead to premature truncation of the period-doubling cascade.

If mixing is imperfect then spatial fluctuations in the concentration field also influence the dynamics. One can imagine the system to be composed of roughly homogeneous patches interacting with each other in a time-dependent fashion. Since the patches have linear dimensions much smaller than those of the entire system, the fluctuations within the patches have larger amplitudes. This feature, combined with the complicated but weak interactions among patches, leads to fluctuation effects that are strong and difficult to understand from simple considerations.

In experiments on macroscopic systems, the number of particles is so large that the effects of global concentration fluctuations are likely to be negligibly small, even in or near the chaotic regime where fluctuations are enhanced by the unstable dynamics. Since any experiment naturally involves some level of external noise, it would be difficult to distinguish internal and external noise effects: both act in a similar manner on the dynamics. [66] However, all experiments involve spatial degrees of freedom and local inhomogeneities. If diffusion is weak these spatial degrees of freedom come into play and the effects of fluctuations can become pronounced. In this regime one has to consider both the diffusion length scale as well as the temporal character of the dynamics. Thus, noise can influence the local structures that form in spatially-distributed systems.

# X. MODELS WITHOUT EXCLUSION

The exclusion principle does impose some limitations on the types of reacting systems that can be treated. For instance, there are limitations on the values of the rate constants that may preclude exploration of certain dynamical regimes. The origin of these restrictions was described in Sec. V where the scheme for the construction of the reaction probability matrix was given. One way to extend the range of applicability of the class of reactive lattice gas models is to modify or abandon the exclusion principle. Naturally there are penalties if this is done and some of the computational elegance is lost but systems can be considered that would be difficult to treat otherwise. There are a number of ways to construct reactive LGA models of this type but we focus on models that retain an integer-valued representation of the species concentrations [77–79] instead of real-valued representations [80] which lead to instabilities and are more akin to other finite-difference methods or coupled map lattices. Models using a Boolean representation but with a modified exclusion principle will not be considered here [31].

The reactive LGA models without exclusion studied thus far are similar. Basically particles reside at the nodes of a lattice with no restriction on the numbers of particles per node. Particles diffuse by hopping to neighboring nodes on the lattice and react according to probabilities that depend on the number of particles of a given species at the node. Since the reaction probabilities depend on the particle numbers and these numbers may in principle increase indefinitely, it is possible that the requirement that the reaction probability be less than or equal to unity will be violated. Conditions of the simulation can always be arranged so that this is an extremely rare event and is not encountered in finite space and time simulations. This restriction is a weaker version of that for the model with exclusion where it is appropriate to scale the dynamics so that the species concentrations rarely reach values corresponding to maximum occupancy of a node where exclusion distorts the transition probabilities and hence the fluctuations.

Briefly the automaton rules may be summarized as follows. The rule used by Chopard et al. [77] may be composed of diffusion and reaction steps. Diffusion is simulated by allowing particles to jump to nearest neighbor sites with equal probabilities for each lattice direction. The diffusion coefficient is varied by the number $\ell_\tau$ of consecutive diffusion steps for a given species $\tau$ so that $D_\tau = \ell_\tau/4$ (for a square lattice). Reactions of a specific type are considered. For a reaction of the form $n_1 X_1 + n_2 X_2 \to n_3 X_3$, all possible reactions composed from $n_1$ $X_1$ and $n_2$ $X_2$ particles have a probability $k$ to create $n_3$ $X_3$ particles. The automaton rule considered by Karapiperis and Blankleider [78] is similar and again consists of diffusion and reaction steps. Diffusion is simulated by allowing particles of species $\tau$ to move to neighboring lattice nodes with a given probability. The probabilities may be different for the various species and are chosen to satisfy isotropy conditions or may be biased depending on the application. Reactions are carried out with probabilities that depend on the numbers of particles of the given species at a node. The "rule II" in [78] is similar in form to that described earlier for reactive LGA and leads to the standard mass action rate law when correlations are neglected.

We illustrate the application of such reactive LGA models without exclusion using the following rule [79]: The dynamics is carried out on a stack of species lattices with coordination number $m_c$. The number of channels at each node is taken to be equal to the coordination number ($m = m_c$; allowed velocities correspond to jumps from node to nearest neighbors node in a discrete time step). Now there is no restriction on the numbers of particles that can occupy a given channel.

- Diffusion is simulated by deterministic motion of particles from a node to its neighbors along the directions specified by their velocities. Following propagation the velocities are randomized by reassigining the particles with equal probabilities to the $m$ velocity directions. As noted above, there is no restriction on the number of particles with a given velocity. These are just the analogs of the propgation and velocity randomization processes described earlier for the model with exclusion.

- Reaction is carried out by probabilistic particle number changes similar to those described for the model with exclusion except that the probability distributions are Poisson instead of binomial. The elements of the probability matrix can be determined from the formulas given earlier by replacing ratios of factors involving $m$ by unity, i.e., in the elements of $P(\boldsymbol{\alpha},\boldsymbol{\beta})$ the limit $m \to \infty$ is taken. As in the abovementioned two models, it is possible in the course of simulation that the computed reaction probability will exceed unity but this can always be arranged to occur with negligible probability and when it does occur the corresponding probability is set equal to zero.

As an illustration of the application of the reactive LGA without exclusion we consider the Willamowski-Rössler model described earlier. [67] Figure 25 shows two noisy limit cycle simulations carried out on small lattices ($100 \times 100$). Panel (a) is the result using an LGA model with exclusion while panel (b) shows the result for the automaton without exclusion.

FIG. 25. Comparison of noisy limit cycles for the WR model, (a) with exclusion and (b) without exclusion. The parameters are: $\kappa_2 = 1.4$, $D = 1$, $h = 1.5 \times 10^{-4}$. The origins of the boxes containing the attractors are at (0.10440,0.02660,0.2600) while the opposite corners lie at (1.24540,1.55900,3.86590)

The fact that the two simulation methods give similar results is not surprising in view of the fact that the average particle number densities are such that maximum occupancy of a node is a rare event. In fact, if in the models with exclusion this is not the case the fluctuations predicted by the model should be suspect. As noted earlier, the model without exclusion allows a larger class of chemical systems to be treated with a corresponding increase of computational complexity. Thus with minor modifications the methods and algorithms developed in the preceding sections can be carried over to models without exclusion should the need arise.

# XI. CELLULAR AUTOMATA

The reactive lattice-gas automaton is only one of a class of discrete models that may be constructed for the description of spatially-distributed chemically reacting systems. Such discrete models include, among others, cellular automata (CA), coupled map lattices, lattice Boltzmann equations and birth-death master equation models. We shall not consider coupled map lattices further since the dynamical variables are continuous; comments on their relation to CA and reactive LGA can be found in Ref. [50] and their application to reactive systems is described in Ref. [81]. Lattice Boltzmann methods [82] also utilize real-valued dynamical variables and in this sense lie in the same class of systems as coupled map lattices. Their application to reactive systems can be found in [83]. As noted in the Introduction and in Secs.VI-IX, master equation methods are closely related to the lattice gas methods described here; discussions of their application to reacting systems are given in Refs. [11–13] as well in the specialized references quoted in Secs.VI-IX. We shall confine our attention to cellular automaton models since these are closest in structure to the LGA method. In this section we contrast the lattice gas automaton scheme with more traditional implementations of cellular automaton ideas.

Cellular automata are abstract discrete dynamical systems devised by Von Neumann to describe evolution in biological systems. [84] A cellular automaton consists of a set of nodes, usually arranged on a regular lattice; each node supports state variables that take on a finite number of possible values. The state variables are synchronously updated at discrete time intervals according to a local deterministic or probabilistic rule that depends on the state variables at and in a neighborhood of a node. It is clear that reactive LGA lie within the general class of models given by this prescription and can formally be called automata (for a different viewpoint see [85]); however, the modeling strategy and rule construction is quite distinct from that of traditional cellular automaton models since the aim of reactive LGA is the construction of a mesoscopic description of the reactive collision dynamics and the rule is composed of propagation and collision steps.

It is now well established that the above simple prescription for the rule construction can lead to CA with a bewildering variety of complex dynamics, even for Boolean state variables and a one-dimensional lattice of nodes with nearest-neighbor interactions. There exists a large literature on the formal properties of such automata. [86] Cellular automata can be constructed as simplified models of reaction-diffusion systems. Often the main features of an apparently very complex dynamics can be captured in a very simple rule. Cellular automaton models have been constructed for a wide variety of chemically reacting systems. This section is not intended to be an exhaustive review of the extensive literature in this field; instead a few examples will be given to illustrate and contrast the point of view taken in the CA model construction with that of reactive LGA. Further references to the chemical CA literature can be found in Ref. [50].

Often the CA modeling strategy can be described as the inverse problem: given a phenomenon, what is the cellular automaton rule that produces it? [87] One generally starts from some notion of the physical ingredients that characterize the system and builds them into the rule. A good rule will yield dynamics that mimics that of the real system. Once a class of rules that depend on parameters, such as the neighborhood size or the number of states, is constructed, it is interesting to attempt to classify the possible types of dynamical behavior that arise from such parameter variations. This is the forward problem in cellular automata theory. Since cellular automaton rules are notorious for their lack of smooth behavior under such parameter variations, it is a topic of considerable interest in applications.

Consider, for example, excitable media discussed in Secs.VI-IX. The reactive lattice gas modeling of such excitable media is really no different from that for any other type of reactive dynamics: one starts with the reaction mechanism and constructs a microdynamics that mimics the reactive and nonreactive collision processes in the system. If the kinetics happens to give rise to excitability then so will a properly constructed reactive lattice gas model. This was illustrated in Secs.VI-IX where the Selkov model was studied. The same general Selkov lattice gas microdynamics was shown to describe excitability, bistability, Turing pattern formation, etc. Thus, no special features of excitability are exploited in order to construct the automaton rules. The situation is different in traditional applications of cellular automata and below we outline some aspects of the way one constructs CA models for such media.

Excitable media were among the first systems to be modeled using cellular automaton methods. In an early paper Wiener and Rosenblueth [88] constructed a cellular automaton-like model for excitable cardiac tissue. They recognized the principle features of an excitable medium described in Secs.VI-IX: an excited state at the wave front, refractory states where the medium can "neither transmit or receive impulses" and a resting state that follows the termination of the refractory state where the system is susceptible to stimulation. While these basic ideas and their variants were used in subsequent studies of excitable media (cf. Farley's work on neural activity [89]), the most extensive early investigations of these models were carried out by Greenberg and Hastings [49]. The subsequent generalizations of their work have been the subject of a great deal of mathematical study and form the basis of recent applications. The Greenberg-Hastings (GH) rule construction illustrates how the above characterization of an excitable medium can be transcribed into a simple CA rule with rich dynamics, only some of which resembles that of real excitable media.

### A. Cellular automata for excitable media

The Greenberg-Hastings model can be written as the following rule: the dynamical variable $u$ describing the state of the system can take on the three values $u = \{-1, 0, 1\}$ where 0 is the resting state, 1 is the excited state and $-1$ is the refractory state. The three-state excitable medium cellular automaton rule is:

$$u(\mathbf{r}, k+1) = \begin{cases} 0, & \text{if } u(\mathbf{r}, k) = -1 \\ -1, & \text{if } u(\mathbf{r}, k) = 1 \\ \max\{0, u(\mathbf{r}', k) : \mathbf{r}' \in \mathcal{N}(\mathbf{r})\}, & \text{if } u(\mathbf{r}, k) = 0 \end{cases} \quad . \tag{246}$$

Here $\mathbf{r}$ labels the sites of a regular lattice and $\mathcal{N}(\mathbf{r})$ is a neighborhood of $\mathbf{r}$. This very simple rule embodies the essential features of excitable medium dynamics. In the context of this rule it is simple to understand the formation of the two principal patterns of excitation, rings and spiral waves, and the initial states that give rise to their formation. It is also possible to appreciate some of the complexities of the dynamics starting from random initial conditions in terms of this rule.

Generalizations of this simple model are termed GH rules. Again, let $\mathbf{r}$ label the nodes of a regular lattice. The discrete dynamical variable $u$ takes $n$ integer values in the set $\{0, \ldots, n-1\}$. A site can change its value either independently of the values of its neighbors or in a manner that depends on the values of the neighboring sites. The interaction with neighboring sites is characterized by two additional parameters: a threshold $\theta$ and a range $\delta$. Let $\mathcal{N}_\delta(\mathbf{r}) = \{\mathbf{r}' : |\mathbf{r}' - \mathbf{r}| \leq \delta\}$ where $|\ldots|$ is a suitably defined norm. Let $N_\delta$ be the cardinality of the set $\{\mathbf{r}' \in \mathcal{N}_\delta(\mathbf{r}) : u(\mathbf{r}', k) = (u(\mathbf{r}, k) + 1) \bmod n\}$. With these definitions the GH rules are

$$u(\mathbf{r}, k+1) = \begin{cases} (u(\mathbf{r}, k) + 1) \bmod n, & \text{if } u(\mathbf{r}, k) \geq 1 \text{ or } N_\delta \geq \theta \\ u(\mathbf{r}, k), & \text{otherwise} \end{cases} \quad , \tag{247}$$

This is the generalization of the simple $n = 3$, $\theta = 1$, $\delta = 1$ rule originally studied by GH. This generalized GH rule has been the subject of extensive mathematical investigations. [90–92] Some aspects of its ergodic properties are known and the broad structure of the phase diagram as a function of $n$, $\theta$ and $\delta$ has been determined. Fisch, Gravner and Griffeath have shown that GH rules can exhibit a variety of asymptotic states depending on the rule parameters and, in addition, the relationship between these models and continuum reaction-diffusion models for excitable media has been examined. [90] Probabilistic versions of excitable medium CA rules have also been constructed and investigated. [90–92] In addition to these simple GH rules, more complex rules that are designed to model specific features of the wave propagation processes in excitable media have been constructed. In particular, we mention the work of Gerhardt, Schuster and Tyson [52] and Markus and Hess. [51]. These CA models can account for the dispersion of the wave velocity and, in the case of the latter model, the wavefront curvature.

### B. Cellular automata for reaction-diffusion systems

CA for reaction-diffusion systems are discrete models which offer an alternative to partial differential equations. They have been shown to produce qualitatively correct behavior. However, they are often restricted to certain RD systems and certain types of phenomena and have been subjected to the criticism of experimentalists and researchers working with PDEs who are concerned with quantitative predictions. A class of CA was developed for modelling many reaction-diffusion systems in a quantitative way [93]. The main idea behind this class of CA is careful discretization. Space and time are discretized as in normal finite difference methods for solving the PDEs. Finite difference methods then proceed to solve the resulting coupled system of $N \times n$ ordinary differential equations ($N$ points in space, $n$ equations in the PDE system) by any of a number of numerical methods operating on floating point numbers. The use of floating point numbers on computers implies a discretization of the continuous variables. The errors introduced by this discretization and the ensuing roundoff errors are often not considered explicitly, but assumed to be small because the precision is rather high (8 decimal digits for usual floating point numbers). In contrast, in the CA approach, all variables are explicitly discretized into relatively small integers. This discretization allows the use of lookup tables to replace the evaluation of the nonlinear rate functions. It is this table lookup, combined with the fact that all calculations are performed using integers instead of floating point variables, that accounts for an improvement in speed of orders of magnitude on a conventional multi-purpose computer. The undesirable effects of discretization are overcome by using probabilistic rules for the updating of the CA.

The state of the CA is given by a regular array of concentration vectors $\boldsymbol{\rho}$ residing on a $d$-dimensional lattice. Each $\boldsymbol{\rho}(\mathbf{r})$ is a $n$-vector of integers, where $n$ is the number of reactive species and each component $\rho_\tau(\mathbf{r})$ can only take integer values between 0 and $b_\tau$, where the $b_\tau$'s can be different for each species $\tau$. The position index $\mathbf{r}$ is a $d$-dimensional vector in the CA lattice (for cubic lattices, $\mathbf{r}$ is a $d$-vector of integers).

The central operation of the automaton consists of calculating the sum of the concentrations in some neighborhood $\mathcal{N}_\tau$

$$\tilde{\rho}_\tau(\mathbf{r}) = \sum_{\mathbf{r}' \in \mathcal{N}_\tau} \rho_\tau(\mathbf{r} + \mathbf{r}'). \tag{248}$$

The neighborhood (which can be specific to each species $\tau$) is specified as a set of displacement vectors, e.g. for a two-dimensional square lattice the neighborhood restricted to first neighbors is given by

$$\mathcal{N}_{sq} = \left\{ \begin{pmatrix} 0 \\ 0 \end{pmatrix}, \begin{pmatrix} 1 \\ 0 \end{pmatrix}, \begin{pmatrix} 0 \\ 1 \end{pmatrix}, \begin{pmatrix} -1 \\ 0 \end{pmatrix}, \begin{pmatrix} 0 \\ -1 \end{pmatrix} \right\}. \tag{249}$$

For some – in particular square and cubic – neighborhoods, the summation operation can be executed very efficiently [93]. It is convenient to introduce the normalized values $\varrho_\tau(\mathbf{r}) = \rho_\tau(\mathbf{r})/b_\tau$ and $\tilde{\varrho}_\tau(\mathbf{r}) = \tilde{\rho}_\tau(\mathbf{r})/(b_\tau |\mathcal{N}_\tau|)$, which are always between zero and one. The resulting fields $\tilde{\varrho}_\tau(\mathbf{r})$ are then the local averages of the $\varrho_\tau(\mathbf{r})$. The averaging has the effect of diffusion as can be seen from a Taylor expansion of $\varrho_\tau(\mathbf{r} + \mathbf{r}')$ around $\varrho_\tau(\mathbf{r})$:

$$\tilde{\varrho}_\tau(\mathbf{r}) = \frac{1}{|\mathcal{N}_\tau|} \sum_{\mathbf{r}' \in \mathcal{N}} \sum_{l=0}^{\infty} \frac{1}{l!} \left( \mathbf{r}' \frac{\partial}{\partial \mathbf{r}'} \right)^l \varrho_\tau(\mathbf{r})$$
$$= \varrho_\tau(\mathbf{r}) + D_\tau \nabla^2 \varrho_\tau(\mathbf{r}) + \cdots. \tag{250}$$

The factors $D_\tau$ can be computed as in [94] and are easily calculated from (250) for square neighborhoods with radius $\ell$ : $D_\tau = \ell(\ell+1)/6$.

The second operation in the cellular automaton is the implementation of the reactive processes described by a rate law. Given the reaction-diffusion equation

$$\frac{\partial \boldsymbol{\rho}}{\partial t} = \mathbf{F}(\boldsymbol{\rho}) + \mathcal{D} \nabla^2 \boldsymbol{\rho}, \tag{251}$$

we discretize the time derivative to obtain

$$\boldsymbol{\rho}^{t+\Delta t} = \boldsymbol{\rho}^t + \Delta t\, \mathbf{F}(\boldsymbol{\rho}^t) + \Delta t\, \mathcal{D} \nabla^2 \boldsymbol{\rho}^t. \tag{252}$$

Changing the time and space scales by setting $k = t/\Delta t$ and $r = x/\Delta x$, and using the variable $\varrho$ for the rescaled set yields

$$\varrho^{k+1} = \varrho^k + \Delta t\, \mathbf{F}(\varrho^k) + \frac{\Delta t}{(\Delta x)^2}\mathcal{D}\nabla^2 \varrho^k. \tag{253}$$

as the equation to be treated by the CA.

Let us define

$$\mathbf{F}^*(\varrho) = \varrho + \Delta t \mathbf{F}(\varrho). \tag{254}$$

From (250) and (254)

$$\begin{aligned}\mathbf{F}^*(\tilde{\varrho}^k) &= \varrho^k + \mathbf{D}\nabla^2 \varrho^k + \cdots + \Delta t \mathbf{F}(\varrho^k + \mathbf{D}\nabla^2 \varrho^k + \cdots) \\ &= \varrho^k + \mathbf{D}\nabla^2 \varrho^k + \Delta t \mathbf{F}(\varrho^k) + O(\Delta t^2).\end{aligned} \tag{255}$$

Then

$$\varrho^{k+1} = \mathbf{F}^*(\tilde{\varrho}^k) \tag{256}$$

is consistently first order accurate in time and, within this limit, (256) can be validly identified with (253) to describe the evolution of the system. The identification yields

$$D_\tau = \frac{\Delta t}{\Delta x^2}\mathcal{D}_\tau \qquad (\tau = 1,\cdots, n) \tag{257}$$

which defines the space scale. As $\tilde{\rho}$ is the result of the diffusion step, the average output of the CA reaction-diffusion process is therefore given by

$$\Gamma_\kappa = b_\kappa F_\kappa^* \left(\left\{\frac{\tilde{\rho}_\tau}{b_\tau |\mathcal{N}_\tau|}\right\}_{\tau=1}^n\right) \tag{258}$$

for species $\kappa$. Probabilistic rules are important. Given an input configuration $\tilde{\rho}^k(\mathbf{r})$, one assigns new values $\rho^{k+1}(\mathbf{r})$ probabilistically in such a way that the *average* result corresponds to the finite difference approximation to the given reaction-diffusion equation, $\Gamma_\tau$ (for further details see [95]).

An interesting application is the mapping of the Ginzburg-Landau equation onto the automaton. Consider the PDE [76]

$$\frac{\partial z}{\partial t} = D\nabla^2 z + az - b|z|^2 z, \tag{259}$$

which can be viewed (for $D$ real) as a two-species reaction-diffusion system by separating real and imaginary parts as $z = x + iy$, $a = \alpha + i\gamma$, and $b = \beta + i\delta$,

$$\frac{\partial x}{\partial t} = D\nabla^2 x + \alpha x - \gamma y + (-\beta x + \delta y)(x^2 + y^2), \tag{260}$$

$$\frac{\partial y}{\partial t} = D\nabla^2 y + \alpha y + \gamma x + (-\beta y - \delta x)(x^2 + y^2), \tag{261}$$

where $x$, $y$ are space- and time-dependent variables. With $z = re^{i\phi}$, the corresponding rate equation can be conveniently written as

$$\frac{\partial r}{\partial t} = \alpha r - \beta r^3 = \beta r\left(\alpha/\beta - r^2\right), \tag{262}$$

$$\frac{\partial \phi}{\partial t} = \gamma - \delta r^2. \tag{263}$$

One can set $\beta = 1$ by changing the time scale. The stable homogeneous solution is $z = r_s e^{i\Omega t}$ with $r_s = \sqrt{\alpha/\beta}$ and $\Omega = \gamma - \delta r_s^2$ (A steady but unstable solution is $z = 0$). Since the equation for $r$ is independent of $\phi$, one can transform the solution for $\Omega = 0$ to any given $\Omega$ by multiplying $z(t)$ with a factor $e^{i\omega t}$, yielding an oscillating solution. The full PDE system, (259), also admits oscillating and rotating spirals, and other inhomogeneous solutions.

Study of the homogeneous solutions shows that the CA behaves like an explicit finite difference method with added noise. In the automaton, the noise is intrinsic (it arises from the discretization) but has little effect on the amplitude of the solution (although it introduces random drifts in the phase). An interesting situation concerns spiral wave solutions. By starting with the initial condition

$$R(r_x, r_y, t^0) = c_1 \sqrt{(r_x - r_x^0)^2 + (r_y - r_y^0)^2} \;, \tag{264}$$

$$\theta(r_x, r_y, t^0) = c_2 \arctan \frac{r_y - r_y^0}{r_x - r_x^0} \;, \tag{265}$$

which creates exactly one phase singularity at $(r_x^0, r_y^0)$, for nonzero $\delta$ smaller than a critical value, a spiral develops and rotates steadily after some time as shown in Fig. 26(left panel). The value of the wavelength (obtained by an Archimedian spiral fit) is in agreement with the theoretical value [96] ( for small enough $\Delta t$).

FIG. 26. Spiral waves in the Ginzburg-Landau system obtained by CA simulation. Left panel: Automaton size $200^2$ sites (Superimposed on the grayscale plot of $x$ is the fitted Archimedian spiral). Right panel: Automaton size $500^2$ sites.

As an example of greater complexity, Fig. 26(right panel) shows the simulation of a large system ($500^2$ sites) initialized in the state $z \approx 0$: Under such conditions many interacting spirals develop. Indeed, the initial state is unstable, and because of the intrinsic (low level) noise, different regions depart from this unstable state with different phase values $\phi$. This situation automatically creates many phase singularities (points with $r = 0$, surrounded by points with all values of $\phi$), which then develop into spirals. These phase singularities can merge and move as they are influenced by each other [97]. Such large system simulations are made possible by the speed advantage that this class of CA – as compared to the class of LGA [27] – offers over other numerical methods for solving PDEs. Further applications include the study of spiral patterns [98] and the spatial coexistence of different patterns [99].

---

[27]Note however that LGA simulations performed on the CAM-8 machine compare in computation speed to PDE solvers.

## C. Remarks

The brief discussion in this section was intended to illustrate how the reactive lattice gas method differs from both traditional and more modern CA modeling of reaction-diffusion systems. In the reactive LGA method the dynamics is modeled at the mesoscopic level where one attempts to construct a fictitious albeit faithful dynamics that describes the reactive and non-reactive collision events. The reaction probability matrix that controls the reactive dynamics may be built on the reaction mechanism and suitably constructed to yield the reaction-diffusion equation in the mean field limit. If this matrix is properly constructed as described in Sec. V, then particle number fluctuations that arise from reaction and diffusion will be described correctly. Features such as wave dispersion and curvature are automatically incorporated and follow from the mesoscopic dynamics. No special additions to the rule are needed to achieve these results.

Of course, since the model building takes place at the level of the mechanism the reactive LGA rule is as complex as the mechanism. If one is interested solely in the macroscopic wave propagation properties then a PDE level of description will suffice and CA models, even rather complex ones, may provide a computationally efficient way to simulate this macroscopic reactive dynamics. The results in the preceding subsection describe one such method. However, as stressed elsewhere in this paper, the reactive LGA allows one to treat fluctuations in a realistic manner and can be used to explore regimes that are inaccessible by PDE or simple CA methods.

# XII. PERSPECTIVES

There are circumstances when only the global behavior of spatially-distributed reacting systems on long time and distance scales is of interest. In this domain macroscopic reaction-diffusion equations provide an appropriate way to model the dynamics and, provided the systems are sufficiently simple, the boundary value problems implied in the solutions of these partial differential equations can be implemented and solved numerically. However such a macroscopic model is not capable of describing the system on short distance and time scales, nor can it account for microscopic fluctuations which can be amplified by the dynamics to macroscopic scale. Since reaction-difusion equations are "mean field" in character they cannot give a full description of the correlations which can lead to the breakdown of standard mass action rate laws. Also, the systems of interest may have complex, inhomogeneous structure, as in porous media, and it may be impossible or very difficult to implement the boundary conditions on the reaction-diffusion equations in such cases.

The reactive lattice gas automata described in this paper are designed to provide a microscopic approach to complex, spatially-distributed, reacting systems and allow an analysis of these systems at a mesoscopic level. Indeed it is an ideal method for the exploration of reactive dynamics on the interesting scales that lie between the microscopic, where full molecular dynamics must be used, and the macroscopic, where reaction-diffusion equations suffice. In addition, it can be used to explore the dynamics of systems near bifurcation points or in chaotic regimes where fluctuations can have important effects.

Since the method is based on a well defined reactive particle dynamics with integer-valued state variables, it does not suffer from any of the instabilities of traditional finite-difference schemes. It is easily applied to systems with complex boundaries since one need only include collisions with the "walls" to account for their presence. [101] Since the method is akin to a molecular dynamics simulation in that particle dynamics is followed, simulations of a given system (with simple geometry) are generally slower than PDE or traditional cellular automaton models – with the advantage that LGA provide a much deeper level of description of the system. Nevertheless the reactive LGA can be implemented very efficiently on parallel machines as well as special purpose machines such as the CAM-8 [102] and in these circumstances the speed of LGA simulations may surpass that of reaction-diffusion PDEs. However, we stress that the real utility of these reactive LGA methods lies in the more fundamental treatment of the system that they provide.

The applications in sections 6 to 9 show the types of new information that can be obtained from the study of reactive LGA models. In particular we should stress a few important points:

– Simulation results and analytical developments confirm the validity of the phenomenological description of Reaction-Diffusion systems when the Boltzmann hypothesis is applicable; when the frequency of reactive collisions becomes large, non-Boltzmann effects are observed and the phenomenological description breaks down;

– LGA provide qualitative insight in nucleation processes and in the early stages of pattern formation;
– the role of fluctuations can be studied quantitatively;
– LGA results on reactive chaos provide a physically motivated basis to phenomenological treatments;
– reactive LGA have the same level of validity and fundamental character as Master Equation models with the advantage that with LGA we perform mesoscopic simulations on large, spatially-distributed systems which could hardly be realized with Master Equation methods and Molecular Dynamics techniques.

Many problems which could be investigated from the LGA approach remain unexplored. For instance, very little work has been done on the application of these methods to reactive flows [28] and extensions of the theory are necessary before this problem can be fully mastered. Another question which remains open is the incorporation of energy levels in the automaton – as in thermal LGA [40] – to account for the kinetic processes at a foundamental level. The class of systems that can be treated by the methods described here is in fact much broader than the specific applications indicated. An interesting extension is the modeling of polymerization through hetrogeneous catalysis. [101] Many diverse problems can actually be cast into the form of a reactive dynamics. For example, the dynamics of the populations of excited atomic or molecular states in lasers can, in some circumstances, be modeled as reactions between different "chemical" species. [103] There is an even broader class of phenomena in biology that lend themselves to such modeling. One may generalize the notion of automaton particles to have them represent, for example, motile cells (bacteria or amoebae [104,105]) undergoing "random walks" yet showing collective behavior at the macroscopic level, just as the molecules in some of the chemical examples treated in this paper. Thus, the reactive lattice gas automaton can provide a theoretical and computational method for the study of diverse classes of phenomena at the mesoscopic scale. In some of these classes there are open questions where the automaton approach offers interesting perspectives.

---

[28]Recent work on nonlinear reactions advected by a flow using the Lattice Boltzmann method shows how flows can modify the resulting effects of the reactive processes. [100]

## Acknowledgements

This work was supported in part by grants to RK and AL from the Natural Sciences and Engineering Research Council of Canada, and to JPB and DD from the European Community (grants ST2J-0190 and SC1-0212). DD acknowledges the Donors of The Petroleum Research Fund, administered by the American Chemical Society, for partial support of this research. JPB and DD acknowledge support from the Fonds National de la Recherche Scientifique (FNRS, Belgium). We would also like to express our gratitude to our colleagues who have participated in the research presented in this paper. In particular we thank K. Diemer, D. Gruner, B. Hasslacher, Y.-X. Li, P. Masiar, S. Ponce-Dawson, D. Rusu, S. Succi, J.P. Voroney, J.R. Weimar and X.-G. Wu.